%% file: CoursCosmo.tex
\documentstyle{article}
\def\beq#1{\begin{equation}#1\end{equation}}
\def\bea#1{\begin{eqnarray}#1\end{eqnarray}}
\def\be{\begin{equation}}
\def\ee{\end{equation}}
\def\ba{\begin{eqnarray}}
\def\ea{\end{eqnarray}}
\def\vx{{\bf x}}
\def\vq{{\bf q}}
\def\vu{{\bf u}}
\def\vr{{\bf r}}
\def\vp{{\bf p}}
\def\vv{{\bf v}}
\def\vw{{\bf w}}
\def\vk{{\bf k}}
\def\vg{{\bf g}}
\def\ii{{\rm i}}
\def\d{{\rm d}}
\def\dr{\partial}
\def\grad{\nabla}
\def\rhob{\overline{\rho}}
\def\gradx{\nabla_{\hbox{\rm x}}\,}
\def\gradq{\nabla_{\hbox{\rm q}}\,}
\def\disp{\displaystyle}
\def\dta{\delta}
\def\mg{\big<}
\def\md{\big>}
\def\mA{{\cal A}}
\def\mH{{\cal H}}
\def\mI{{\cal I}}
\def\mS{{\cal S}}
\def\mG{{\cal G}}
\def\mGd{{\cal G}_{\delta}}
\def\mGt{{\cal G}_{\theta}}
\def\WTHt{W\left(\vert\vk_1+\vk_2+\vk_3\vert\ R\right)}
\def\gm{\gamma}
\def\ort{\bot}
\def\xib{\overline{\xi}}

\def \aa    #1 #2   {{\em Astr. \& Astrophys. \/} {\bf #1}, {#2}}

\def \aas   #1 #2   {{\em Astr. Astrophys. Suppl. Ser. \/} {\bf #1}, {#2}}
\def \aj    #1 #2   {{\em Astron. J. \/} {\bf #1}, {#2}}
\def \apj   #1 #2   {{\em Astrophys. J. \/} {\bf #1}, {#2}}

\def \apjs  #1 #2   {{\em Astrophys. J. Suppl. Ser. \/} {\bf #1}, {#2}}
\def \araa  #1 #2   {{\em Annual Review of Astr. \& Astrophys. \/} {\bf #1}, {#2}}
\def \mnras #1 #2   {{\em Mon. Not. R. astr. Soc. \/} {\bf #1}, {#2}}

\def \nat   #1 #2   {{\em Nature \/} {\bf #1}, {#2}}
\def \prevd   #1 #2   {{\em Phys. Rev. D \/} {\bf #1}, {#2}}

\begin{document}
\parskip = 0 pt
\rightline{\large SPhT Saclay, F\'evrier-Mars 1998}
\vspace{1. cm}
\centerline{\huge Cosmologie, la formation }
\centerline{\huge des grandes structures de l'Univers}
\vspace{.5 cm}
\centerline{par}
\vspace{.5 cm}
\centerline{\large\bf Francis Bernardeau}
\vspace{.1 cm}
\centerline{\large Service de Physique Th\'eorique, C.E. de Saclay}
\centerline{\large F-91191 Gif-sur-Yvette Cedex, France}
\date{}

\input{Cours123.tex}

\input{Cours456.tex}

\section{ Application aux propri\'et\'es statistiques de quantit\'es
observables}

Dans cette partie je vais illustrer les r\'esultats des 
th\'eories des perturbations sur des quantit\'es observables,
catalogues de galaxies tri-dimensionnel ou bi-dimensionnel,
champ de distorsion gravitationnelle.

\subsection{Les catalogues de galaxies}

A priori on peut toujours supposer que les fluctuations du nombre
local de galaxies sont une mesure des fluctuations de densit\'e
cosmique. En g\'en\'eral on suppose que
les galaxies forment une repr\'esentation discr\`ete r\'esultant d'un
processus Poissonnien d'un champ de densit\'e continu sous-jascent.
(Ce qui revient \`a dire que les mesures sont affect\'ees d'un
bruit Poissonnien).
 
\begin{figure}
\vspace{5.5 cm}
\special{hscale=40 vscale=40 voffset=-70 hoffset=24 psfile=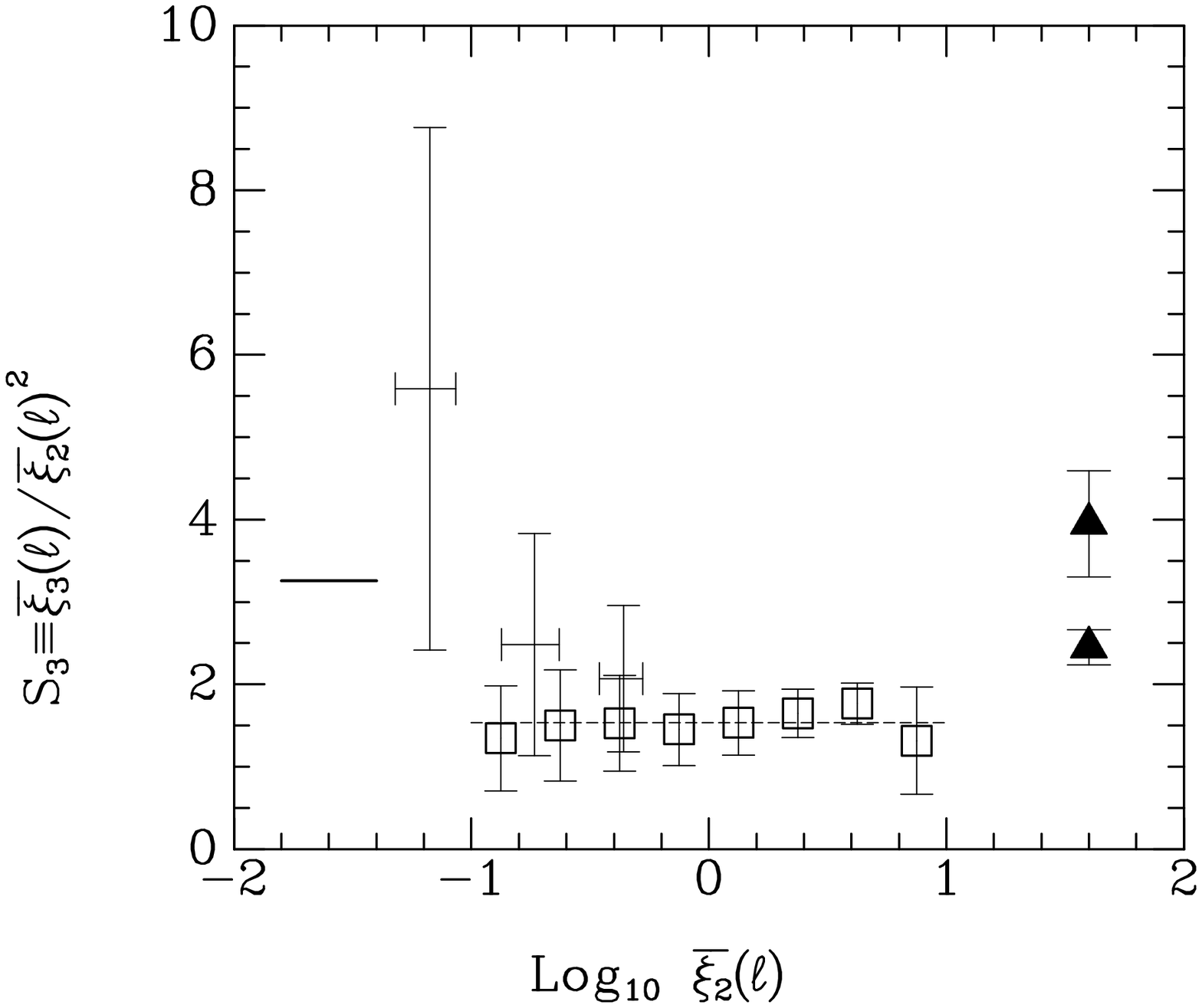}
\caption{La skewness mesur\'ee dans le catalogue IRAS (carr\'es)
compar\'ee \`a des mesures faites pour les galaxies optiques
(triangles). La ligne continue \`a gauche est la pr\'ediction 
de la th\'eorie des perturbations (figure extraite de Bouchet et
al. 1993).
}
\label{s3IRASfit}
\end{figure}

\begin{figure}
\vspace{5.5 cm}
\special{hscale=50 vscale=50 voffset=-50 hoffset=0 psfile=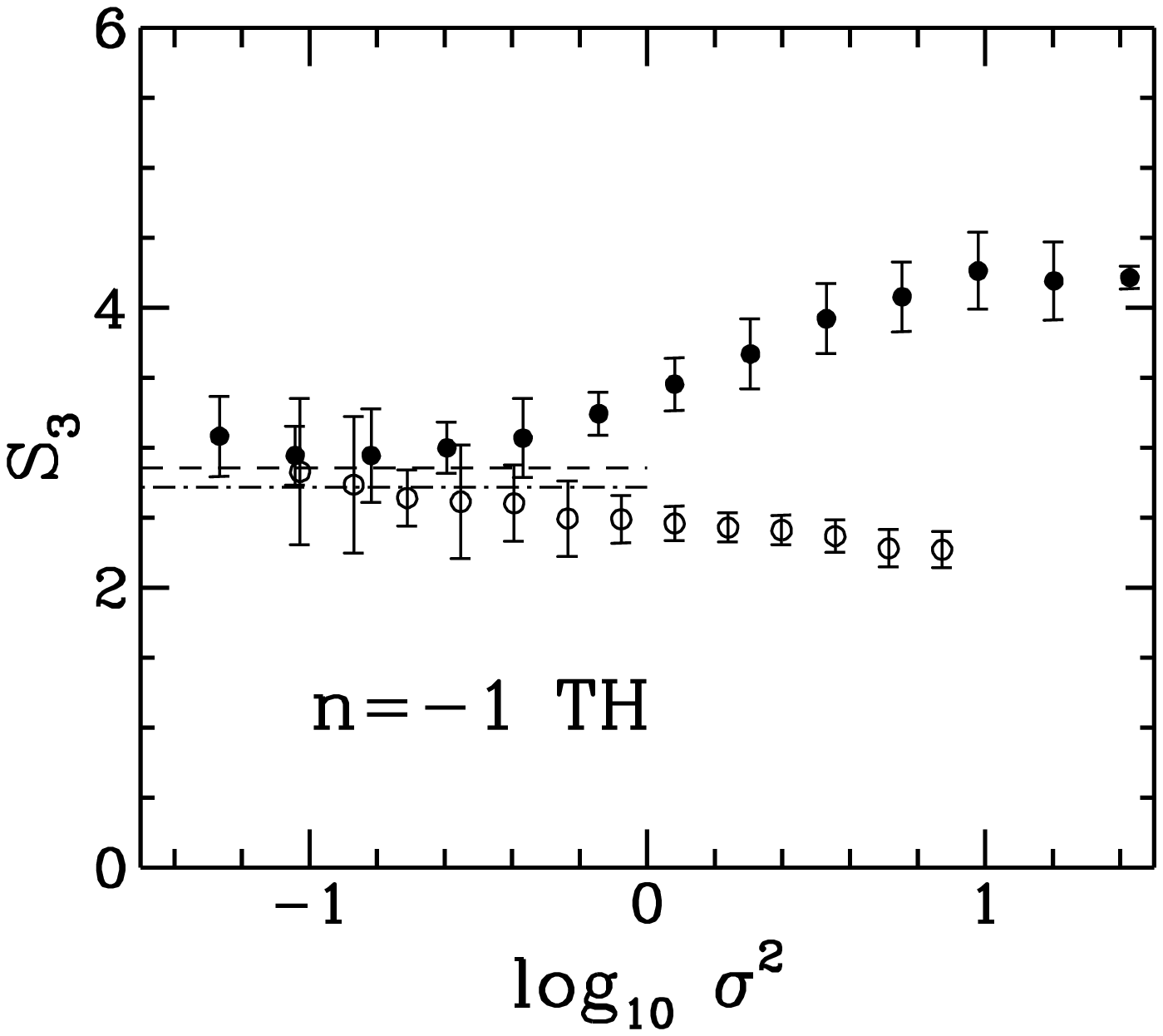}
\caption{La skewness calcul\'ee dans l'espace r\'eelle (ligne
pointill\'ee) et calcul\'ee dans l'espace des redshifts (ligne continue).
Ces pr\'edictions sont compar\'ees \`a des r\'esultats de simulations
num\'eriques, espace r\'eel (symboles pleins) espace des redshift 
(symboles vides). La cas pr\'esent\'e est $n=-1$ pour un filtre
top-hat (figure extraite de Hivon et al. 1995).
}
\label{HivonetalS3}
\end{figure}

Dans les catalogues tri-dimensionnels il faut tenir compte d'un effet
ignor\'e jusqu'\`a pr\'esent qui est d\^u au fait que
les positions des galaxies sont estim\'ees \`a partir de
la vitesse de recession qui combine
la distance r\'eelle (mouvement de Hubble) et la vitesse
propre le long de la ligne de vis\'ee. C'est ce qu'on appelle
la position dans l'espace des redshifts.
Il faut en principe en tenir compte dans le calcul
des $S_p$. Hivon et al. (1995) ont montr\'e que
cette correction ne changeait que tr\`es peu la valeur
de $S_3$ \`a grande \'echelle (enfin tant qu'on peut n\'egliger les 
'doigts de Dieu', voir figure \ref{HivonetalS3}).

\begin{figure}
\vspace{10 cm}
\special{hscale=50 vscale=50 voffset=-20 hoffset=0 psfile=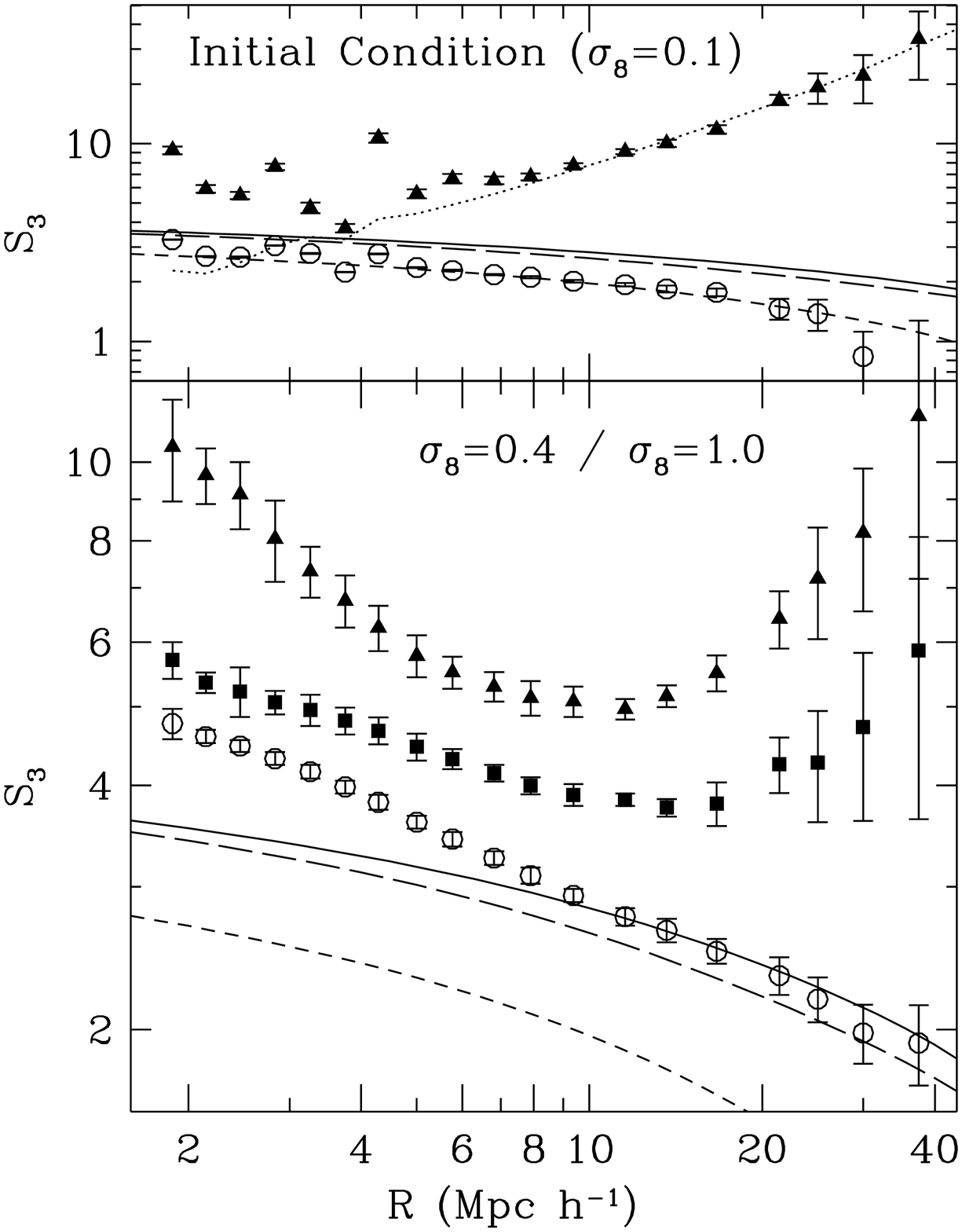}
\caption{La skewness pour des mod\`eles \`a conditions initiales
Gaussiennes (symboles vides) compar\'e \`a des mod\`eles
avec conditions initiales non-Gaussiennes de type textures
(symboles pleins) pour diff\'erentes valeurs de la variance.
Figure extraite de Gazta\~naga \& M\"ah\"onen
(1996).
}
\label{NGInitCond}
\end{figure}

Des mesures de ces param\`etres ont \'et\'e faites
dans tous les grands catalogues de galaxies \`a notre
disposition par Bouchet et al. (1992, 1993 pour le survey de galaxies
infrarouges IRAS), Gazta\~naga (1992, 1993, et 1994 pour les surveys 
optiques CfA, SSRS et APM).

Toutes ces mesures montrent
que les comportements hi\'erarchiques attendus sont bien
v\'erifi\'es. C'est en soit un r\'esultat extr\`emement important.
Cela implique que les conditions initiales pour la formation des
grandes structures \'etaient sinon Gaussiennes du moins proche
d'\^etre Gaussiennes. Par exemple les donn\'ees excluent
a priori les mod\`eles de textures (Gazta\~naga \& M\"ah\"onen
1996, voir figure \ref{NGInitCond}). 
En effet on s'attend \`a ce que la hi\'erarchie soit
pr\'eserv\'ee m\^eme si les galaxies sont biais\'ees. 
(du moins tant que les int\'eractions non-gravitationnelles
ayant donn\'e naissance aux galaxies sont locales).

Toutefois les valeurs des $S_p$ mesur\'es ne sont pas
en accord avec les pr\'edictions th\'eoriques. C'est \`a ce niveau
l\`a que se pose le probl\`eme des biais. c'est \`a ce jour
l'interpr\'etation la plus raisonnable pour expliquer les
diff\'erences persistantes qui subsistent.

\subsection{Les catalogues bi-dimensionnels}

\begin{figure}
\vspace{10 cm}
\special{hscale=50 vscale=50 voffset=-70 hoffset=0 psfile=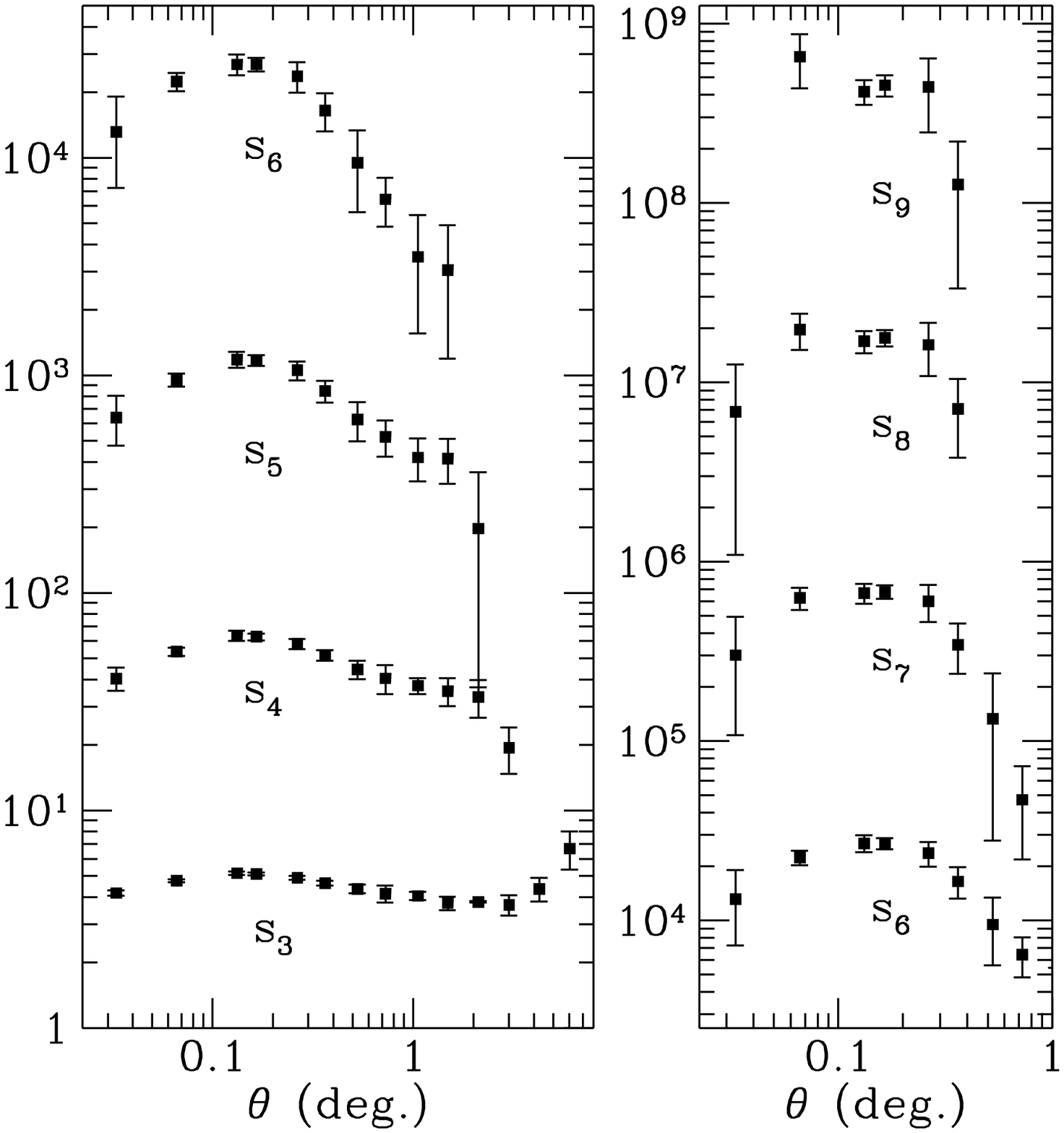}
\caption{Les param\`etres $s_p$ mesur\'es dans le catalogue
angulaire APM (Gazta\~naga 1993).
}
\label{spAPM}
\end{figure}

Les donn\'ees les plus riches actuellement sont dans les
catalogues angulaires. Ils donnent la position angulaire sur
la vo\^ute c\'eleste de plus d'un million d'objets.
De ce fait il est possible de mesurer les param\`etres $s_p$
jusqu'\`a de tr\`es grands ordres (voir figure \ref{spAPM}).

On peut tout aussi bien se poser la question du calcul des
probabilit\'es de comptages dans ce type de catalogues.
Cela correspond \`a un filtrage qui n'est plus \`a sym\'etrie
sph\'erique: la fen\^etre de filtrage a maintenant la forme d'un
c\^one dans l'espace r\'eel. Plus pr\'ecis\'ement 
la densit\'e projet\'ee $w(\gamma)$ dans la direction $\gamma$
vaut,
\beq{
w(\gamma)=\int_0^{\infty}\,r^2\d\,r\,F(r)\,\rhob\,(1+\delta(r,\gamma)),
}
et la densit\'e filtr\'ee \`a une \'echelle angulaire $\theta$ est
\beq{
w_{\theta}(\gamma)=\int_{\vert\gamma'\vert<\theta}\d^2\gamma'
\int_0^{\infty}\,r^2\d\,r\,F(r)\,\rhob\,(1+\delta(r,\gamma+\gamma')).
}

La fonction
$F$ est directement li\'ee au processus de s\'election des galaxies
du catalogues (limite en magnitude, etc..). Typiquement pour des
s\'elections en magnitude apparente en luminosit\'e
optique on a, 
\beq{
F(r)\sim {1\over r^{0.5}}\,\exp(-r^2/D^2),
}
o\`u $D$ est la profondeur du catalogue. Notons que jusqu'\`a
pr\'esent les catalogues sont peu profonds d'un point de vue
cosmologique et donc on peut supposer que la distance lumineuse
est une simple distance Euclidienne.
On peut toujours normaliser $F$ de telle mani\`ere que 
\beq{
\int_0^{\infty}\,r^2\d\,r\,F(r)\,\rhob=1.
}

Je cherche maintenant \`a exprimer les diff\'erents moments de
la fonction de probabilit\'e de densit\'e angulaire locale.
La fonction de corr\'elation \`a deux points
est donn\'ee par,
\bea{
\mg w(\gamma_1)\,w(\gamma_2)\md=\int r_1^2\d\,r_1\,F(r_1)
\int r_2^2\d\,r_2\,F(r_2)\int{\d^3\vk\over (2\pi)^3}\,P(k)\times
\nonumber\\
\exp\left[\ii k_r(r_1-r_2)+\ii\vk_{\ort}.(r_1\gamma_1-r_2\gamma_2)\right],
}
o\`u $\vk$ est d\'ecompos\'e en sa partie radiale $k_r$
et sa partie ortho-radiale $\vk_{\ort}$.
L'approximation des petits angles consiste \`a remarquer
que la distance angulaire $\vert\gamma_1-\gamma_2\vert$ est faible alors
$k_r$ va \^etre tr\`es petit devant la norme de $\vk_{\ort}$,
du coup,
\beq{
P(k)\approx P(k_{\ort}).
}
Il en r\'esulte que
\beq{
\mg w(\gamma_1)\,w(\gamma_2)\md=\int r^4\d\,r\,F^2(r)\int
{\d^2\vk\over (2\pi)^2}\,P(k)\,
\exp\left[\ii\vk_{\ort}.(\gamma_1-\gamma_2)r\right].
}
En effet l'int\'egrale sur $k_r$ introduit une fonction delta
en $r_1-r_2$. L'approximation des petits angles signifie
simplement que les paires qui contribuent le plus \`a la variance
de la densit\'e projet\'ee viennent de points qui \'etaient \`a la 
m\^eme distance de l'observateur. Ce r\'esultat correspond \`a
l'approximation dite de Limber (1954) sur la focntion de corr\'elation
angulaire. Cette approximation n'est d'ailleurs valable que si la
fonction de corr\'elation est suffisamment pentue.

Pour avoir la variance de la densit\'e filtr\'ee, il suffit 
de prendre la TF de la fen\^etre donc,
dans l'espace de Fourrier, multiplier par la fen\^etre\footnote{A deux
dimensions la fen\^etre top-hat de l'espace r\'eel s'\'ecrit en espace
de Fourier, $W(k)=2J_1(k)/k$} $W(k)$ et on
obtient,
\beq{
\mg w_{\theta}^2\md=
\int r^4\d\,r\,F^2(r)\int {\d^2\vk\over (2\pi)^2}\,P(k)\,
W^2(k\,\theta\,r).
}

Le calculs de la skewness peut se faire de mani\`ere
similaire et on peut utiliser pour un filtrage angulaire top-hat
de propri\'et\'es similaires \`a celles utilis\'ees pour le cas 3D.

Il en r\'esulte que,
\bea{
s_3=&
\disp{\int r^6\d\,r\,F^3(r)\int {\d^2\vk\over (2\pi)^2}\,P(k)\,
W^2(k\,\theta\,r)\over
\left[\int r^4\d\,r\,F^2(r)\int {\d^2\vk\over (2\pi)^2}\,P(k)\,
W^2(k\,\theta\,r)\right]^2}
\times\nonumber\\
&\disp{\left[{36\over 7}
\int {\d^2\vk\over (2\pi)^2}\,P(k')\,
W^2(k'\,\theta\,r)+{3\over 2}{\d \over \log \theta}
\int {\d^2\vk\over (2\pi)^2}\,P(k')\,
W^2(k'\,\theta\,r)\right]}.
}

Cette expression se simplifie quelque peu quand on suppose
qu'on a une loi de puissance pour le spectre $P(k)$ ($P(k)\sim k^n$),
\beq{
s_3=
{\int r^6\d\,r\,F^3(r)\,r^{-2(n+2)}\over
\left[\int r^4\d\,r\,F^2(r)\,r^{-(n+2)}\right]^2}\,
\left[{36\over 7}-{3\over 2}(n+2)\right].
}
La hi\'erarchie compl\`ete des $s_p$ se calcule de mani\`ere
analogue au cas 3D. On a simplement en plus un facteur
g\'eom\'etrique li\'e aux effets de projection. 
Les coefficients purement dynamiques
sont en fait identiques \`a ceux qu'on aurait obtenus si
on \'etait parti d'une dynamique bidimensionnelle.

\begin{figure}
\vspace{10 cm}
\special{hscale=50 vscale=50 voffset=-0 hoffset=0 psfile=projSp.ps}
\caption{Les param\`etres $s_3$ et $s_4$ obtenus par la th\'eorie
des perturbations (lignes continues) compar\'es \`a des r\'esultats
de simulations num\'eriques (Gazta\~naga \& Bernardeau 1998).
}
\label{projSp}
\end{figure}

Encore une fois, les comparaisons avec les catalogues de galaxies montrent un 
d\'esaccord avec les pr\'edictions de la th\'eorie des perturbations,
probablement d\^u aux effets de biais.

\subsection{Les effets de lentille gravitationnelle}

Une motivation pour aller chercher de nouveaux traceurs du champ de
densit\'e est justement
de s'affranchir des effets de biais. Un moyen de tracer les champs de
densit\'e locaux est d'utiliser les effets de distorsion gravitationnelle.
Pour une description g\'en\'eral de ces effets voir Sachs (1961) et
Schneider, Ehlers \& Falco (1992).

\subsubsection{Les lentilles gravitationnelles}

Je vais commencer par rappeler des concepts simples des effets
de lentilles.
Une lentille induit une d\'eflection $\partial\alpha$ des rayons
lumineux pau unit\'e de distance $\partial s$ \'egale \`a,
\beq{
\partial\alpha/\partial s=-2\grad_{\ort}\phi
}
o\`u $\phi$ est le potentiel de la lentille et le gradient
est pris dans le plan orthogonal \`a la trajectoire des photons.
Quand on regarde des objets lointains ces d\'eflections induisent
des d\'eplacements des positions apparentes des objets d'arri\`ere
plan. Le d\'eplacement apparent d\'epend des diff\'erentes
longueurs qui interviennent dans le banc optique.
Il en r\'esulte que  la position apparente d'un objet $\gamma^I$
(dans le plan dit image) est li\'ee \`a sa position
r\'eelle dans le plan source $\gamma^S$ par,
\beq{
\gamma^I=\gamma^S+\xi(\gamma^I)
}
avec
\beq{
\xi(\gamma^I)=-2{D_{LS}\over D_{OS}\,D_{OL}}\grad_{\gamma}\Phi.
}
o\`u $D_{LS}$, $D_{OS}$ et $D_{OL}$ sonr respectivement
les distances angulaires entre la lentille et les sources,
l'observateur et les sources et l'observateur et la lentille.
D'autre part $\grad_{\gamma}\phi$ est la d\'eriv\'ee angulaire du
potentiel projet\'e $\Phi$. 

Cependant le champ de d\'eplacement n'est pas une observable
directe (sauf dans les cas d'images multiples). Un objet (faiblement)
\'etendu comme une galaxie va \^etre non seulement d\'eplac\'e
mais aussi d\'eform\'e. La matrice de d\'eformation est donn\'ee
par la matrice 2x2 des d\'eriv\'ees secondes du champ de
d\'eplacement,
\beq{
\mA\equiv \xi_{i,j}=\delta_{ij}-2{D_{LS}\over D_{OS}\,D_{OL}}
\Phi_{,ij}.
}
On voit que la matrice $\mA$ est une matrice sym\'etrique.
On l'\'ecrit en g\'en\'eral,
\beq{
\mA=\left(
\begin{tabular}{cc}
$1-\kappa-\gamma_1$&$-\gamma_2$\\
$-\gamma_2$&$1-\kappa+\gamma_1$
\end{tabular}
\right).
}
La trace de la matrice donne la convergence $\kappa$, les autres
composantes donnent le cisaillement. On va avoir
des lignes critiques (et des images multiples) quand $\kappa\to 1$.

\subsubsection{Les observables}

Pour chaque galaxie on peut calculer une matrice de forme, $\mS$,
\`a partir de la distribution de lumi\`ere de cette galaxie.
Plus pr\'ecis\'ement on d\'efinit,
\beq{
\mS=\disp{\disp{\int\d^2\theta\,\theta_i\,\theta_j\,\mI(\theta)}\over
\disp{\int\d^2\theta\,\theta^2\,\mI(\theta)/2}}
}
En se souvenant que la densit\'e de luminosit\'e d'un 
objet n'est pas affect\'ee par les effets de lentille
on peut relier la matrice de forme d'une galaxie d'arri\`ere-plan
telle qu'elle peut \^etre mesur\'ee dans le plan Image $\mS^I$
\`a celle qu'on aurait obtenue en l'absence de lentilles,
\beq{
\mS^S={\mA\cdot\mS^I\cdot\mA\over \det(\mA)}.
}
Chaque galaxie d'arri\`ere plan permet donc une estimation de la
matrice de distorsion, mais normalis\'ee de telle sorte que
son d\'eterminant vaut 1.
Du coup les matrices de forme donnent acc\`es  aux quantit\'es suivantes,
\beq{
\vg={\gamma \over 1-\kappa}.
}
Cependant une galaxie d'arri\`ere plan seule ne permet pas de faire 
une mesure (sauf cas d'arcs critiques ou quasi-critiques).
Il faut moyenner un certain nombre de galaxies d'arri\`ere-plan 
en supposant que les ellipticit\'es intrins\`eques des galaxies
sont distribu\'ees de mani\`ere al\'eatoire et ind\'ependante.

\subsubsection{La pr\'ecision des mesures possibles}

Le bruit intrins\`eque des mesures est un bruit de type Poissonnien
qui d\'epend de l'ellipticit\'e intrins\`eque des galaxies
d'arri\`ere-plan, donc
\beq{
Bruit\approx {0.3 \over \sqrt{N_g}}
}
Typiquement on d\'etecte environ 50 gal/arcmin$^2$. \`A une \'echelle
de l'ordre de 10 arcmin on a donc un bruit de mesure intrins\`eque
de l'ordre de $5\,10^{-3}$. Autrement dit on est 
en mesure de d\'etecter des signaux cosmologiques \`a cette \'echelle
d\`es qu'ils atteignent l'ordre du \%.

Un amas de galaxies g\'en\`ere des effets de l'ordre de quelques \%
\`a sa p\'eriph\'erie. Les effets de distorsion qu'il induit, m\^eme
en dehors des r\'egions critique, sont donc parfaitement d\'etectables.
En fait m\^eme les grandes structures de l'Univers en g\'en\'erale
sont parfaitement d\'etectables comme l'ont d\'emontr\'e
Blandford et al. (1991), Miralda-Escud\'e (1991), Kaiser (1992) et comme
on va le voir dans la suite.

\subsubsection{la relation entre cisaillement et convergence}

\begin{figure}
\vspace{5 cm}
\special{hscale=50 vscale=50 voffset=-130 hoffset=20 psfile=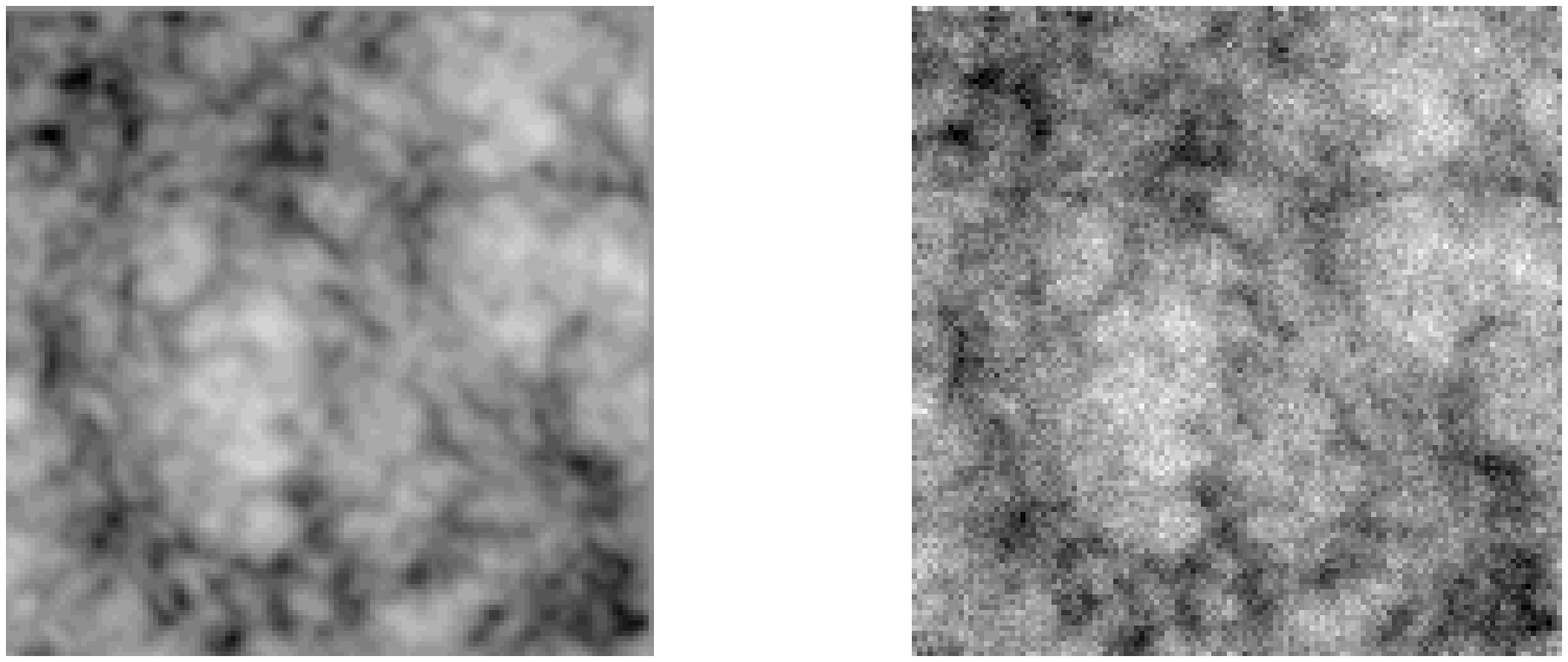}
\caption{Exemple de reconstruction de carte de convergence. La carte
de gauche est une carte simul\'ee de 5x5 degr\'es. La carte
de droite est la carte reconstruite en introduisant du bruit
de mesure pour chaque pixel. On voit \`a petite \'echelle
le bruit Poissonnien des mesures mais \`a grande \'echelle on
retrouve bien les grandes structures (figure extraite de 
Van Waerbeke et al. 1998).
}
\label{kappa_maps}
\end{figure}

En tirant profit du fait que les \'el\'ements de la matrice $\mA$
sont des d\'eriv\'ees secondes d'un m\^eme champ scalaire $\phi$
on peut trouver une relation non-locale entre la convergence locale
et le champ de distorsion. Cette relation explicit\'ee par
Kaiser (1995) s'\'ecrit,
\beq{
\grad \log(1-\kappa)=
\left(\begin{tabular}{cc}
$1-g_1$&$g_2$\\
$g_2$&$1+g_1$
\end{tabular}\right)^{-1}
\cdot
\left(\begin{tabular}{cc}
$\partial_1$&$\partial_2$\\
-$\partial_2$&$\partial_1$
\end{tabular}\right)
\cdot
\left(\begin{tabular}{c}
$g_1$\\
$g_2$
\end{tabular}\right)
}

De cette relation il r\'esulte que le champ de convergence locale 
est a priori observable. En fait, en pratique on ne fait pas une
int\'egration
directe de cette relation mais on recherche le potentiel
$\phi$ qui reproduit au mieux le champ de distorsion mesur\'e
en minimisant un $\chi^2$. La figure (\ref{kappa_maps})
donne un exemple d'une telle reconstruction.

\subsubsection{Variance et skewness de la convergence locale}

Dans un contexte cosmologique la convergence locale est donn\'ee
par l'ensemble des concentrations de masse sur les lignes de vis\'ee.
Ainsi $\kappa$ dans la direction $\gamma$ est donn\'ee
par (dans le cas d'une lentille mince ou par lin\'earisation),
\beq{
\kappa(\gamma)=-{3\over 2}\Omega_0
\int\d z_s\,n(z_s)\,
\int\d\chi{D(z,z_s)\,D(z)\over D(z_s)}\,\delta_{\rm
masse}(\chi,\gamma)\,(1+z(\chi)) 
}
o\`u $z_s$ est le redshift des sources, et $n(z_s)$ leur
distribution en redshift (normalis\'ee \`a l'unit\'e),
$\chi$ est la distance radiale et $D$ les distances angulaires
qui interviennent (exprim\'ees en unit\'es $c/H_0$). 
En g\'en\'eral les distances cosmologiques
qui interviennent sont diff\'erentes, sauf dans le cas d'une
section spatiale plane.

On voit que la convergence locale est formellement 
tr\`es similaire \`a la
densit\'e projet\'ee dans un catalogue angulaire. La diff\'erence
est que la fonction de s\'election n'est pas ici normalis\'ee.
En utilisant l'approximation des petits angles on peut calculer les
premiers moments de cette distribution.

\begin{figure}
\vspace{12 cm}
\special{hscale=60 vscale=60 voffset=240 hoffset=-20 psfile=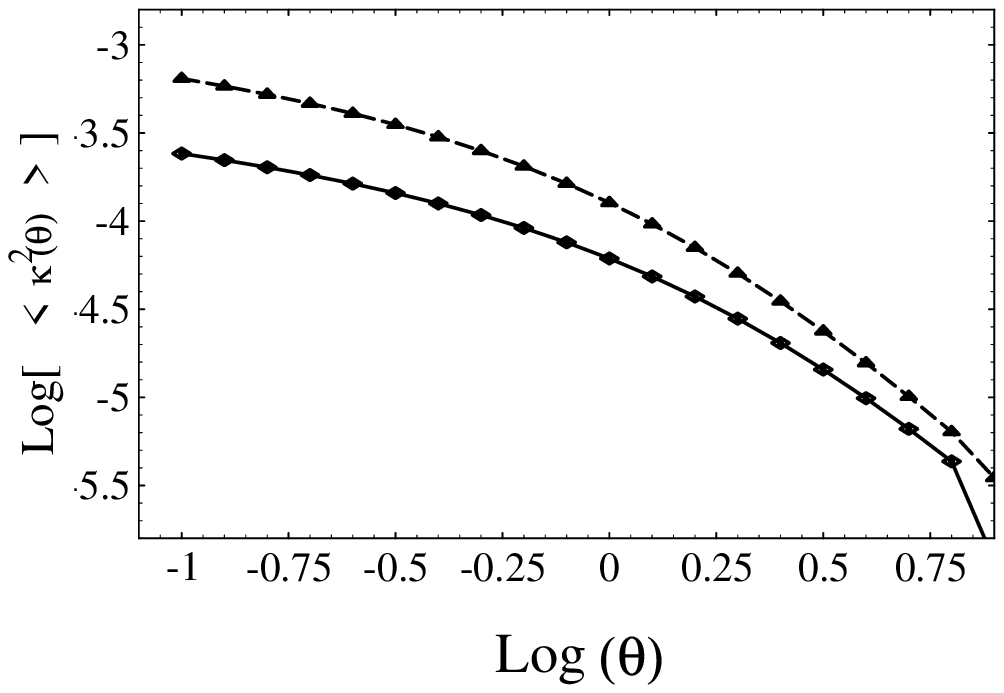}
\special{hscale=60 vscale=60 voffset=120 hoffset=-10 psfile=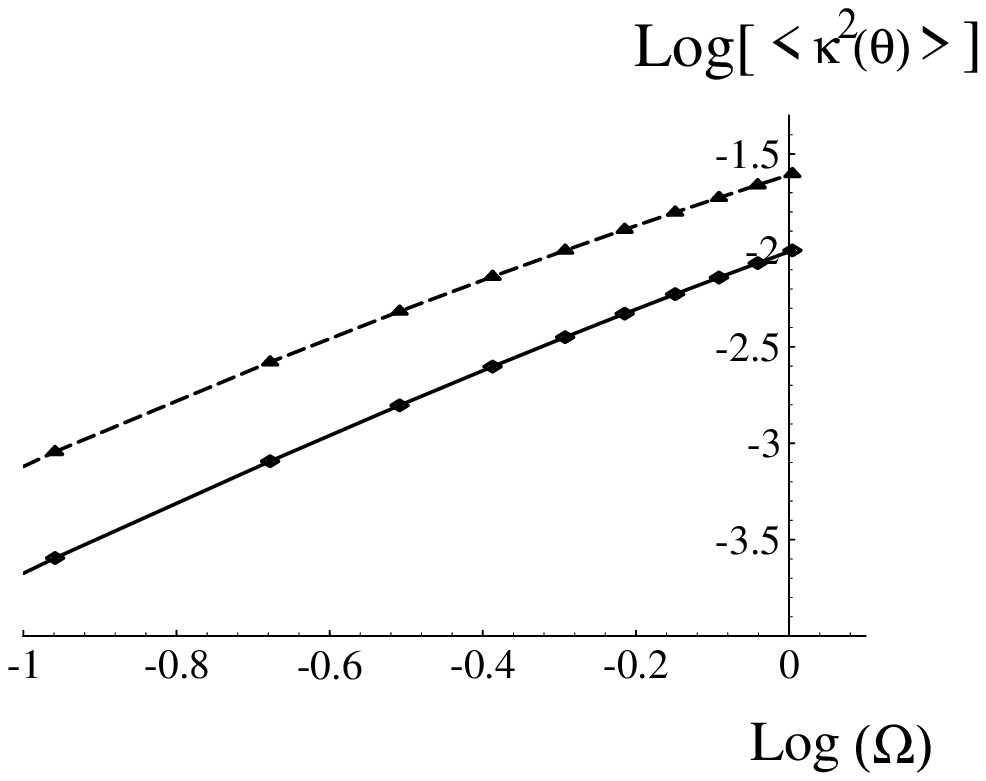}
\special{hscale=60 vscale=60 voffset=-20 hoffset=-20 psfile=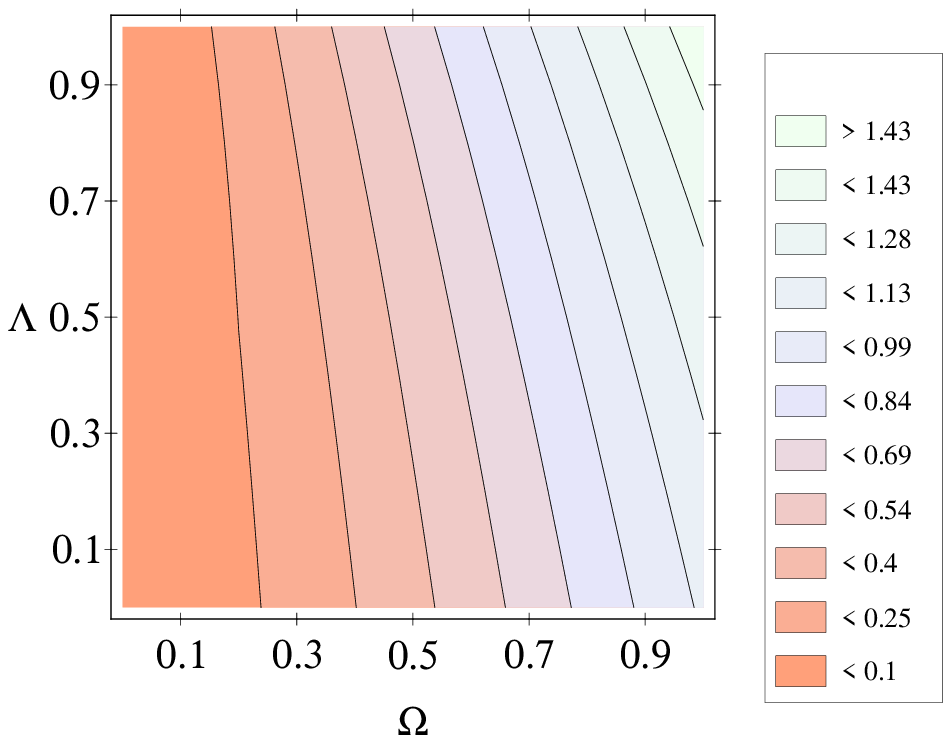}
\special{hscale=60 vscale=60 voffset=240 hoffset=170 psfile=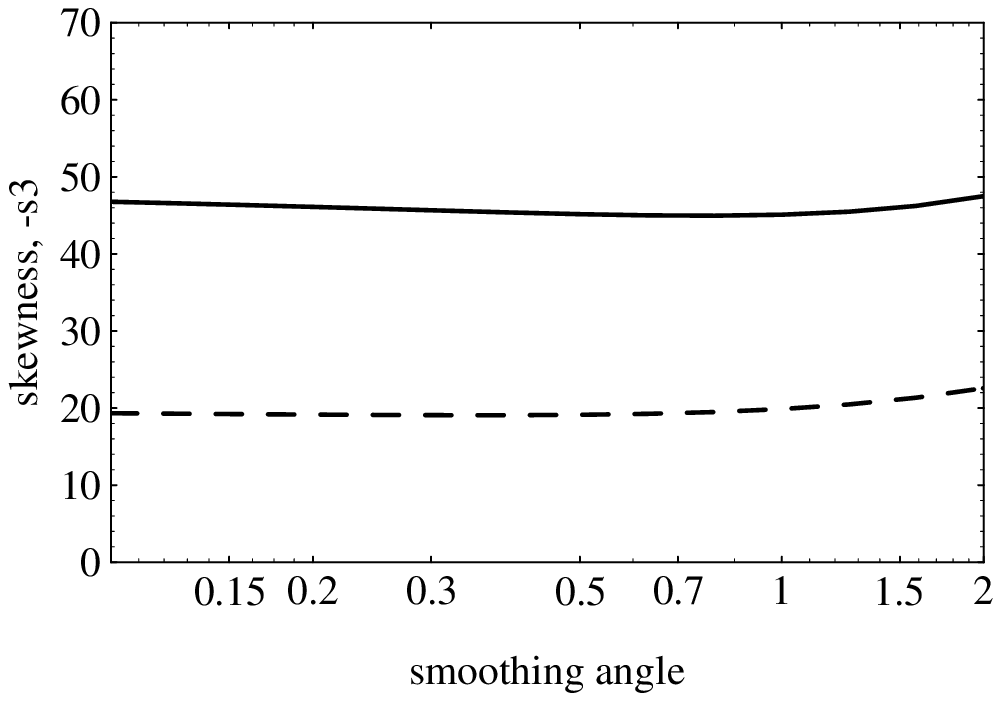}
\special{hscale=60 vscale=60 voffset=120 hoffset=190 psfile=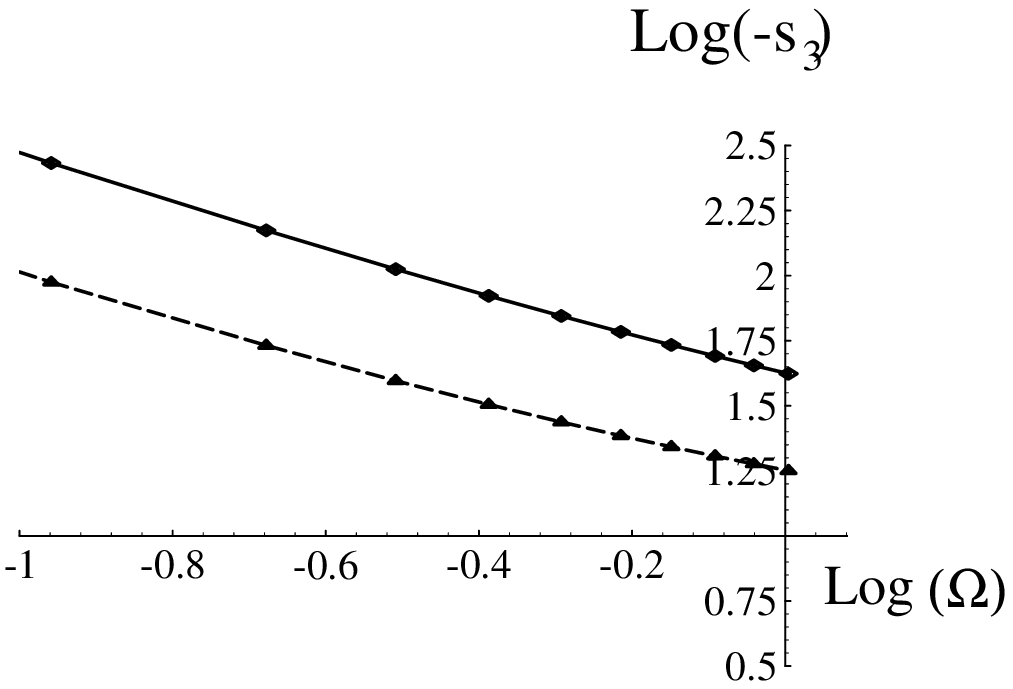}
\special{hscale=60 vscale=60 voffset=-20 hoffset=170 psfile=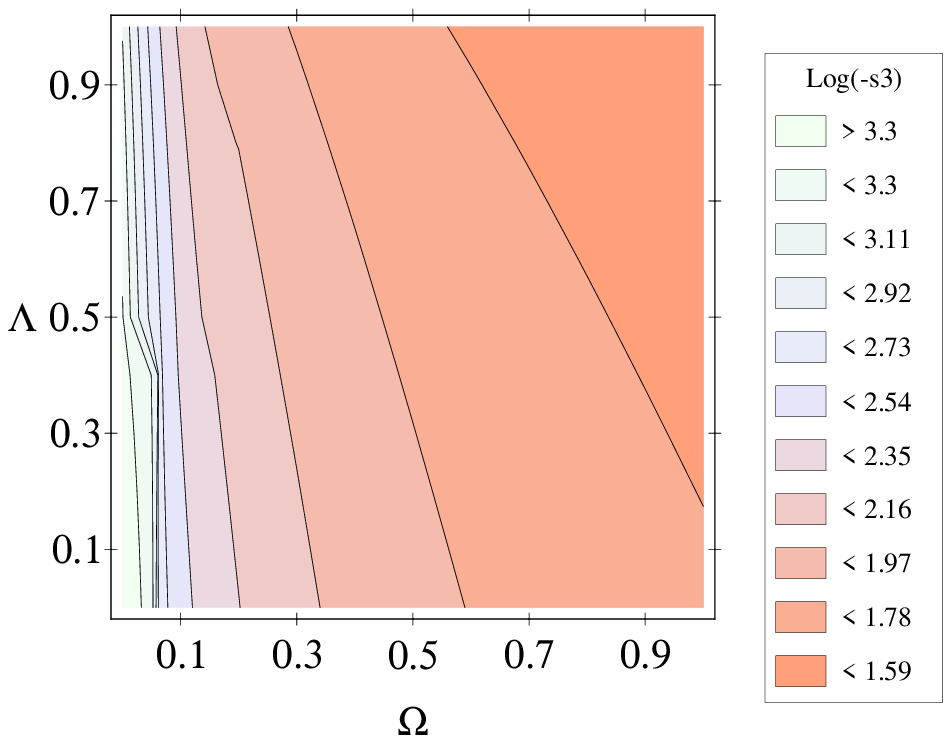}
\caption{La variance (panneaux de gauche) et la skewness (panneaux de
droite) de la convergence locale. En haut est donn\'ee la d\'ependance
avec l'\'echelle angulaire pour deux populations diff\'erentes de
sources ($z_s=1$, lignes continues), ($z_s=2$, lignes pointill\'ees). 
Au centre est repr\'esent\'ee la d\'ependance avec $\Omega_0$ et en bas la
d\'ependance conjointe avec $\Omega_0$ et $\Lambda_0$ pour
une population de sources \`a $z_s=1$ (figures extraites de Bernardeau
el al. 1997).
}
\label{kappaStats}
\end{figure}

On trouve que (Bernardeau, van Waerbeke \& Mellier 1997)
\beq{
\mg\kappa^2_{\theta}\md^{0.5}\approx 0.01\ \sigma_8\ \Omega_0^{0.75}\ 
z_s^{0.75}\ \left({\theta\over 1\deg}\right)^{-(n+2)/2},
}
et
\beq{
s_3(\theta)\equiv{\mg\kappa^3_{\theta}\md\over
\mg\kappa^2_{\theta}\md^2}=
-40\ \Omega_0^{-0.8}\ z_s^{-1.35}.
}

Il est sans doute utile de faire un certain nombre de commentaires,
\begin{itemize}
\item La variance d\'epend non seulement de l'amplitude
des fluctuations (caract\'eris\'ee par $\sigma_8$) mais
aussi de $\Omega_0$.
\item La skewness est ind\'ependante de l'amplitude des fluctuations
(comme on s'y attend pour des conditions initiales Gaussiennes), mais
d\'epend de $\Omega_0$. La mesure conjointe de la variance et de $s_3$
est n\'ecessaire pour lever la d\'eg\'en\'erescence entre $\sigma_8$ et
$\Omega_0$.
\item Ces deux quantit\'es d\'ependent de la distribution en redshift
de la population de galaxie sources utilis\'ee pour faire les mesures.
\end{itemize}

Pour l'instant aucune mesure de ces quantit\'es n'a \'et\'e faite.
On en est \`a des \'etudes de faisabilit\'e et des investigations sur
les erreurs syst\'ematiques possibles (projet DESCART avec
Y. Mellier). Parmi les sources d'erreurs syst\'ematiques possibles on
peut noter,
\begin{itemize}
\item les couplages non-lin\'eaires des lentilles entre-elles en cas
de lentilles multiples;
\item l'existence de couplages observationels suppl\'ementaires dues
en particulier \`a l'effet d'amplification qui peut changer la
population d'objets s\'electionn\'es;
\item les effets de clustering des galaxies sources.
\end{itemize}
Ces sources d'erreur affectent la robustesse de l'interpr\'etation
cosmologique des cartes produites. A cela il faut encore ajouter
une bonne d\'etermination des redshifts des sources, mais aussi
s'assurer que les cartes produites ont bien une origine cosmologique,
autrement dit que la qualit\'e optique instrumentale, le suivi du
t\'elescope, etc, permettent bien la r\'ealisation de cartes fiables.
C'est \`a ce jour l\`a que les efforts les plus importants sont faits.

\section{Vers le r\'egime fortement non-lin\'eaire}

\subsection{ Donn\'ees exp\'erimentales}

Les donn\'ees exp\'erimentales sugg\`erent que la hi\'erarchie
obtenue avec la th\'eorie des perturbations se prolonge
dans le r\'egime non-lin\'eaire. C'est particuli\`erement vrai dans 
l'espace des redshift (voir la figure \ref{HivonetalS3}).

Par exemple on sait que la fonction de corr\'elation \`a trois
points des galaxies peut se d\'ecrirent par le produit de paires 
de fonctions de corr\'elation \`a deux points (Groth \& Peebles 1977), i.e.,
\beq{
\xi(\vr_1,\vr_2,\vr_3)=Q\,
\left[\xi(\vr_1,\vr_2)\xi(\vr_2,\vr_3)+{\rm sym.}\right]
}
o\`u $Q$ est de l'ordre de l'unit\'e. Dans le cas de la th\'eorie des perturbations
$Q$ est en fait une fonction sans dimension des angles et des distances
entre points. Cette d\'ependance semble dispara\^\i tre au profit de cette
forme plus simple dans le r\'egime non-lin\'eaire. Des tentatives ont
\'et\'e faites pour mesurer des fonctions de corr\'elation d'ordre plus 
\'elev\'e mais cela devient tr\`es ardu \`a cause des d\'ependances g\'eom\'etriques.

Existe-t-il une explication naturelle \`a ces propri\'et\'es?

\subsection{ Les solutions auto-similaires}

On aborde ici un r\'egime qui est tr\`es non-lin\'eaire et une description
avec un fluide avec un seul flot serait donc inadapt\'ee.
Il faut revenir \`a l'\'equation d'\'evolution dans l'espace des phases,
\bea{
{\partial \over \partial t}\,f(\vx,\vp,t)+{\vp\over m\, a^2}
{\partial\over \partial\vx}
\,f(\vx,\vp,t)-m\nabla_{\vx}.\Phi(\vx)\,{\partial\over \partial\vp}
\,f(\vx,\vp,t)&=&0\label{diffdef}\\
\Delta\Phi(\vx)={4\pi\,G\,m\over a}\int\,f(\vx,\vp,t)\d^3\vp.&&\nonumber
}
On peut facilement montrer qu'il existe une famille de solutions possibles
dites auto-similaires qui ont la d\'ependance en temps suivante,
\beq{
f(\vx,\vp,t)=t^{-3\beta}\,\hat{f}\left(\vx/t^{\alpha},\vp/t^{\beta}\right)
}
o\`u les deux constantes satisfont,
\beq{
\beta=\alpha+1/3.
}
Cette relation s'obtient assez facilement en r\'e\'ecrivant
l'\'equation (\ref{diffdef}) avec le changement de variable
$\vu=\vx/t^{\alpha}$ et $\vv=\vp/t^{\beta}$. C'est la condition
pour que la d\'ependance en temps se factorise.
Il est impotant de remarquer qu'alors les fonctions de
corr\'elations ne sont fonctions que de $\vu$: elles ne d\'ependent plus
explicitement du temps.

Je donne ici les relations de similarit\'e pour le cas d'un Univers
Einstein-de Sitter (Davis \& Peebles 1977). Il existe des formes
similaires pour le cas d'un univers ouvert dans la limite $\Omega\ll1$
(Balian \& Schaeffer 1988). 
On a vu que la virialisation des objets entra\^\i nait un d\'ecouplage 
de leur dynamique par rapport \`a l'expansion g\'en\'erale. Cette propri\'et\'e
se retrouve au niveau de la fonction de corr\'elation \`a deux
points. On s'attend en effet \`a ce que lorsque cette fonction 
devient grande elle tende simplement \`a s'opposer directement \`a
l'expansion. Le nombre de voisins d'un objet donn\'e doit en effet
rester constant d\`es que l'on se trouve au sein d'un objet tr\`es
compact, en \'equilibre gravitationnel (comme une galaxie..). Dans la
limite non-lin\'eaire ($\xi_2(r)\gg 1$), la densit\'e au voisinage
d'un point s'\'ecrit, 
\beq{
\rho'(\vx)=\rhob(t)\ \xi_2(x,t).
}
Il faut alors que $\xi(x,t)$ se comporte comme $a^3$ {\it \`a une
distance r\'eelle constante} pour compenser les effets de dilution
caus\'es par l'expansion, soit, 
\beq{
\xi_2(x,t)\sim a^3\,\hat{\xi}(a\,x).
}
On peut g\'en\'eraliser cette id\'ee  aux fonctions de corr\'elation
de plus grand ordre et on obtient naturellement,
\beq{
\xi_p(\vx_1,\dots,\vx_p,t)\sim
a^{3(p-1)}\,\hat{\xi_p}(a\,\vx_1,\dots,a\,\vx_p). 
}
Si on combine ces lois de conservation qui traduisent les hypoth\`eses
de corr\'elation stable avec la forme g\'en\'erale attendue pour les
solutions auto-similaires, on obtient des lois d'\'echelle pour ces
fonctions de corr\'elation, 
\bea{
\xi_2(\lambda\,x,t)&=&\lambda^{-\gamma}\,\xi_2(x,t)\\
\xi_p(\lambda\,\vx_1,\dots,\lambda\,\vx_p,t)&=&\lambda^{-(p-1)\gamma}
\xi_p(\vx_1,\dots,\vx_p,t)
}
o\`u $\gamma$ est li\'e \`a l'exposant $\alpha$ par 
\beq{
\gamma=6/(2+3\alpha).
}
Il suffit d'\'ecrire que $a^3\hat{\xi}(ax)=\hat{\xi}(x/t^{\alpha})$.
De plus si on suppose que la solution auto-similaire est toujours
valable, du r\'egime lin\'eaire au r\'egime non-lin\'eaire, (ce qui
stricto sensus implique le spectre de puissance initial suit une loi
de puissance), alors on peut relier l'exposant $\gamma$ \`a $n$:
(Davis \& Peebles 1977), 
\beq{
\hat{\xi}\left({x\over t^{\alpha}}\right)=x^{-(n+3)}\,t^{4/3}
\ \Rightarrow\ 
\alpha={4\over 3}{1\over n+3}
\ \Rightarrow\ 
\gamma={9+3\,n\over 5+n}.
}
Cette relation est finalement assez bien v\'erifi\'ee et elle traduit
une propri\'et\'e importante qui est que la d\'ependance spatiale de
la fonction de corr\'elation non-lin\'eaire doit pouvoir se
reconstruire \`a partir du spectre initial. 

\subsection{La transform\'ee lin\'eaire-nonlin\'eaire \`a la Hamilton et al.}

Cette id\'ee a \'et\'e exploit\'ee \`a son maximum par Hamilton et
al. (1991) qui ont montr\'e qu'il existait une relation, au moins \`a
peu pr\`es universelle, qui reliait la fonction de corr\'elation
lin\'eaire \`a la fonction de corr\'elation non-lin\'eaire. Pour
\'ecrire cette relation il faut \'ecrire une esp\`ece de relation de
conservation de la mati\`ere,
\beq{
r^3\,(1+\xi(r,t))=constante
}
alors $\xi(r,t)$ est une fonction de $\xi_{\rm lineaire}$ au point
$r_l$ de l'espace r\'eel auquel on a
\beq{
r_l^3=r^3(1+\xi(r,t)).
}
Dans le r\'egime lin\'eaire, quand $\xi$ est petit, $r_l$ et $r$ sont
\'evidemment confondus, mais ce n'est plus le cas par la suite. L'id\'ee
v\'erifi\'ee par Hamilton et al. est qu'il existe une fonction $\mH$ 
(une simple fonction pas une fonctionnelle) telle que,
\beq{
\xi(r,t)=\mH\left[\xi_{\rm lineaire}(r_l,a_{\rm initial})\,{a^2(t)\over
a^2(t_{\rm initial})} \right].
}
Num\'eriquement ces auteurs ont obtenus,
\beq{
\mH(x)={x+0.358\ x^3+0.0236\ x^5\over
1+0.0134\ x^3+0.00202\ x^{9/2}}.
}
Cette id\'ee a \'et\'e plus r\'ecemment abondamment reprise pour obtenir
des transform\'ees plus pr\'ecises, et valable pour toutes les
cosmologies, s'appliquant \`a $P(k)$ (Peacock \& Dodds 1994, Jain et
al. 1995).

\subsection{ Les mod\`eles hi\'erarchiques}

On a l\`a une description ph\'enom\`enologique du comportement
de la fonction \`a deux points (ou de mani\`ere \'equivalente du spectre
de puissance), mais \'evidemment cela ne constitue pas une description
compl\`ete du champ non-lin\'eaire. Je reviens alors aux propri\'et\'es
d'\'echelle trouv\'ees pour les fonctions de corr\'elation des grands
ordres. 

\subsubsection{D\'ecomposition en arbres des fonctions de corr\'elation}

Un moyen de les r\'ealiser '\`a la main' est de supposer que
la fonction de  corr\'elation \`a l'ordre $p$ s'\'ecrit comme des
produits de $p-1$ fonctions de corr\'elation \`a deux points. C'est
d'ailleurs, comme on l'a vu, un moyen assez juste de d\'ecrire la
fonction de corr\'elation \`a 3 points. 

Cela revient \`a dire que,
\beq{ 
\xi_p(\vx_1,\dots,\vx_p)= \sum_{{\rm config.}\ \alpha}
Q_{p,\alpha}\prod_{\rm liens, (ij)}\xi_2(\vx_i,\vx_j),
\label{xipech}
} 
o\`u les configurations sur lesquelles cette somme est faite
correspondent \`a toutes les mani\`eres possibles de connecter $p$
points ensemble avec des  diagrammes en arbre, les liens sont
l'ensemble des liens qui appara\^\i ssent entre  2 points (i,j).

\subsubsection{Contraintes sur les $Q_{p,\alpha}$}

Un des probl\`emes auquel on est confront\'e quand on veut
construire explicitement de tels mod\`eles hi\'erarchiques
est qu'on n'est pas libre de prendre les param\`etres
$Q_{p,\alpha}$ qu'on veut. Il y a des conditions de positivit\'e
qui donnent un certains nombre de contraintes. Par exemple
il faut que la fluctuation du nombre de voisins d'un point donn\'e
soit positive. Quand $\xi$ est grand le nombre moyen de voisins
dans un volume $V$ est essentiellement, $n\,V\,\xib$
si la densit\'e de point est $n$ tandis que le nombre moyen carr\'e
est donn\'e par la fonction de corr\'elation \`a trois points
moyenn\'ee,
\bea{
\mg N_v\md &\approx& n\,V\,\xib\\
\mg N_v^2\md &\approx& n^2\,V^2\,\xib_3.
}
Si on \'ecrit que 
$\mg N_v^2\md-\mg N_v\md^2$ est positif on obtient,
\beq{
Q_3 > {1\over 3}.
}
On peut g\'en\'eraliser ce genre de contraintes et obtenir des
conditions sur l'ensemble des $Q_{p,\alpha}$ (Fry 1984b), mais
ce ne sera toujours que des conditions n\'ecessaires et non
suffisantes. Autrement dit on est jamais s\^ur que le mod\`ele
qu'on choisit est r\'ealisable par un processus al\'eatoire
quelconque. Si on part de mod\`eles de fractales, par exemple
celui donn\'e par les points occup\'es par une marche al\'eatoire
avec une loi de Levy, on n'obtient pas un mod\`ele r\'ealiste.

\subsubsection{Les mod\`eles}

A priori les coefficients $Q_{p,\alpha}$ peuvent d\'ependre tout aussi
bien  de l'ordre $p$ que de la g\'eom\'etrie $\alpha$ de l'arbre.
Diff\'erentes hypoth\`eses ont \'et\'e explor\'ees dans la
litt\'erature.  Saslaw et Szapudi ont fait de nombreuses analyses en
supposant que les param\`etres $Q$ ne d\'ependent que de $p$ mais dans
la suite je vais plut\^ot  suivre les hypoth\`eses de Bernardeau \&
Schaeffer (1992)!  Pour r\'eduire le nombre de param\`etres on peut
faire l'hypoth\`ese que les param\`etres $Q_{p,\alpha}$ s'\'ecrivent
comme des produits de vertex $\nu_p$, un facteur $\nu_p$ par vertex
qui appara\^\i t dans la repr\'esentation diagrammatique de la
fonction \`a $p$-points.
Alors toutes des fonctions de corr\'elation peuvent \^etre construites
\`a partir de la fonction de corr\'elation \`a 2 points et de la
fonction g\'en\'eratrice des vertex $\mG$. Je la note de cette fa\c on
par analogie avec les r\'esultats de la th\'eorie des
perturbations. La fonction $\mG_{\delta}$ de la th\'eorie des
perturbations peut guider notre choix (c'est d'ailleurs un choix pas
si mauvais) mais il n'y a a priori aucune de raison de penser que ces
fonctions sont identiques.

\subsection{Les probabilit\'es de comptage}

Pour calculer les probabilit\'es de comptage on a besoin de conna\^\i
tre la fonction g\'en\'eratrice des cumulants.  Il est facile de voir
que la relation d'\'echelle (\ref{xipech}) induit des propri\'et\'es
dites hi\'erarchiques entre les moyennes g\'eom\'etriques\footnote{les
moyennes g\'eom\'etriques des fonctions de corr\'elation s'identifient
naturellement avec les cumulants de la distribution de densit\'e
filtr\'ee dans cette m\^eme bo\^\i te} des fonctions de corr\'elation,
$\xib_p(R)$, dans des bo\^\i tes de rayon $R$,
\beq{ 
\xib_p(R)\sim S_p\,\xib^{p-1}(R),
} 
o\`u $S_p$ sont des param\`etres ind\'ependants
de $R$.  On d\'efinit alors la fonction $\varphi$ comme la fonction
g\'en\'eratrice des param\`etres $S_p$ de la m\^eme mani\`ere que pour
le cas perturbatif, 
\begin{equation}
\varphi(y)=\sum_{p=1}^{\infty}-S_p{(-y)^{p-1}\over p!}.  
\end{equation}
La relation entre les $S_n$ et les $\nu_p$ 
n'est la m\^eme que celle donn\'ee dans le cadre des calculs
perturbatifs que dans une approximation de type champ moyen
(qui est tr\`es bien v\'erifi\'ee).
L'hypoth\`ese de comportement hi\'erarchique implique que la
fonction g\'en\'eratrice $\varphi$ est ind\'ependante de $\sigma$
du moins quand celui-ci est grand (on a vu que la th\'eorie des
perturbations permettait de calculer
la limite $\sigma\to 0$ de cette quantit\'e).
Dans une longue \'etude Balian \& Schaeffer (1989) ont examin\'e les
propri\'et\'es g\'en\'eriques induites par une telle hypoth\`ese sur
les propri\'et\'es des probabilit\'es de comptage.  

\subsubsection{Propri\'et\'es g\'en\'eriques de la distribution de densit\'e}

On sait que la fonction de probabilit\'e de densit\'e est 
reli\'ee \`a la fonction g\'en\'eratrice $\varphi(y)$ \`a travers,
la transform\'ee de Laplace inverse,
\beq{
p(\delta)\d\delta=\int_{-\ii\infty}^{+\ii\infty}
{\d y\over 2\pi\ii\sigma^2}\exp\left[-{\varphi(y)\over \sigma^2}
+{y\dta\over \sigma^2}\right] \d\delta.\label{invLap}
}
Dans la suite je ne vais pas regarder les effets discrets, mais
supposer qu'on travaille dans la limite continue (on peut
de toute mani\`ere retrouver les expressions des probabilit\'e
de comptage par simple convolution avec une distribution de Poisson).
La difficult\'e maintenant est de trouver des contraintes 
sur $\varphi(y)$. Cette fonction est loin de pouvoir prendre 
une forme arbitraire: il faut que toutes les quantit\'es
physiques qu'on peut consid\'erer et calculer avec cette fonction
aient un sens.

Un cas particulier important est la probabilit\'e
de vide $P(0)$, probabilit\'e
de ne trouver aucun objet dans un volume de rayon $R$ si le champ de
densit\'e est d\'ecrit par des particules de densit\'e $n$.
Cette probabilit\'e est donn\'ee par
\bea{
P(0)&=&\int\d\delta\,p(\delta)\,\exp\left[-n\,V\,(1+\delta)\right],\\
V&=&{4\,\pi\over 3}R^3.
}
En utilisant (\ref{invLap}) on obtient une expression 
qui s'int\`egre directement sur $\delta$ et qui donne,
\beq{
P(0)=\int_{-\ii\infty}^{+\ii\infty}
{\d y\over 2\pi\ii}\ {1\over y-n\,V\,\sigma^2}\ 
\exp\left[-{\varphi(y)\over \sigma^2}\right].
}
Finalement, $P(0)$ est donn\'e par la valeur du r\'esidu
et s'exprime simplement en fonction
de la fonction g\'en\'eratrice $\varphi(y)$ (White 1979, Balian
\& Schaeffer 1989)
\beq{
P(0)=\exp\left(-n\,V\,{\varphi(y)\over y}\right),\ \ 
y\equiv n\,V\,\sigma^2.
}
Cette expression\footnote{On retrouve le cas Poissonnien
quand $\varphi(y)=y$, c'est \`a dire quand on a annul\'e
toutes les fonctions de corr\'elation} 
montre en tout cas que $\varphi(y)/y$ doit \^etre
fini et positif pour des valeurs arbitrairement grandes de $y$.
Il est alors naturel de supposer que,
\beq{
\varphi(y)\sim a\,y^{1-\omega}\ \ y\gg 1,
}
avec $0<\omega<1$. Les valeurs de $y$ positif correspondent \`a une
s\'erie altern\'ee pour $\varphi$. Par contre quand $y$ est n\'egatif
tous les termes de la s\'erie ont le m\^eme signe et il n'y a pas de
raison que $\varphi(y)$ converge toujours. On note $y_s$ la plus
petite (en valeur absolue) valeur de $y$ o\`u $\varphi(y)$ est
singulier, et on peut param\'etriser la forme de $\varphi(y)$ par,
\beq{
\varphi(y)-\varphi_s \sim a_s\,(y-y_s)^{\omega_s}.
}

\begin{figure}
\vspace{5 cm}
\special{hscale=50 vscale=50 voffset=-0 hoffset=0 psfile=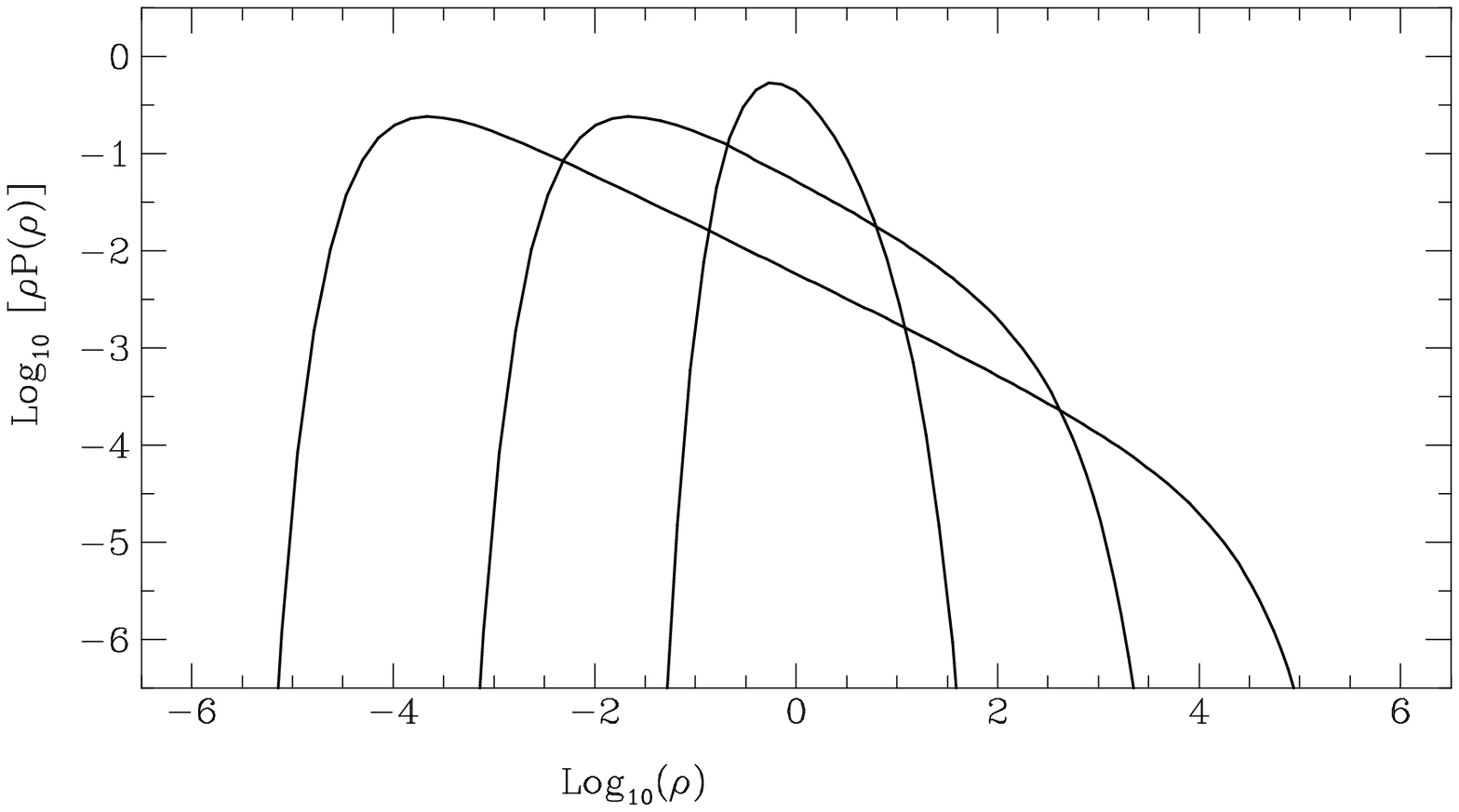}
\caption{Evolution de la forme de la fonction de distribution de
probabilit\'e de la densit\'e locale en fonction de la variance.
Les cas $\sigma=1,10$ et $100$ ont \'et\'e trac\'es
pour des cumulants correspondant \`a $n=-2$.
}
\label{pdfnl}
\end{figure}

Avec ces hypoth\`eses on peut entreprendre le calcul de $p(\delta)$.
Il appara\^\i t deux r\'egimes, l'un correspondant aux r\'egions
plut\^ot sous-denses, l'autre aux r\'egions sur-denses.
Ici je vais m'int\'eresser exclusivement au deuxi\`eme. Il s'obtient 
on supposant que $y$ ne prend que des valeurs finies ou petites
ce qui permet de d\'evelopper une partie des termes sous l'exponentiel
et d'avoir,
\beq{
p(\delta)\d\delta=-
\int_{-\ii\infty}^{+\ii\infty}
{\d y\over 2\pi\ii\sigma^4}{\varphi(y)\over \sigma^2}\exp\left[
{y\dta\over \sigma^2}\right] \d\delta.
}
On voit que la function de distribution peut s'exprimer
en fonction d'une unique int\'egrale, $h(x)$, 
\bea{
p(\delta)\d\delta&=&{\d\delta\over
\sigma^2}\,h\left({\delta\over\sigma^2}\right);\\
h(x)&=&-
\int_{-\ii\infty}^{+\ii\infty}{\d y\over 2\pi\ii}\varphi(y)\exp(y\,x).
}
Les propri\'et\'es de $\varphi(y)$ induisent les
propri\'et\'es suivantes pour $h(x)$,
\bea{
h(x)&=&a\,{1-\omega\over \Gamma(\omega)}\,x^{\omega-2}, \ \ x\ll1\\
h(x)&=&-{a_s\over \Gamma(-\omega_s)}\,x^{-\omega_s-1}\,\exp(y_s\,x),\
\ x\gg1
}
On voit donc que la fonction $h(x)$ (et par l\`a m\^eme
la fonction $p(\delta)$) est construite par assemblage d'une loi
de puissance quand $x$ est petit, et d'une coupure exponentielle
\`a grand $x$ (voir figure \ref{pdfnl}).

Il y a bien s\^ur une part d'arbitraire dans ce r\'esultat (existence
et forme de la singularit\'e pour $\varphi(y)$, ...), aussi
il est int\'eressant de voir ce qui peut \^etre construit de mani\`ere
g\'en\'erique.

On a vu que dans le cas de la th\'eorie des perturbations,
$\varphi(y)$ pouvait \^etre  construite \`a partir d'une fonction
g\'en\'eratrice de vertex, $\mG(\tau)$. 
Le comportement assymptotique de $\varphi(y)$ est li\'e au comportement
asymptotique de $\mG(\tau)$ pour $\tau$ positif. Plus sp\'ecifiquement
si
\beq{
\mG(\tau)\sim c\,\tau^{-\kappa}
}
alors
\bea{
\omega&={\kappa \over \kappa+2},\\
(1-\omega)a&=\kappa^{-\omega}\,c^{1-\omega}.
}
Par exemple pour $\mG$ obtenu \`a partir de l'effondrement sph\'erique
on a $\kappa={3/ 2}$, ce qui donne $\omega={3/7}\approx 0.43$.

L'hypoth\`ese d'une composition en arbres se r\'ev\`ele tr\`es
f\'econde pour la partie singuli\`ere de $\varphi(y)$. En effet
l'\'equation implicite en $\tau$ induit tr\`es naturellement
une singularit\'e en $\varphi$ sans faire d'hypoth\`ese particuli\`ere
sur $\mG(\tau)$. Ce point singulier est donn\'e par la solution
du syst\`eme,
\bea{
y_s&=-{1 \over \mG''(\tau_s)},\\
\tau_s&={\mG'(\tau_s) \over \mG''(\tau_s)}.
}
Alors on a 
\beq{
\varphi(y)-\varphi_s\sim {2\over3}(y-y_s)^{3/2}
\mG'(\tau_s)\mG''(\tau_s)\left({2\mG'(\tau_s)\mG''(\tau_s)
\over \mG'''(\tau_s)}\right)^{1/2}\,.
}
Dans le cas du mod\`ele de th\'eorie des perturbations cette
singularit\'e semble en effet jouer un r\^ole actif. On a vu dans la
section pr\'ec\'edente quelles \'etaient les valeurs d\'ecrivant
cette singularit\'e en fonction de l'indice spectral.

\subsubsection{La th\'eorie des perturbations '\'etendue'}

\begin{figure}
\vspace{6 cm}
\special{hscale=30 vscale=30 voffset=-40 hoffset=-10 psfile=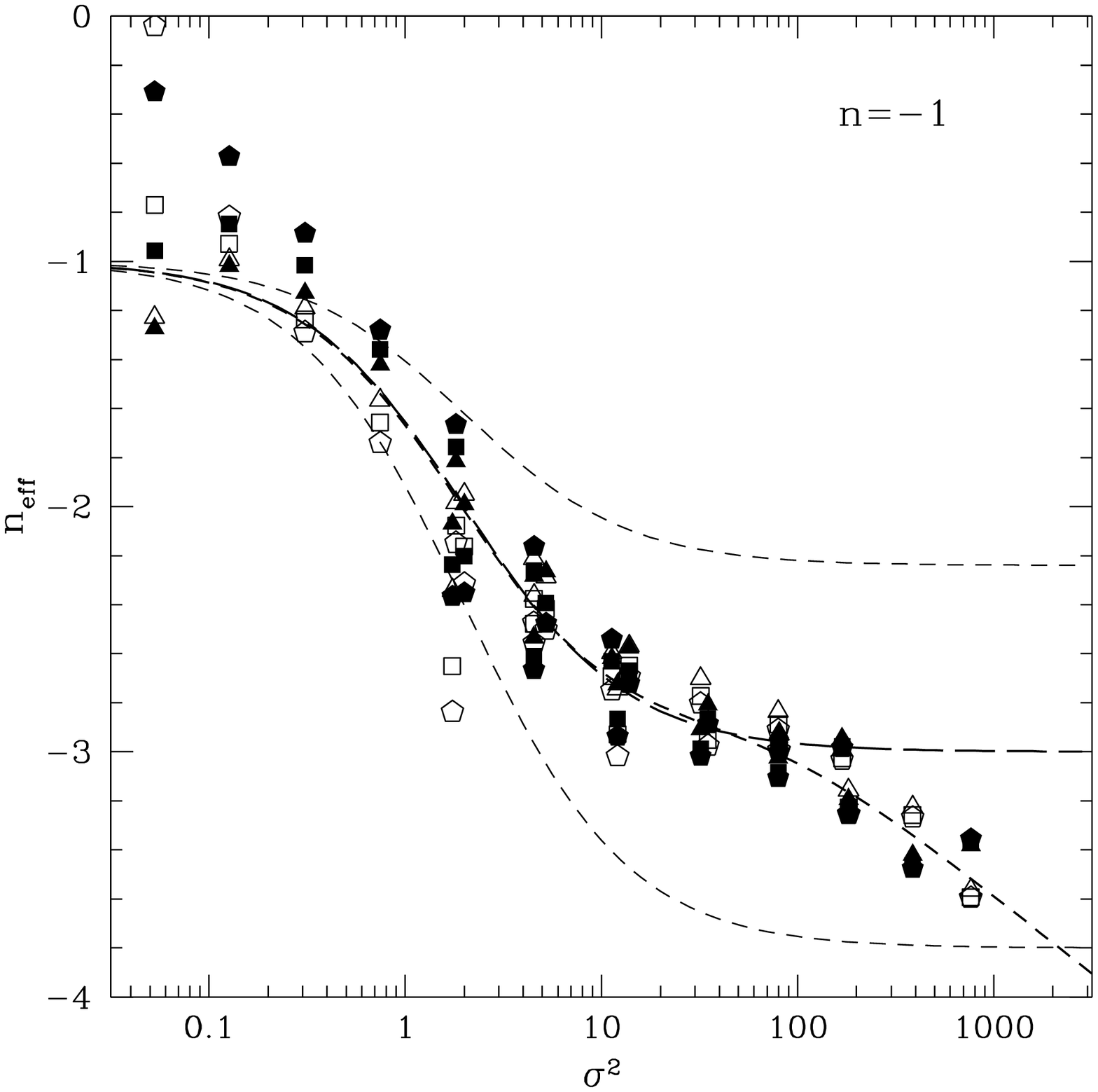}
\special{hscale=36 vscale=36 voffset=-50 hoffset=140 psfile=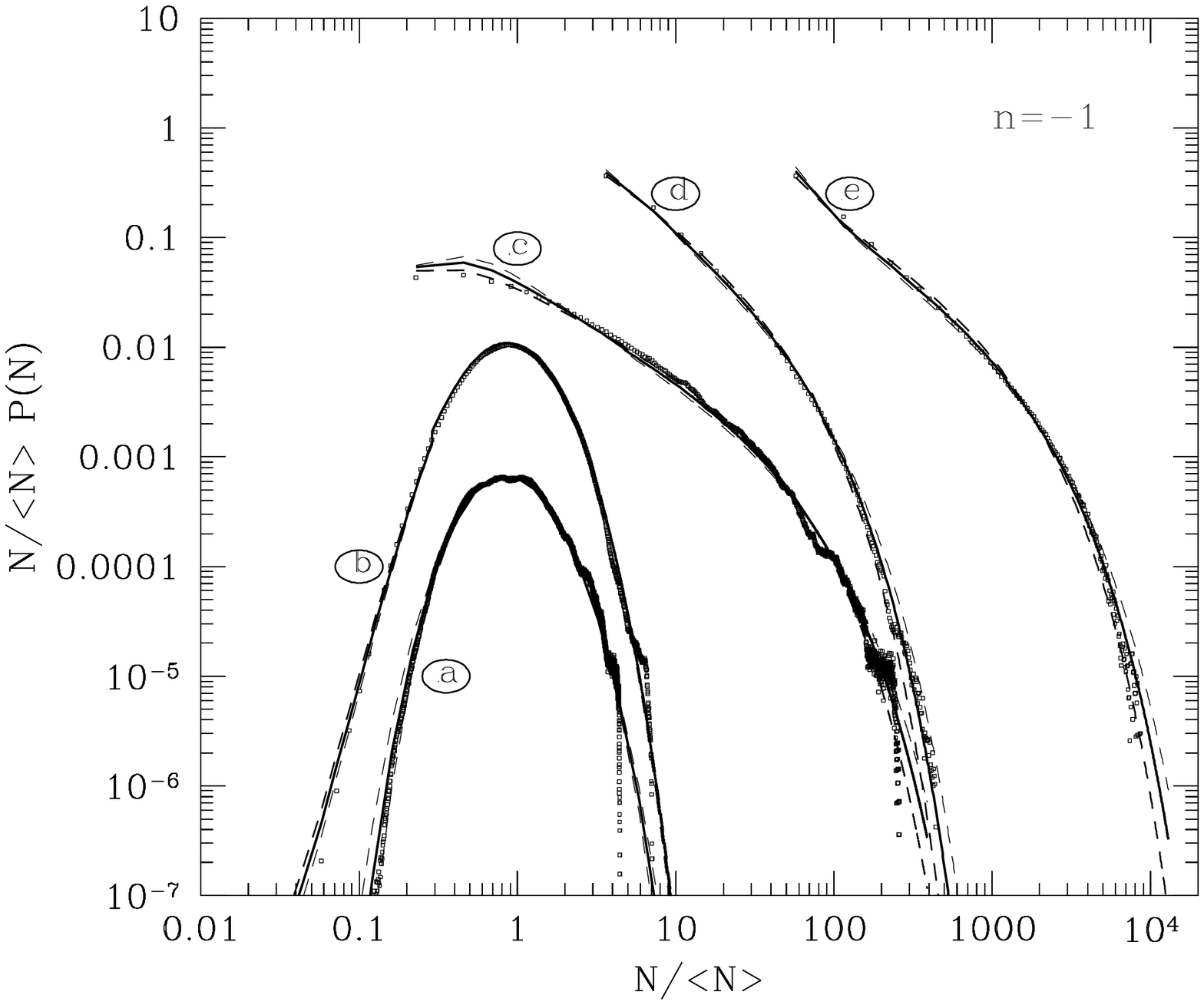}
\caption{A gauche, d\'etermination de l'indice effectif $n_{\rm eff}$
\`a partir des premiers moments $S_p,\ p=3,4,5$ pour $n_{\rm lin}=-1$.
La courbe pointill\'ee correspond au fit propos\'e dans le texte.
A droite les probabilit\'es de comptage mesur\'ees $P(N)$
sont compar\'ees aux pr\'edictions th\'eoriques obtenues avec ce
$n_{\rm eff}$. Les simulations ont \'et\'e examin\'ees \`a
diff\'erents temps et avec diff\'erents rayons de filtrage.
}
\label{Colombim1}
\end{figure}

Quel mod\`ele pour $\varphi(y)$?  Au del\`a du r\'egime
quasi-lin\'eaire une id\'ee est de suivre au plus
pr\`es le mod\`ele sugg\'er\'e par la th\'eorie des
perturbations. Formellement on peut voir la forme fonctionnelle de
$\varphi(y)$ comme \'etant param\'etr\'ee par l'indice spectral $n$.
On peut donc 'oublier' la signification physique de ce param\`etre et
le supposer libre pour voir s'il est possible de l'ajuster pour
d\'ecrire les r\'esultats num\'eriques, aussi bien au niveau des
param\`etres $S_p$ que de la forme des $P(N)$. 
Ce travail a \'et\'e men\'e \`a bien par  Colombi et al (1997).
L'id\'ee de d\'epart est la suivante: les mesures des premiers
$S_p$ fournissent une estimation de $n_{\rm eff.}$ en supposant que
\beq{
S_p(n_{\rm eff.})\equiv S_p^{\rm mesure}.
}
Le r\'esultat remarquable de Colombi et al. est que les $n_{\rm
eff.}$ ainsi mesur\'es ne d\'ependent pas (ou peu) de l'ordre $p$
(pour un spectre initial de forme donn\'e). On le voit sur
la partie gauche de la figure (\ref{Colombim1}).
Du coup les probabilit\'es de comptages qu'on obtient 
en utilisant le $n_{\rm eff}$ ainsi mesur\'e reproduisent
tr\`es bien celles qui sont obtenues num\'eriquement (partie droite de 
la figure \ref{Colombim1}).

Le param\`etre $n_{\rm eff}$ est donn\'e par
\bea{
n_{\rm eff}&=&n_{\rm lin}+(n_{\rm non-lin}-n_{\rm lin})\,
{x^{\tau}\over x^{\tau}+x^{-\tau}}\label{fitneff}\\
x&=&\exp[\log_{10}(\sigma^2/\sigma_0^2)]\nonumber.
}
Cette transition est d\'ecrite par la valeur de $n_{\rm non-lin}$
atteinte pour les grandes valeurs de $\sigma$, le moment
de la transition $\sigma_0$ et la largeur de cette
transition, $\tau$. Les param\`etres trouv\'es sont donn\'es
dans la table (\ref{tableneff}).

\begin{table}
\caption[ ]{Param\`etres utilis\'es dans le fit (\ref{fitneff}).
}
\label{tableneff}
\begin{tabular}{cccccc}
\hline
$n_{\rm linear}$ & 
$n_{\rm nonlinear}$ & $n_{\rm nonlinear}^-$ & 
$n_{\rm nonlinear}^+$ & $\sigma_0$ & $\tau$ \\ \hline
   -2            &  -9.5  & -12.4 & -7.22 &  1.6  & 1.4       \\
   -1            &  -3    & -3.8  & -2.24 &  1.4  & 1.2       \\
    0            &  -1.2  & -1.6  & -0.86 &  1.25 & 0.6       \\
   +1            &  -0.85 & -1.17 & -0.57 &  0.7  & 0.3       \\ \hline
\end{tabular}
\end{table}
Les figures (\ref{Colombim1}) montrent que la descriptions des moments
et des $P(N)$ en terme d'un simple indice effectif sont effectivement
prometteuses. Il n'y a cependant aucune justification th\'eorique
en faveur d'un tel comportement.

\subsection{ Distribution de mati\`ere et distribution de lumi\`ere: vers
une th\'eorie unifi\'ee}

Ultimement ce qu'on veut c'est une description compl\`ete des objets
pr\'esents dans l'Univers, en particulier les propri\'et\'es de
corr\'elation des puits de potentiel qui se sont cr\'e\'es. Ils sont
implicitement contenus dans la description des fonctions de
corr\'elation, aussi compliqu\'ees qu'aient pu \^etre leurs histoires
(accr\'etion, merging, ...). Potentiellement on doit donc faire
beaucoup mieux que des th\'eories \`a la Press et Schechter (1977) qui
font des hypoth\`eses drastiques sur la dynamique.

La mise en \oe uvre de ce calcul est cependant assez
complexe. L'identification math\'ematique d'un puits de potentiel est
une chose qui est loin  d'\^etre ais\'ee. Cependant on est maintenant
dans une situation assez favorable: les objets que l'on recherche ont
des contrastes en densit\'e tr\`es grands et leur d\'etection ne doit
d\'ependre que tr\`es peu du crit\`ere de s\'election utilis\'e. Et en
fait on va supposer que la simple probabilit\'e d'avoir localement un
contraste en densit\'e suffisamment grand va suffire \`a d\'efinir un
objet.

En cons\'equence de quoi la fonction $h(x)$ donne une description de
la forme de la distribution de masse d'objets de taille donn\'ee. Les
propri\'et\'es de corr\'elation de ces objets seront donn\'ees par les
probabilit\'es de comptage conjointes (\`a deux points si on veut la
fonction de corr\'elation \`a deux points, \`a 3 points si on veut
celle \`a 3 points, etc...).

L'hypoth\`ese d'une d\'ecomposition en arbres est alors encore
extr\`emement
f\'econde pour faire le calcul de ces probabilit\'es conjointes.

\subsubsection{Le biais 'dynamique'}

Il faut commencer par calculer la distribution conjointe, 
$p(\delta_1,\delta_2)$,
probabilit\'e d'avoir la densit\'e $\delta_1$ et la densit\'e $\delta_2$
conjointement en deux points aux positions $\vx_1$ et $\vx_2$.
Pour obtenir une telle distribution il faut calculer la fonction 
g\'en\'eratrice
$\varphi(y_1,y_2)$ des cumulants conjoints pris en deux points distincts.
Plus pr\'ecis\'ement on peut d\'efinir $C_{pq}$ par
\beq{
\mg\delta_R^p(\vx_1)\,\delta_R^q(\vx_2)\md_c=
C_{pq}\,\mg\delta_R^2\md^{p+q-2}\,\mg\delta_R(\vx_1)\,\delta_R(\vx_2)\md.
}
Cette relation d\'efinit des param\`etres finis $C_{pq}$ dans la limite
o\`u la distance $\vert\vx_1-\vx_2\vert$ est beaucoup plus grande que
l'\'echelle de filtrage $R$. A partir de l\`a on d\'efinit la fonction
g\'en\'eratrice double,
\beq{
\varphi(y_1,y_2)=\sum_{p=0}^{\infty}\sum_{q=0}^{\infty}
C_{pq}\,{(-y_1)^{p-1}\over p!}\,{(-y_2)^{q-1}\over q!}.
}
en ayant pos\'e,
\beq{
C_{p0}=S_p,\ \ \ C_{00}=0.
}

La d\'ecomposition exacte en arbres implique que les facteurs $C_{pq}$ ont
en plus une propri\'et\'e de factorasibilit\'e, c'est \`a dire que,
\beq{
C_{pq}=C_p\,C_q\ \ {\rm pour}\ \ p>1\ \ {\rm et}\ \ q>1.
}
C'est une propri\'et\'e assez facile \`a voir d'un point de vue diagrammatique:
le poids qu'on mets d'un c\^ot\'e du diagramme ne d\'epend pas de la topologie
qu'on met de l'autre cot\'e. De plus, compte tenue de la remarque qui a \'et\'e
faite plus t\^ot, la fonction $\tau(y)$ s'identifie \`a la fonction 
g\'en\'eratrice
des vertex \`a une patte externe. Plus pr\'ecis\'ement on a,
\beq{
\tau(y)=-\sum_{p=1}^{\infty}C_{p}\,{(-y)^{p-1}\over p!},
}
quand $\tau(y)$ est solution du syst\`eme donnant $\varphi(y)$.
Cette relation fonctionelle est extr\`emement int\'eressante  mais
d\'epend de mani\`ere explicite de l'hypoth\`ese de d\'ecomposition en
arbres des fonctions de corr\'elation.

Si on exploite cette propri\'et\'e on peut calculer la probabilit\'e conjointe
$p(\delta_1,\delta_2)$ qui s'\'ecrit,
\beq{
p(\delta_1,\delta_2)=
{1\over \sigma^8}\,h(x_1)\,h(x_2)\,\left[1+b(x_1)\,b(x_2)\,
\xi_2(\vx_1,\vx_2)\right]
}
avec
\beq{
h(x)\,b(x)=-\int_{-\ii\infty}^{\ii\infty}{\d y\over 2\pi\ii}\tau(y)\,
\exp(x\,y).
}
Autrement dit les r\'egions les plus denses continuent \`a suivre une
loi hi\'erarchique. Le facteur $b(x)$ s'identifie au facteur de biais
entre les pics. Cela implique que le biais est a priori une fonction
de $x\equiv {\delta\over \sigma^2}$ uniquement (et pas de $\delta$ par
exemple). La forme de $b(x)$ n'a pas a priori une forme ind\'ependante
des param\`etres des mod\`eles. Cependant la singularit\'e qui
appara\^\i t  naturellement pour $\varphi(y)$ induit un comportement
sp\'ecifique pour $b(x)$, 
\beq{
b(x)\sim x\ \ {\rm quand}\ \ x\gg1.
}
Le r\'esultat g\'en\'erique est donc que le biais croit avec la
raret\'e. Ce r\'esultat obtenu \`a partir d'une description aussi
r\'ealiste que possible du r\'egime non-lin\'eaire prolonge d'une
certaine mani\`ere les calculs de Bardeen et al. (1986) men\'es dans
un champ Gaussien. La principale cons\'equence de ces calculs est qu'a
priori les puits de potentiel qui se forment sont effectivement
biais\'es. 

\subsubsection{La hi\'erarchie des corr\'elations pour les pics de densit\'e}

Le paragraphe pr\'ec\'edent pr\'esente les r\'esultats pour la
fonction de probabilit\'e 
\`a deux points. On peut en fait g\'en\'eraliser ces r\'esultats pour
des fonctions de corr\'elation 
d'ordre donn\'e. 
Dans l'hypoth\`ese de fonctions de corr\'elation en arbre, on peut d\'efinir
la fonction g\'en\'eratrice des arbres \`a un nombre arbitraire de pattes. 
Cela permet de calculer l'ensemble des propri\'et\'es de corr\'elation
jusqu'\`a un ordre donn\'e. 
Les r\'esultats sont les suivants,
\begin{itemize}
\item Les propri\'et\'es hi\'erarchiques des fonctions de corr\'elation aux
grands ordres sont conserv\'ees.
\item Les param\`etres $S_p$ correspondant sont des fonctions de $x$
uniquement.
\item Dans la limite $x\to\infty$, ces param\`etres ont une limite finie
donn\'ee par 
\beq{
S_p(x)\to p^{p-2}\ \ {\rm quand}\ \ x\to\infty.
}
\end{itemize}
Ces r\'esultats sont assez g\'en\'eraux et d\'ependent finalement peu
des hypoth\`eses pr\'ecises sur les fonctions de corr\'elation. Ils supposent
des propri\'et\'es de r\'egularit\'e de la fonction g\'en\'eratrice des vertex.
Une cons\'equence, peut-\^etre importante, de ces r\'esultats est
que les $S_p$ des pics rares devraient \^etre les m\^emes qu'on soit
dans le r\'egime quasi-lin\'eaire ou dans le r\'egime fortement
non-lin\'eaire. C'est peut-\^etre le d\'ebut d'une explication pour
les observations qui sont faites.

Ces r\'esultats montrent que m\^eme en l'absence de m\'ecanismes
non-gravitationnels il est possible d'avoir des effets de biais. Il
faudrait, pour faire mieux,  conna\^\i tre les solutions explicites du
r\'egime non-lin\'eaire. 

\section{Conclusions}

Dans ce petit tour d'horizon de la cosmologie
j'ai \'et\'e loin d'\^etre exhaustif. Dans le domaine de l'\'evolution
des grandes structures je n'ai pas mentionn\'e bon
nombre de domaines de recherche en plein d\'eveloppement, comme
\begin{itemize}
\item L'\'etude des flots cosmiques \`a grande \'echelle, et leur
utilisation pour mesurer les param\`etres cosmologiques;
\item L'\'etude des amas de galaxies et de leur contenu aussi
bien en mati\`ere noire (\'etudes dynamique, reconstruction de masse
par effets de lentilles gravitationnelles..), qu'en mati\`ere
baryonique (galaxies, rayonnement X..);
\item L'\'etude des absorbants et des objets de grand $z$;
\item ...
\end{itemize}

Du point de vue du physicien th\'eoricien un domaine de recherche de
pr\'edilection est inconstablement la cosmologie primordiale, lieu de
rencontre entre entre la physique des hautes \'energies et la
cosmologie observationnelle. 

Cependant la formation des grandes structures recelle un certain
nombre de probl\`emes ouverts digne d'int\'er\^et,
\begin{itemize}
\item La th\'eorie des perturbations est loin d'avoir livr\'e tous ses
secrets. En particulier il serait extremement utile de comprendre les
corrections en boucles.
\item De nombreux aspects du r\'egime non-lin\'eaire ne sont pas
compris. Par exemple on ne sait pas d\'ecrire la transition vers le
r\'egime multiflots qui conduit \`a la virialisation de la mati\`ere
dans les puits de potentiel. Il serait aussi tr\`es int\'eressant de
pouvoir exhiber une solution explicite des \'equations dynamiques dans
le r\'egime nonlin\'eaire (m\^eme si ce n'est qu'une forme
assymptotique). 
\item L'exploitation des catalogues tridimensionnel ou bidimensionnel 
notamment avec la mise en \'evidence de propri\'et\'es non-Gaussiennes
n'est pas encore optimale. La question se pose par exemple pour les
cartes de distorsion gravitationnelle. 
\item La relation entre baryons et mati\`ere noire est un domaine qui
n'a pratiquement pas \'et\'e explor\'e. On ne dispose pour l'instant
que d'exp\'eriences num\'eriques pour essayer de comprendre ce qui se passe.
\end{itemize}

\end{document}

%% file: Cours123.tex
\section{Introduction}

La d\'ecouverte en 1965 par Penzias et Wilson
du fond de rayonnement cosmologique  diffus
\`a 3 K a d\'efinitivement fait du mod\`ele du Big Bang chaud le cadre
g\'en\'eral de la cosmologie moderne.  Ces trois derni\`eres
d\'ecennies ont \'et\'e marqu\'ees par de  nombreux progr\`es tant
th\'eoriques qu'observationnels qui n'ont fait que confirmer sa
validit\'e, et de pr\'eciser petit \`a petit les d\'etails de ce
sc\'enario.

Parmi les avanc\'ees notables de ces derni\`eres ann\'ees, citons
pour m\'emoire: 

\begin{itemize}
\item La r\'ealisation de grands catalogues de galaxies. On a
maintenant des
catalogues tri-dimensionnels qui contiennent des milliers
d'objets. Ces catalogues permettent de faire une v\'eritable
cosmographie
de l'univers local. De nouveaux moyens sont apparus r\'ecemment
pour compl\'eter ce type de catalogues, la d\'etermination des
champs de vitesse cosmiques, permet ainsi d'acc\'eder \`a des
informations cin\'ematiques tr\`es pr\'ecieuses. Et depuis la
fin des ann\'ees 80, un nouveau moyen d'investigation est en train
d'\'emerger, il
s'agit des cartes de distorsion gravitationnelle. Elles permettent
de visualiser les lignes de potentiel de la masse projet\'ee. 

\item le mod\`ele de Mati\`ere Noire Froide (CDM pour Cold Dark
Matter)
qui est apparu au d\'ebut des ann\'ees 80 et qui a servi de 
point de r\'ef\'erence (\`a d\'efaut de devenir un mod\`ele
standard) pour tous les travaux sur le probl\`eme de la formation
des grandes structures.

\item Le d\'eveloppement des th\'eories inflationnaires.
C'est encore un terrain tr\`es sp\'eculatif, mais c'est le lieu
tr\`es excitant o\`u les concepts de la physique des hautes \'energies
rencontrent des exigences observationnelles de plus en plus
fiables.

\item Enfin, la d\'etection en 1992 des fluctuations de temp\'erature
du fond de rayonnement cosmologique par l'exp\'erience
satellitaire COBE/DMR (Smoot et al. 1992) 
a marqu\'e un tournant pour la cosmologie:
pour la premi\`ere fois on avait une preuve directe de
l'origine des grandes structures de l'univers.

\end{itemize}

Dans ce cours je vais principalement m'int\'eresser au probl\`eme de
la formation des grandes structures. 

Quelques ouvrages de r\'ef\'erence,
\begin{itemize}
\item Relativit\'e G\'en\'erale: Weinberg, {\it 
Gravitation and Cosmology}, 1972; 
Landau et Lifschitz, {\it Classical Theory of Fields}, 1975
\item Univers primordial: Kolb et Turner {\it The Early Universe},
1990, mais il est peu pr\'ecis; revue de Brandanberger, {\it
Inflation and Cosmic Strings: two Mechanisms for producing Structure
in the Universe}, Int. J. Mod. Phys. A2 : 77, 1987.
\item Inflation: Linde, {\it Particle Physics and Inflationary
Cosmology}, 1990; Liddle et Lyth, {\it The Cold Dark Matter density
perturbation}, 1993, Physics Reports, 231, 1;
Lidsey et al. {\it Reconstructing the inflaton potential - an overview},
1997, Reviews of Modern Physics, 69, 2
\item Formation des grandes structures: Peebles {\it The Large Scale
Structure of the Universe}, 1980; {\it Principle of Physical
Cosmology}, 1993; V. Sahni \& P. Coles, {\it Approximation Methods for
Non-linear Gravitational Clustering}, 1995, Physics Reports, 262, 1
\end{itemize}

\section{\'El\'ements de cosmologie, l'univers homog\`ene}
\subsection{Les principes cosmologiques et cosmographie}

Le principe fondamental sur lequel repose la cosmologie,  est
l'hypoth\`ese que l'univers est, au moins \`a grande \'echelle,
homog\`ene et isotrope.  Cette id\'ee est n\'ee au d\'ebut de ce
si\`ecle. Elle est en fait, \`a l'origine, davantage motiv\'ee par des
consid\'erations philosophiques que des raisons physiques
observationnelles. Einstein a \'et\'e un des premiers, d\`es la
naissance de le Relativit\'e G\'en\'erale, \`a tenter de construire
une m\'etrique susceptible de d\'ecrire l'ensemble de
l'univers. L'id\'ee g\'en\'erale de d\'epart \'etait que l'univers
\'etait non seulement homog\`ene, isotrope mais aussi immuable.
L'abandon de cette derni\`ere hypoth\`ese est d'abord due \`a la mise
en \'evidence en 1929, par un travail acharn\'e de E. Hubble, du
mouvement de r\'ecession des galaxies. Cela dit, il faudra quand
m\^eme attendre 1965, avec la d\'ecouverte du rayonnement fossile \`a
3K par Penzias \& Wilson pour que l'id\'ee que l'univers ait
travers\'e une phase chaude, donc le mod\`ele du Hot Big Bang,
s'impose face aux tenants d'un Univers immuable qui d\'efendaient le
mod\`ele 'steady state' (mod\`ele stationnaire).

On a aujourd'hui des \'evidences assez solides en faveur de
l'homog\'en\'eit\'e et de l'isotropie. Pendant longtemps les relev\'es
locaux d'\'etoiles ou de galaxies ont montr\'e de fortes 
h\'et\'erog\'en\'eit\'es. Les plus grands catalogues de galaxies 
arrivent enfin \`a mettre en \'evidence une homog\'en\'eit\'e \`a grande
\'echelle. Le {\it Las-Campanas Redshift Survey} 
qui contient plusieurs milliers d'objets l'illustre assez bien.

\begin{figure}
\vspace{14 cm}
\special{hscale=80 vscale=80 voffset=-70 hoffset=-80 psfile=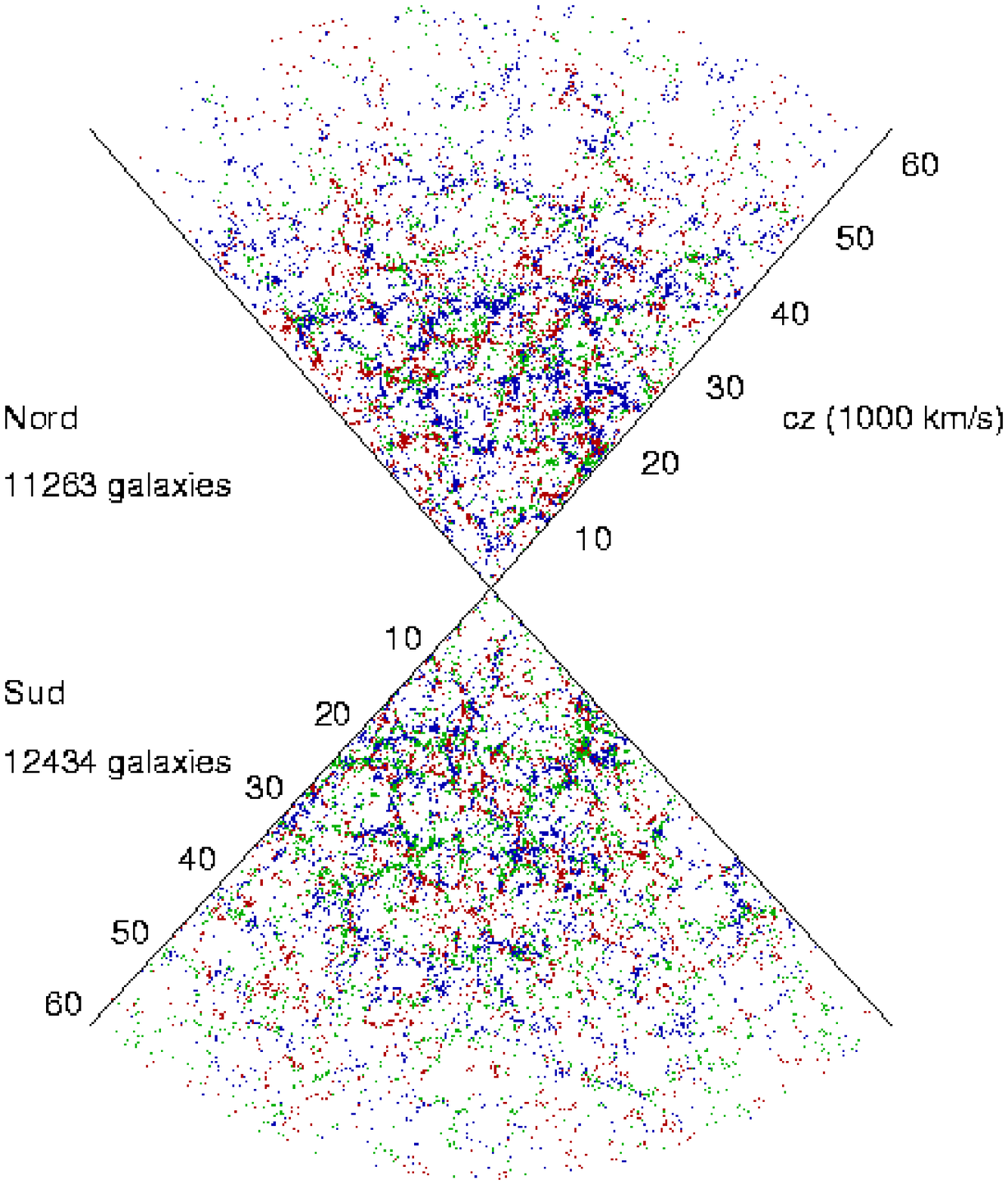}
\caption{Le {\it Las-Campanas Redshift Survey}, Shectman et al. 1996 }
\end{figure}

L'\'evidence la plus forte reste sans doute l'isotropie du fond de
rayonnement \`a 3K. Les \'ecarts \`a l'isotropie ont maintenant \'et\'e
d\'etect\'es. Le mouvement propre de notre galaxie induit un
\'ecart de l'ordre de $10^{-3}$, mais les anisotropies propres
sur cette surface de derni\`ere diffusion ne sont que de l'ordre de
$10^{-5}$. 

Tout cela contribue \`a valider le mod\`ele d'un Univers homog\`ene
et isotrope \`a grande \'echelle.

Dans le cadre de la Relativit\'e G\'en\'erale le principe cosmologique
d\'etermine une forme g\'en\'erale de la m\'etrique. On peut se r\'ef\'erer
pour ce calcul au livre de Weinberg (1972). La forme g\'en\'erale de
la m\'etrique qui r\'epond \`a ces conditions est donn\'ee par 
la m\'etrique dite de Robertson-Walker, 
\be \d
s^2=-\d t^2+a^2(t)\left[{\d x^2\over 1-k\,x^2}+x^2\d\theta^2+
x^2\,\sin^2\theta\d\phi^2\right] 
\ee o\`u $k$ est une constante. Si
$k$ est nul la partie spatiale de la m\'etrique est plate. Sinon on a
une courbure non nulle.

On voit que cette m\'etrique est ind\'ependante de  la position et
qu'il n'y a aucune direction privil\'egi\'ee.  Mais cela ne pr\'ejuge
pas a priori de la topologie {\it globale} de l'univers. M\^eme quand $k$
est nul ou n\'egatif, on peut tr\`es bien  avoir un Univers de taille
finie en imaginant une topologie p\'eriodique  (voir une revue
r\'ecente de Lachi\`eze-Rey et Luminet, gr-qc/9605010, Phys.
Rept. 254 (1995) 135). Notons que cela brise tout de m\^eme
l'isotropie globale de l'Univers. De telles conditions p\'eriodiques
n'ont pas de cons\'equences sur la dynamique, sur la r\'esolution des
\'equations d'Einstein, qui sont purement locales (c'est vrai aussi
bien pour le facteur d'expansion, que pour la formation des
structures, au moins tant que ces structures sont plus petites que la
maille \'el\'ementaire).  Par contre cette hypoth\`ese a des
cons\'equences observationnelles d\`es que l'on s'int\'eresse aux
propri\'et\'es statistiques des champs cosmiques: la m\^eme portion
se retrouve dans des directions diff\'erentes induisant de ce fait
des propri\'et\'es statistiques sp\'ecifiques.  Dans la suite de cet 
expos\'e je vais n\'egliger une telle  possibilit\'e.

Une cons\'equence du facteur d'expansion $a$, c'est
la loi de Hubble. En effet la vitesse relative de deux particules
\`a la distance physique, $\vr\equiv a\,\vx$,
est donn\'ee par
\be
\vv={\d \over \d t} (a\,x)={\dot{a}\over a}\vr+a\,\dot{x}.
\ee
Cette relation montre que la vitesse moyenne des objets les uns
par rapport aux autres est proportionnelle \`a la distance de  ces
objets. On appelle $H$ cette quantit\'e, dite 'constante de
Hubble'\footnote{ce terme est bien mal appropri\'e puisque cette
quantit\'e d\'epend en fait du temps}, 
\be
H={\dot{ a}\over a}.
\ee
Il y a un terme suppl\'ementaire qui est la vitesse
particuli\`ere des particules. Cette vitesse est en moyenne
nulle. Pour une galaxie elle est de l'ordre de 300 $km/s$.

Cette relation l\`a est purement locale, si on veut faire
des calculs plus pr\'ecis il faut tenir compte de la courbure
spatio-temporelle. Imaginons qu'on d\'etecte un signal lumineux
(d'une galaxie lointaine) \`a un instant $t_0$. Il a \'et\'e \'emis
\`a l'instant $t_1$ et \`a la distance $x_1$. Pour relier $t_1$ \`a
$x_1$ il faut int\'egrer le long de la g\'eod\'esique suivie par les
photons. Dans ce cas c'est tr\`es simple,
\be
\d s^2=0=\d t^2-a^2(t){\d x^2\over 1-k\,x^2}.
\ee
On a donc,
\be
\int_{t_1}^{t_0}{\d t\over a(t)}=\int_0^{x_1}{\d x\over \sqrt{1-k\,x^2}}.
\ee
Si la galaxie \'emettrice est au repos, $x_1$ ind\'ependant du temps,
alors deux \'ev\'enements s\'epar\'es par un intervalle de temps
$\delta\,t_1$ au moment de l'\'emission sont s\'epar\'es d'un
intervalle de temps $\delta\,t_0$ \`a la r\'eception avec,
\be
{\delta\,t_0\over a(t_0)}={\delta\,t_1\over a(t_1)},
\ee
ce qui implique que les longueurs d'onde \'electromagn\'etiques
$\lambda_1$ et $\lambda_0$ sont reli\'ees par,
\be
{\lambda_0\over \lambda_1}={a(t_0)\over a(t_1)}.
\ee

Or dans un spectre d'objets lointains comme les galaxies,
des raies sp\'ecifiques, d'absorption par exemple,
se retrouvent assez facilement. L'expansion (donc \'eloignement
relatif des objets) se
traduit par un d\'ecalage vers le rouge de ces raies,  (que je vais
dans toute la suite appel\'e redshift). 
En mesurant le rapport d'une longueur d'onde mesur\'ee avec
la longueur d'onde correspondante au repos, on
mesure donc directement un rapport de facteurs d'expansion entre
l'\'emission et la r\'eception,
\be
1+z(t)={a(t_0)\over a(t)}.
\ee
Notons que si on interpr\'etait ce d\'ecalage spectral comme
une vitesse d'\'eloignement, cette vitesse serait de $cz/(1+z)$.

Le redshift est finalement la mesure de distance par excellence en
cosmologie.  Si on a tendance \`a utiliser les Mega-parsecs 
(un parsec vaut approximativement 3.5 ann\'ee-lumi\`ere), on 
r\'eintroduit finalement la
constante de Hubble dedans et c'est bien de vitesses
relatives que l'on parle. Pour cela on introduit la
variable r\'eduite,
\be 
h={H\over 100\ {\rm km}\,{\rm s}^{-1}\,{\rm Mpc}^{-1}}.
\ee 
 Ainsi la distance typique des galaxies
entre elles est de l'ordre de $5h^{-1}$Mpc, 
c'est \`a dire de $500\ km/s$.
La difficult\'e qu'on a \`a mesurer la constante de
Hubble vient du fait qu'il faut trouver un moyen de calibrer une
relation. Il faudrait une "r\`egle", et on en a pas de tr\`es bonne
\`a notre disposition.

La relation de proportionnalit\'e entre la vitesse et la distance n'est
vraie que dans l'Univers local. D\`es que le redshift approche
l'unit\'e il faut pr\'eciser de quelle distance on parle.
\begin{itemize}
\item La distance angulaire. Un objet de taille propre $D$ donn\'ee
sera vu sous un angle $\delta$. Si cet angle est petit, alors on peut
d\'efinir la distance $d_A$ par
\be
\delta\equiv {D\over d_A}.
\ee
Si on \'ecrit $\d s^2=0$ pour la distance transversale d'un objet,
donc \`a $x_1$ fixe, on a $\d t=a(t_1)\,x_1(t_1)\,\delta$ d'apr\`es la
forme de la m\'etrique. On a donc simplement 
\be
d_A(t_1)=a(t_1)\,x_1(t_1).
\ee

\item La distance lumineuse. Cela correspond au fait que physiquement
la luminosit\'e apparente d'une source d\'ecro\^\i t avec la
distance. Plus pr\'ecis\'ement si $L_{\rm abs.}$ est la luminosit\'e absolue
d'un objet et $L_{\rm app.}$ sa luminosit\'e apparente, alors on peut d\'efinir
$d_L$ par,
\be
L_{\rm abs.}=L_{\rm app.}\,4\,\pi\,d_L^2,
\ee
et on trouve
\be
d_L(t_1)={a^2(t_0)\over a^2(t_1)}\,d_A(t_1).
\ee
La meilleure mani\`ere d'obtenir cette relation est d'\'ecrire
la conservation de l'\'energie des photons \'emis.

\item La distance du mouvement propre. Si l'angle, $\Delta \delta$, 
parcouru par un objet de vitesse perpendiculaire $V_{\perp}$ pendant
le temps (de l'observateur) $\Delta t_0$, on d\'efinir la distance
$d_M$ par,
\be
\Delta \delta={V_{\perp}\,\Delta t_0\over d_M}.
\ee
On a 
\be
d_M(t_1)={a(t_0)\over a(t_1)}\,d_A(t_1).
\ee

\item La distance parallaxe. Cela correspond \`a la mesure d'une
distance par effet de parallaxe (changement de la position de
l'observateur). C'est un peu le contraire du cas de la distance
angulaire. On trouve
\be
d_P(t_1)=a(t_0)\,{x_1\over \sqrt{1-k\,x_1^2}}.
\ee
Ce serait int\'eressant, mais cette m\'ethode ne permet absolument pas
d'acc\'eder \`a des distance cosmologiques!

\end{itemize}

Un param\`etre qu'il peut \^etre utile de consid\'erer est le
param\`etre de d\'ec\'el\'eration $q_0$,
\be
q_0=-{\ddot{a}\,a\over \dot{a}^2}(t_0).
\ee
On peut alors faire le d\'eveloppement perturbatif \`a $t=t_0$,
\be
a(t)=a_0\,\left[1+H_0\,(t-t_0)-{1\over
2}\,q_0\,H_0^2\,(t-t_0)^2+\dots\right]
\ee
qui donne une relation perturbative entre la distance $x_1$ et
le redshift $z$,
\be
x_1={1\over a_0\,H_0}\,\left[z-{1\over 2}\,(1+q_0)\,z^2+\dots\right].
\ee
La comparaison distance-redshift permet donc en premier lieu de
mesurer le param\`etre de d\'ec\'el\'eration $q_0$. Ce qu'on vient de
faire n'est que de la cosmographie \`a partir de la m\'etrique
de R.W.. Pour faire mieux il faut \'ecrire les \'equations dynamiques.

\subsection{\'Evolution du facteur d'expansion}

Les \'equations d'Einstein s'\'ecrivent,  
\be R_{\mu \nu}-{1\over 2} g_{\mu \nu} R_{\rho}^{\rho}=-
{8 \pi\ G}\ T_{\mu \nu}+\lambda\,g_{\mu \nu}.
\ee 
Je garde ici la
possibilit\'e d'avoir une constante cosmologique non nulle. Une telle
m\'etrique impose que le  tenseur \'energie-impulsion du fluide
cosmique est isotrope (pas de pression anisotrope).  Autrement dit,
$T_{\mu \nu}$ peut s'\'ecrire  
\be T_{\mu
\nu}=\left(
\begin{tabular}{cccc}
$\rho$&&&\\
&$p$&&\\
&&$p$&\\
&&&$p$
\end{tabular}
\right).
\ee 

A partir de la conservation de l'\'energie impulsion on peut \'ecrire 
(composante (0,0)), 
\be 
a^3\,{\d p\over \d t}={\d \over \d t}\left[a^3\,(\rho+p)\right] 
\ee 
qui peut se r\'e\'ecrire, 
\be 
{\d \over \d t}(\rho\,a^3)=-3\,p\,a^3.  
\ee 
Pour simplifier les
discussions \`a venir je vais supposer qu'il existe une \'equation
d'\'etat de la forme,  
\be p=w\,\rho, \ee 
qui relie la pression locale
\`a la densit\'e. $w$ est suppos\'e constant. \'Evidemment cette
hypoth\`ese, vraie pour chacun des fluides pris s\'epar\'ement, 
n'est pas vraie en g\'en\'eral pour le fluide total mais c'est vrai une grande
partie du temps (sauf pendant les p\'eriodes de transition). Quand la
densit\'e d'\'energie du rayonnement (ou des particules
ultra-relativistes) domine sur celle des particules non-relativistes,
on a alors $w=1/3$.  Si c'est la mati\`ere qui domine alors on a $w=0$
(poussi\`eres sans pression).

Notons quand m\^eme que la conservation d'\'energie s'applique pour
chaque fluide cosmique s\'epar\'ement (si ils ne sont effectivement 
coupl\'es que par gravitation). Cela donne l'\'evolution de la
densit\'e  de chacun des fluides en fonction du facteur d'expansion. 
\ba
\rho_{\gamma}&\propto & a^{-4}\ \ {\rm pour\ le\ rayonnement}\\
\rho_{\rm mat.}&\propto & a^{-3}\ \ {\rm pour\ la\ matiere}.
\ea

Les \'equations d'Einstein donnent alors les \'equations d'\'evolution
du facteur d'expansion $a$.
\ba
3\, \ddot{a}&=&-4 \pi G\,\rho\,(1+3w)\,a+\lambda\,a\label{Evola1};\\
\dot{a}^2&=&{1\over a}
\left[{8\pi\,G\over 3}\,\rho\,a^3-k\,a
+{\lambda\,a^3\over 3}\right]\label{Evola2}.
\ea
Si on interpr\`ete $\lambda$ comme un fluide cosmique,
on a une densit\'e associ\'ee de
$\rho_{\lambda}=-p_{\lambda}=\lambda/(8\pi G)$.
Noter que, compte tenues des \'equations de conservation,
l'\'equation (\ref{Evola2}) est l'int\'egrale de la premi\`ere
et $k$ joue l\`a le r\^ole d'une constante d'int\'egration.
Cela permet de donner une interpr\'etation plus dynamique \`a $k$.
Plus pr\'ecis\'ement, la premi\`ere \'equation correspond \`a 
l'application du th\'eor\`eme de Gauss (ou plut\^ot de
Birkhoff dans le cadre de la relativit\'e g\'en\'erale)
\`a une sph\`ere de densit\'e
$\rho+\lambda$, alors, si $\lambda=0$, cette sph\`ere peut cro\^\i tre
jusqu'\`a atteindre un rayon infini uniquement si $k\le 0$. 
Si $k>0$, il y a un rayon d'extension maximale.

\begin{figure}
\ba
\vspace{2 cm}
\hspace{6 cm}{\d^2a\over\d t^2}&=&-{G\,M(<a)\over a^2},\nonumber\\
\hspace{6 cm}M(<a)&=&\left(\rho+3\,p-{\lambda\over4\,\pi\,G}\right)\,
{4\,\pi\over 3}\,a^3\nonumber
\ea
\vspace{2 cm}
\special{hscale=80 vscale=80 voffset=-50 hoffset=10 psfile=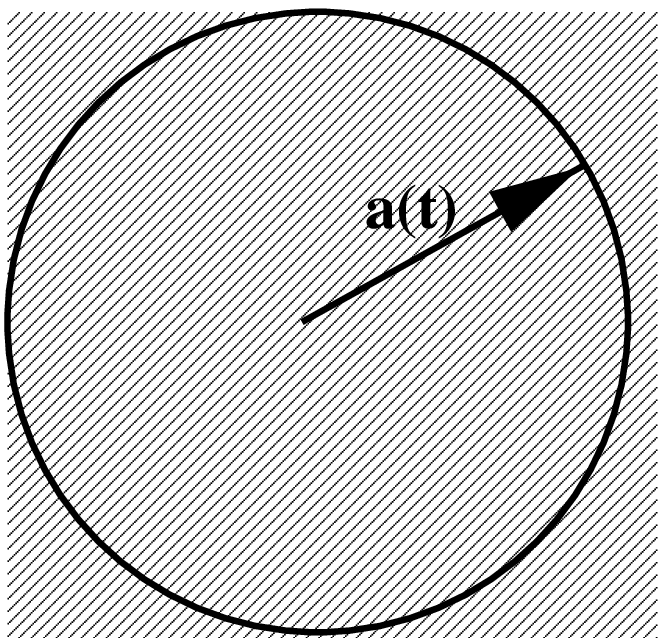}
\caption{Application du th\'eor\`eme de Birkhoff au facteur d'expansion}
\end{figure}

Pour avoir la d\'ependance en temps du facteur d'expansion il faut
finalement r\'esoudre les \'equations compl\`etes.

Si $k=0$ et $\lambda=0$, on trouve 
\be 
a(t)\propto t^{2/3},\ \ {\rm puisque}\ \ \dot{a}^2\sim {1\over a}
\ee 
pour un Univers domin\'e par la mati\`ere et
\be 
a(t)\propto t^{1/2},\ \ {\rm puisque}\ \ \dot{a}^2\sim {1\over a^2}
\ee 
pour un Univers domin\'e par le rayonnement.
Pour un Univers domin\'e par la mati\`ere quand $\lambda=0$
on a une solution explicite pour la croissance du facteur d'expansion.
Si $k=+1$ ($\Omega_0>1$) on a\footnote{Dans toute la suite les indices 0 font
r\'ef\'erence \`a des quantit\'es prises au temps pr\'esent}
\ba
a(t)&=&{a_0\Omega_0\over 2\,(\Omega_0-1)}\,[1-\cos(\theta)];\\
H_0\,t&=&{\Omega_0\over 2\, (\Omega_0-1)^{3/2}}\,[\theta-\sin(\theta)];
\ea
et si $k=-1$ on passe \`a des fonctions hyperg\'eom\'etriques en
faisant $\theta\to\ii \psi$.

\begin{figure}
\vspace{6 cm}
\special{hscale=80 vscale=80 voffset=-20 hoffset=10 psfile=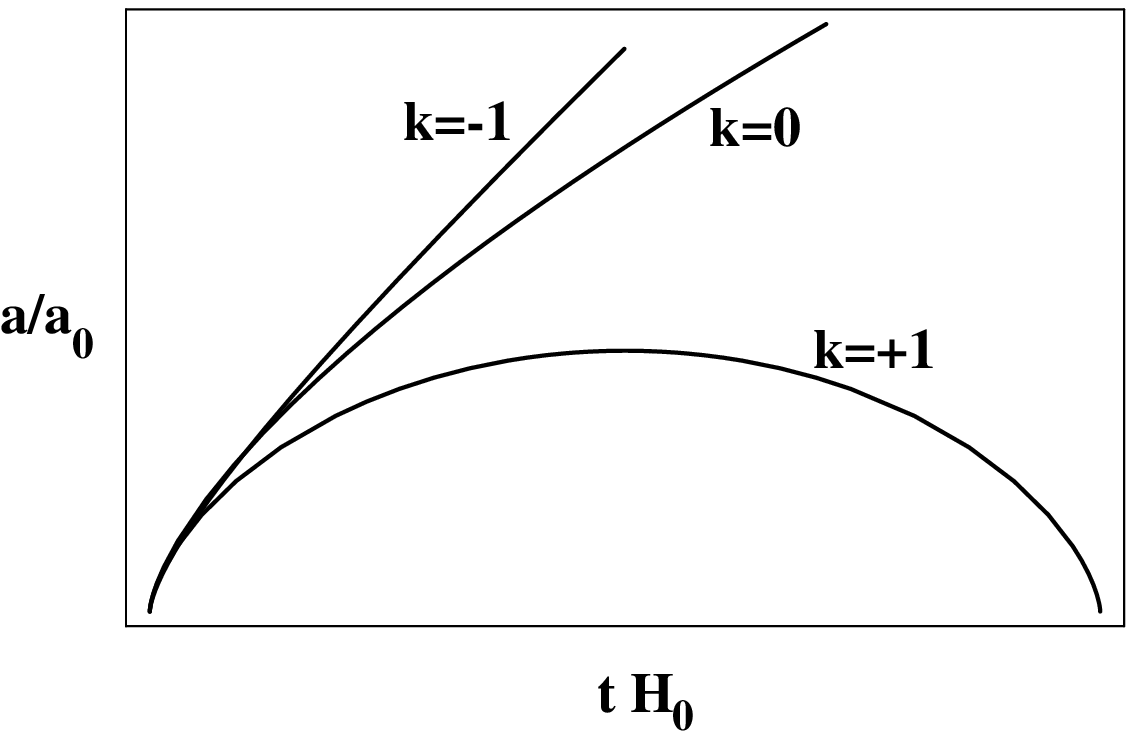}
\caption{\'Evolution du facteur d'expansion dans les cas $k=-1$, $k=0$
ou $k=+1$.}
\end{figure}

Je reviens tout de suite sur l'\'equation d'\'evolution de $a$, 
Eq. (\ref{Evola2}), et des
diff\'erents termes qui la composent. Si on suppose a priori que $a$
croit avec le temps, on voit que les termes en $\rho$, $k$ puis
$\lambda$ dominent successivement le terme de droite.

Pour simplifier la discussion on introduit deux quantit\'es
r\'eduites, 
\be \Omega=\rho/\rho_c\ \ \ {\rm avec}\ \
\rho_c={3\,H^2\over 8\pi\,G}.  
\ee 
et  
\be \Lambda={\lambda\over3\,H^2}.  
\ee
On a alors 
\be
\Omega+\Lambda=1+{k\over \dot{a}^2}.
\ee
Si $k=0$ la somme des 2 doit faire 1 par construction.
Remarquons aussi que ces deux quantit\'es sont en g\'en\'erale d\'ependantes
du temps, $\Omega=1$ et $\lambda=0$ \'etant  le seul point statique
(int\'eressant). On peut remarquer que le param\`etre de 
d\'ec\'el\'eration $q_0$ s'identifie \`a, 
\be
q_0={\Omega_0\over 2}-\Lambda_0.
\ee

Dans le plan ($\Omega$, $\Lambda$) on a, certes, de plus en plus
d'indications sur le point qu'occuperait actuellement l'Univers mais
il faut bien avouer qu'aucune mesure de $\Omega$ ou $\Lambda$ n'est
vraiment probante et indiscutable. On sait quand m\^eme que $\Omega$
doit \^etre compris entre 0.2 et 1. Pour $\Lambda$, c'est encore moins
pr\'ecis, disons qu'il est de l'ordre de quelques unit\'es. Si
$\Lambda$ est trop grand la distance lumineuse n'est plus monotone, et
finalement on a plus de Big-Bang. Si $\Lambda$ est trop petit
(n\'egatif) l'Univers est tr\`es jeune.

\begin{figure}
\vspace{8 cm}
\special{hscale=60 vscale=60 voffset=0 hoffset=-10 psfile=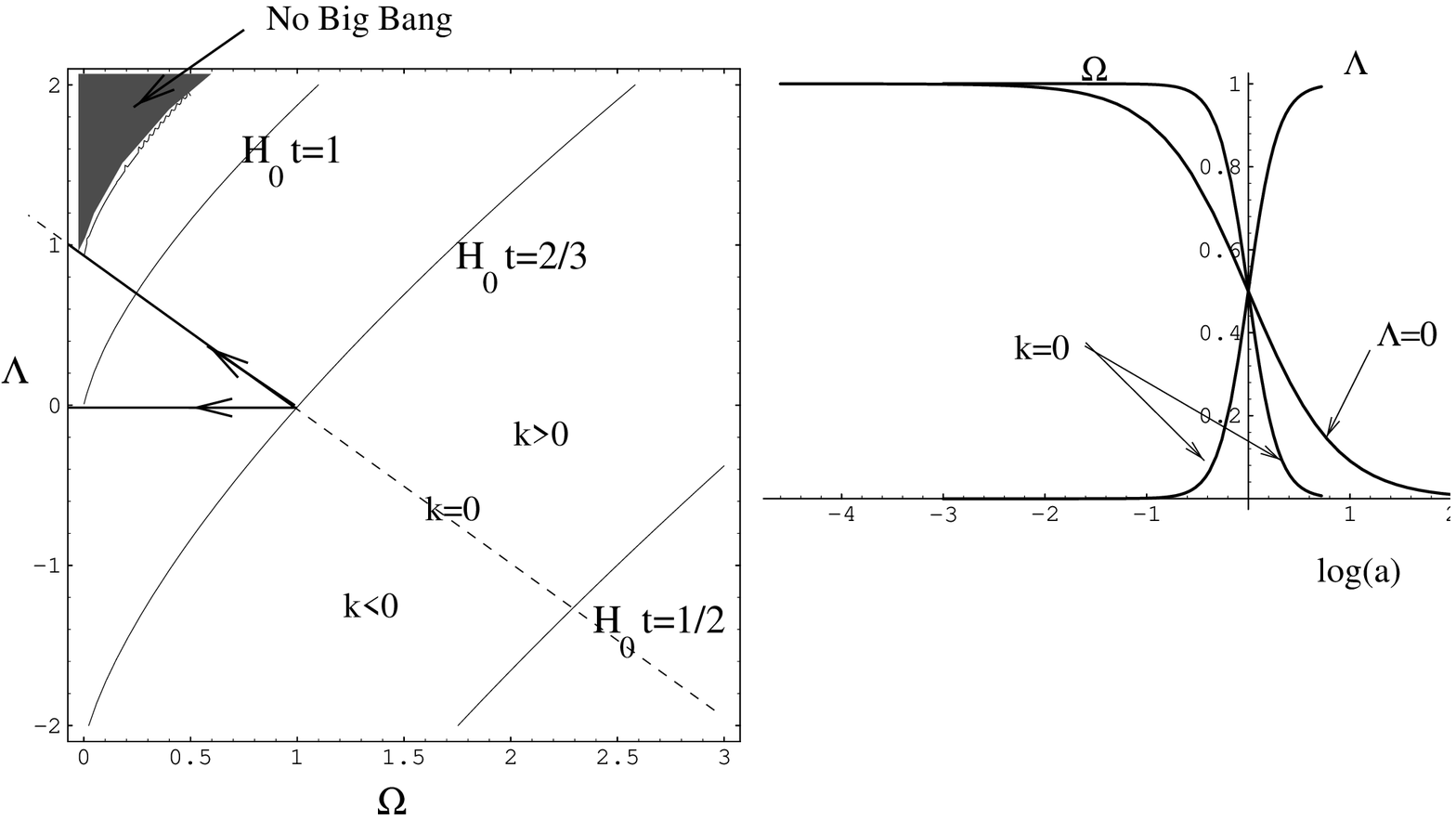}
\caption{Le plan $\Omega$-$\Lambda$ et \'evolution des param\`etres
cosmologiques r\'eduit $\Omega$ et $\Lambda$ dans les cas particulier
o\`u $\lambda=0$ ou $k=0$.}
\label{OmLam}
\end{figure}

Dans ce plan, il y a deux segments qui sont plus particuli\`erement
envisag\'es et qui correspondent \`a des a priori th\'eoriques diff\'erents,
\begin{itemize}
\item Le cas $\lambda=0$;
\item Le cas $\Omega+\lambda=1$ donc $k=0$;
\end{itemize}
Sur la figure (\ref{OmLam}), j'ai repr\'esent\'e les deux directions d'instabilit\'e,
et la d\'ependance de $\Omega$ et $\Lambda$ avec le temps dans ces deux
directions.

Avec une solution explicite pour le comportement du facteur
d'expansion, on peut expliciter la d\'ependance des
distances avec le redshift, 
\be
x_1(z)={z\,q_0+(q_0-1)(-1+\sqrt{2\,q_0\,z+1})\over
H_0\,a_0\,q_0^2\,(1+z)}.
\ee
Cette expression est valable pour $\lambda=0$. \'Evidemment on retrouve
le d\'eveloppement pr\'ec\'edent qui est valable pratiquement
jusqu'\`a $z=1$. Pour $\Omega_0=1$, on a $q_0=1/2$, on a,
\be
x_1={2\over H_0}\,\left(1-{1\over\sqrt{1+z}}\right)
\ee

Un concept int\'eressant \`a consid\'erer est l'horizon d'une
particule. Si on a un observateur \`a un temps $t_0$, on peut
se demander qu'elle elle la distance maximale peut se trouver
un \'ev\'enement observable, donc l'"horizon de la particule". Cet
horizon, $x_H$, est naturellement d\'efini par,
\be
\int_0^{x_H}{\d x\over\sqrt{1-k\,x^2}}=\int_{t\to 0}^t{\d t\over a(t)}.
\ee
Cette int\'egrale converge si le facteur d'expansion ne cro\^\i t pas 
trop vite \`a l'origine. On a,
\be
x_H=3\,\left({t\over t_0}\right)^{1/3}
\ee
Avec $H_0\,t_0=2/3$ on a,
\be
x_H={2\over H_0}\,\left({a\over a_0}\right)^{3/2}.
\ee
La taille de notre propre horizon est donc de l'ordre de 6000 
$h^{-1}$Mpc. Une quantit\'e auquelle on peut s'int\'eresser
c'est la taille apparente de l'horizon d'une particule sur
la surface de derni\`ere diffusion. Cette angle est donn\'ee par,
\be
\delta_H={x_H\over d_A(x_H)}=\left({a_*\over a_0}\right)^{1/2}
\approx 0.03=2\deg.
\ee
La taille de l'horizon sur le ciel est donc de l'ordre de $2\deg$:
si le facteur d'expansion a \'evolu\'e de mani\`ere standard depuis
l'origine du Big-Bang, des r\'egions du ciel s\'epar\'ees de plus
de cette distance n'ont donc jamais \'et\'e en contact causal.

\subsection{Contraintes actuelles sur les param\`etres cosmologiques}

Dans ce paragraphe je vais essayer de faire le point sur la
d\'etermination des param\`etres cosmologiques.

\begin{itemize}
\item La constante de Hubble. La mesure de la constante de Hubble est
une entreprise tr\`es difficile. La ma\^\i trise des biais
syst\'ematiques  est en particulier tr\`es ardue. Le satellite
Hipparcos a apport\'e quelque lumi\`ere \`a la premi\`ere \'etape de
l'\'echelle cosmique, en mesurant la distance par parallaxe de
c\'eph\'eides de notre voisinage. Le Hubble Space telescope s'est
\'evertu\'e \`a rechercher des c\'eph\'eides dans des galaxies
lointaines (Virgo).  Les valeurs les plus souvent admises sont de
l'ordre de 60 \`a 70 km/s/Mpc. Globalement les r\'esultats r\'ecents indiquent
quand m\^eme une r\'eduction du probl\`eme de l'\^age de l'Univers.

\item Le param\`etre de densit\'e, $\Omega$. 
Les estimations dynamiques donnent des valeurs de l'ordre 
de $0.2$ (pour les amas) \`a l'unit\'e (flots cosmiques \`a grande 
\'echelle, voir Dekel 1994). 
Je reviendrai sur ces m\'ethodes qui sont bas\'ees sur
des \'etudes de structures.

\item La constante cosmologique. En plus des contraintes mentionn\'ees
plus haut, on s'attend \`a ce que $\Lambda$ ne d\'epasse pas
l'unit\'e, sinon le volume de l'Univers serait tel qu'on aurait
\'enorm\'ement d'objets lointains (quasars) qui subiraient des effets
de lentille.

\item le param\`etre de d\'ec\'el\'eration $q_0$. La relation
distance-redshift, \`a  grand redshift, 
est sensible au premier ordre au param\`etre de 
d\'ec\'el\'eration $q_0$.  
L'id\'ee est de se servir de Super-Novae, qui sont \`a des
redshift connus, et de d\'eterminer
leur distance en s'appuyant sur leur courbe de lumi\`ere (voir figure).
Pour l'instant les contraintes ne sont pas forc\'ement tr\`es solides.

Pour donner un aper\c cu des derni\`eres tendances, globalement il semble
qu'un mod\`ele o\`u 
\ba
\Omega_0&=&0.3\\
\Lambda_0&=&0.7
\ea
avec une courbure nulle n'est pas une hypoth\`ese ridicule et commence \`a
\^etre prise s\'erieusement en compte.

\begin{figure}
\vspace{13 cm}
\special{hscale=80 vscale=80 voffset=-100 hoffset=-75 psfile=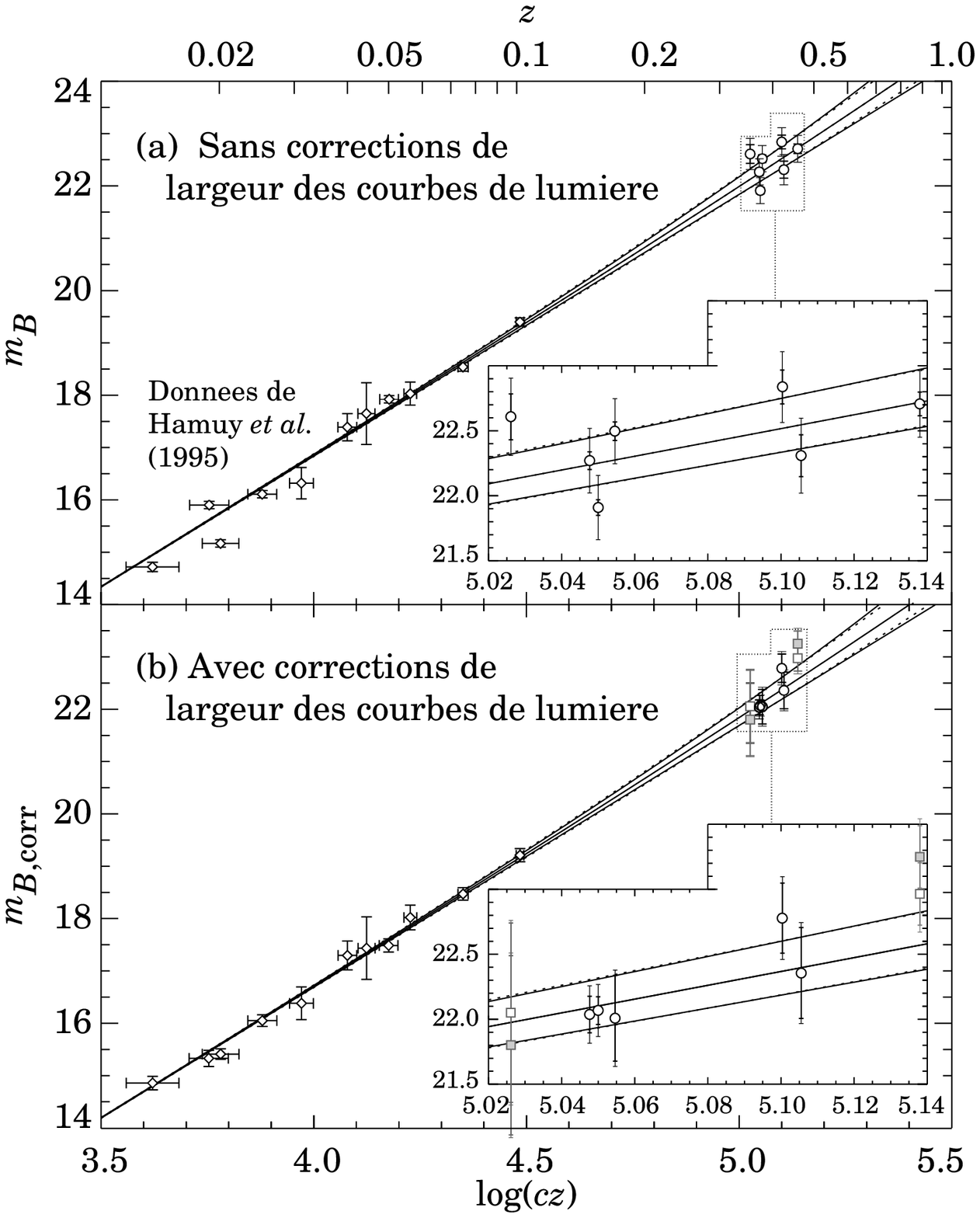}
\caption{La relation luminosit\'e-redshift et son application pour
mesurer le param\`etre de d\'ec\'el\'eration $q_0$. Les courbes
th\'eoriques 
correspondent respectivement \`a $q_0=0$ (en haut), $q_0=1/2$ (au
milieu) et $q_0=1$ (en bas) obtenues en supposant soit $\Lambda=0$
(courbes continues) soit $k=0$ (courbes pointill\'ees). Ces mesures
sont susceptibles d'\^etre affect\'ees par des erreurs syst\'ematiques
non prises en compte ici (extinction ...)
Figure tir\'ee de Perlmutter et al. 1996.}
\end{figure}

\end{itemize}

\subsection{ Histoire thermique de l'Univers}

\begin{figure}
\vspace{15 cm}
\special{hscale=70 vscale=70 voffset=0 hoffset=-20 psfile=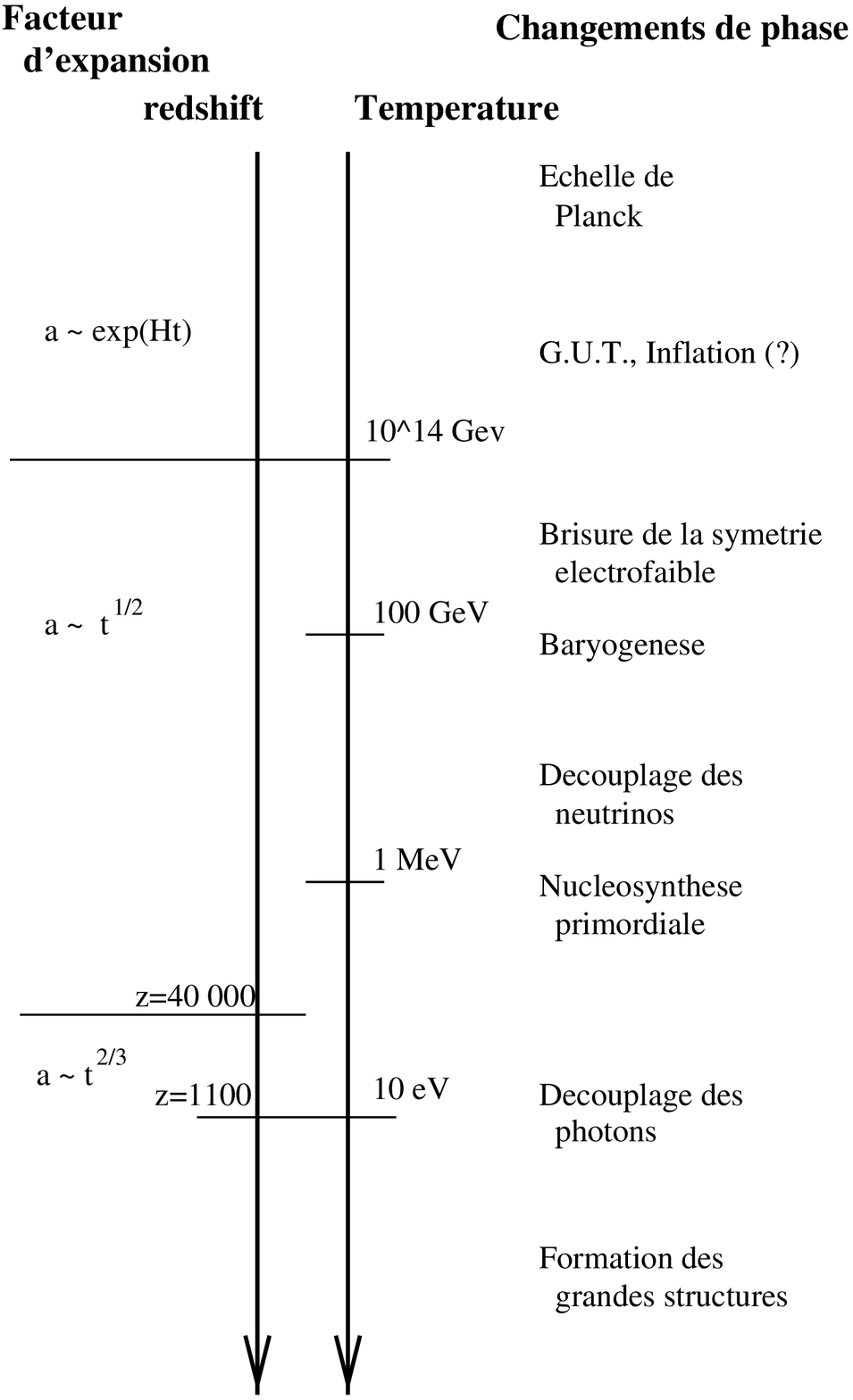}
\caption{Histoire thermique de l'Univers. 
Si les derni\`eres \'etapes sont relativement
bien comprises, d'autres sont beaucoup plus sp\'eculatives. Par exemple
il a \'et\'e envisag\'e que la baryogen\`ese ait pu avoir lieu aussi bien au
moment de la brisure de sym\'etrie \'electrofaible que des \'energies
de la GUT. Notons que la nucl\'eosynth\`ese primordiale
donne une contrainte sur la quantit\'e de baryons,
$\Omega_b\,h^2=0.013\pm 0.0003$.}
\end{figure}

Je poursuis mes rappels sur le mod\`ele du bing-bang chaud en faisant
un bref descriptif de l'histoire thermique de l'univers.
Dans les mod\`eles qui nous int\'eressent le facteur d'expansion est 
\'evidemment une fonction
croissante du temps, ce qui implique que la densit\'e et
\'eventuellement la
pression croissent quand on remonte dans le pass\'e.

On peut identifier un certain nombre d'\'etapes clefs.
On peut comparer la densit\'e de rayonnement \`a la densit\'e de
mati\`ere
(ou du moins ce que l'on en sait).

\begin{figure}
\vspace{8 cm}
\special{hscale=50 vscale=50 voffset=-80 hoffset=-0 psfile=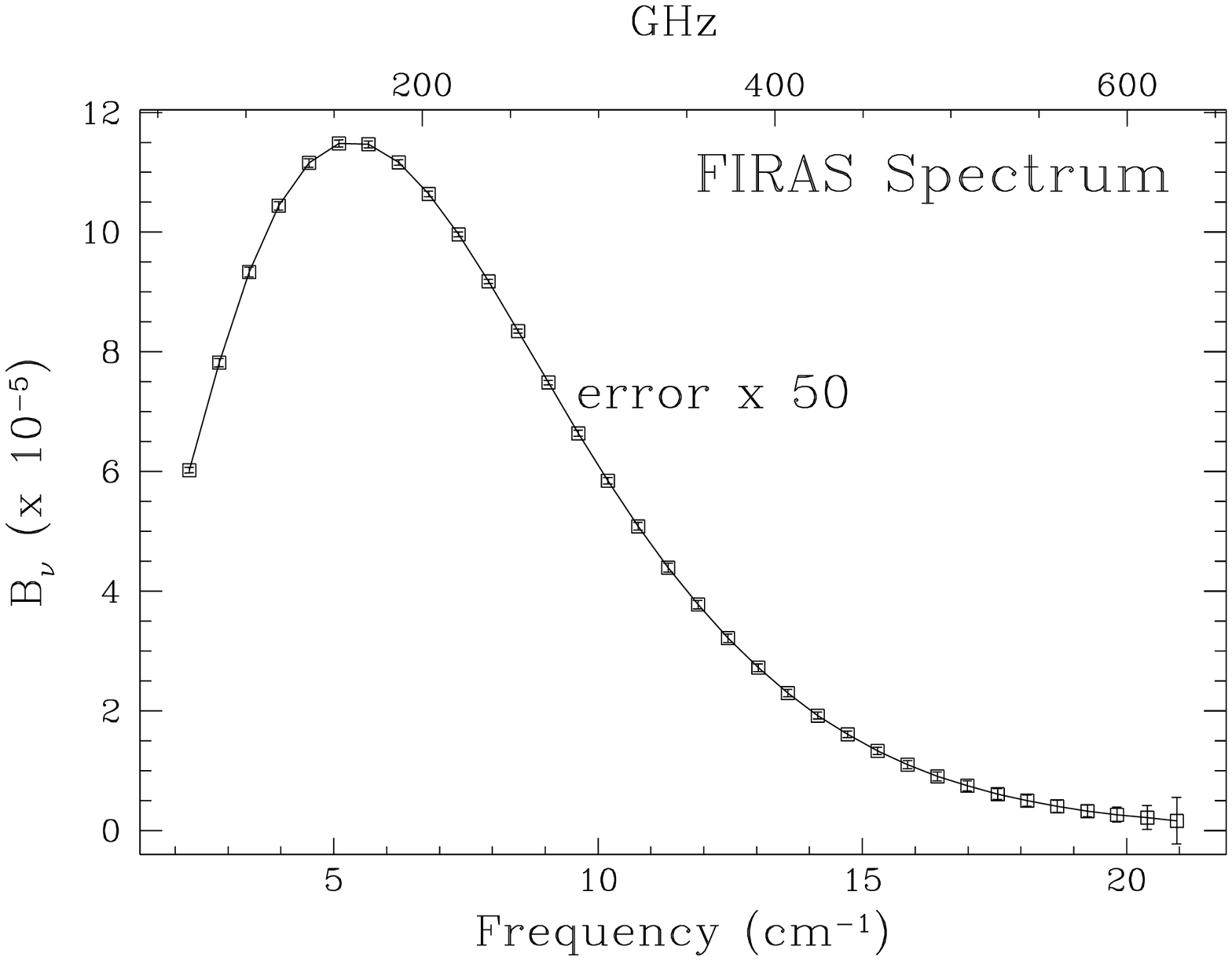}
\caption{Le spectre de corps noir observ\'e par l'exp\'erience
COBE/FIRAS (Mather et al. 1990).}
\end{figure}

Pour le rayonnement, c'est la densit\'e du corps noir cosmologique,
dont la  densit\'e
d'\'energie est donn\'ee par la loi de Stephan,
\be
\rho_{\gamma}=a\,T^4,
\ee
o\`u $a$ est la constante de Stephan.
Notez que c'est l\`a une quantit\'e tr\`es bien connue! (voir figure
montrant  l'\'emission du corps noir).
La densit\'e de mati\`ere, c'est par d\'efinition
\be
\rho_{\rm mat.}=\Omega_0\,\rho_c.
\ee
Celle-ci est nettement moins bien connue, \`a cause de $\Omega$ et de
$H_0$. Il n'emp\^eche,
on est aujourd'hui largement domin\'e par la mati\`ere, mais quand on
remonte dans le pass\'e il y a un moment o\`u $\rho_{\gamma}=\rho_{\rm mat}$.
Cette \'equivalence a lieu quand
\be
a=a_{\rm eq.}
\ee
avec
\be
a_{eq}= {1\over 40\,000\,\Omega_0\,h^2}.
\ee
A ce moment l\`a la temp\'erature est \`a peu pr\`es de 100\,000 K.

C'est un moment crucial pour l'\'evolution du facteur d'expansion, 
mais ce qui est aussi int\'eressant c'est ce qui se passe pour les
baryons quand
la temp\'erature et la pression montent. A une temp\'erature de
quelques milliers
de degr\'es la `recombinaison' a eu lieu: au del\`a de cette
temp\'erature \'electrons et protons se d\'ecouplent.  
Il reste de cette \'epoque le fond de rayonnement \`a 3 K.

Je vais illustrer ce qui se passe en calculant explicitement le redshift
de la recombinaison. Cela va illustrer la m\'ethode g\'en\'erale
mise en \oe uvre pour faire ce genre de calculs.

On suppose qu'on a un \'equilibre chimique entre les diff\'erentes esp\`eces
qui interviennent,
\be
e\ \ +p\ \ \leftrightarrow\ \ H\ \ +\gamma.
\ee
On \'ecrit l'\'equilibre des potentiels chimiques et on obtient,

\be
\mu_{e}+\mu_{p}=\mu_{H}+\mu_{\gamma}.
\ee

Pour les photons on a \'evidemment $\mu_{\gamma}=0$. Le photon est sa propre 
antiparticule et de plus on sait par l'observation du spectre du CMB qu'il
n'y a pas eu de d\'epos significatif d'\'energie dans le corps noir
comme ce pourrait \^etre le cas si on avait des particules instables.

Si on appelle $x$ une espace non-relativiste de masse $m_x$ et de
d\'eg\'en\'erescence $g_x$ on a,
\be
n_x=g_x\ \left({m_x\,T\over2 \pi}\right)^{3/2}\ \exp[(\mu_x-m_x)/T].
\ee
L'\'equilibre des potentiels chimiques implique alors que
(on suppose $m_H=m_p$ dans le pr\'e-facteur),
\be
{n_e\,n_p\over n_H\,n_b}={1\over n_b}\,\left({m_e\,T\over 2\pi}\right)^{3/2}\ 
\exp[-(m_e+m_p-m_H)/T].
\ee
Ce rapport s'exprime facilement en fonction de la fraction de
ionisation, $x_e$, avec 
\be
{n_e\,n_p\over n_H\,n_b}={x_e^2\over 1-x_e}.
\ee
La temp\'erature de la recombinaison qui correspond disons \`a $x_e=0.5$
a donc une d\'ependance logarithmique avec la densit\'e de baryons parce que
le pr\'efacteur est tr\`es grand: $\eta=n_b/n_{\gamma}\approx n_b/T^3\approx10^{-10}$.
Sachant que $m_e+m_p-m_H=13.6\ eV=160\,000\ K$,
on trouve plus sp\'ecifiquement que 
\be
T\approx 3500K\ \ {\rm soit}\ \ z_*\approx 1100.
\ee
Le moment de la recombinaison est donc relativement bien contr\^ol\'e.

A des temp\'eratures de l'ordre
du Mev, ce sont les neutrinos qui se sont d\'ecoupl\'es: il en reste 
un fond de neutrinos (non-d\'etect\'e!) dont il faut tenir compte
quand on calcule la densit\'e de rayonnement. A peu pr\`es au m\^eme
moment (\`a une temp\'erature l\'eg\`erement inf\'erieure, tout de
m\^eme) a eu lieu la nucl\'eosynth\`ese primordiale: formation des noyaux
de deut\'eriun, d'h\'elium 3 et 4 et de Lithium 7. Dans les conditions de densit\'e 
de l'univers primordiale, il est impossible de former des noyaux plus 
lourds (ces r\'eactions n'atteignent jamais l'\'equilibre
thermodynamique, ce qui donnerait des noyaux autour du fer) parce que les
noyaux de masse nucl\'eaire 5 et 8 sont instables. 
La nucl\'eosynth\`ese explosive saute
cette \'etape gr\^ace \`a une r\'eaction \`a trois corps, la triple
$\alpha$  qui donne directement du carbone: cette r\'eaction est 
\'evidemment tr\`es d\'efavorable en milieu t\'enu.

Ces \'etapes sont tr\`es bien comprises, et reposent sur de la physique
de laboratoire (les sections efficaces des r\'eactions nucl\'eaires 
par exemple) mais ce qui se passe \`a temp\'erature plus \'elev\'ee est plus
sp\'eculatif. Baryogen\`ese, m\'ecanismes
de brisure de sym\'etrie sont autant de sujet de recherche actifs. 

Quand on remonte encore en \'energie, \`a l'approche des \'energies de
GUT on a des th\'eories tr\`es sp\'eculatives, de plus en plus
\'eloign\'ees de la physique de laboratoire. Une id\'ee s'est
impos\'ee  depuis le d\'ebut des ann\'ees 80,
c'est le m\'ecanisme de l'inflation.

\subsection{ Inflation, motivations et principe de base}

L'inflation permet d'expliquer pas mal de chose dans un m\^eme
sc\'enario. Son plus grand m\'erite est sans doute
de mettre  dans le champ de la physique 
une probl\'ematique g\'en\'erale, les conditions initiales,
qui semblait jusqu'\`a lors en dehors de son champ d'investigation.
Du m\^eme coup l'Univers observable n'est plus qu'une infime
petite partie de l'Univers au sens plus philosophique du mot.

\begin{itemize}

\item 
Le probl\`eme de la platitude; Aujourd'hui la densit\'e est sinon
critique, du moins tr\`es proche de la densit\'e critique. Cela
implique que dans le pass\'e la densit\'e \'etait infiniment proche
de la densit\'e critique (voir figure 3).

\item 
Le probl\`eme de l'horizon. On observe que des r\'egions du ciel qui,
dans une cosmologie standard, n'ont jamais \'et\'e en relations causales
ont la m\^eme temp\'erature (\`a $10^{-5}$ pr\`es). 
\item
Absence de monopoles. On s'attend naturellement \`a avoir un reliquat 
de monopoles magn\'etiques \`a cause de transitions de phases en
th\'eorie des Grande Unification. Ce serait des objets massifs qui 
produiraient une densit\'e de 15 ordres de grandeur plus grande que
la densit\'e critique.

\item 
La formation des grandes structures: m\'ecanisme pour la
g\'en\'eration des fluctuations.

\end{itemize}

\begin{figure}
\vspace{12 cm}
\special{hscale=70 vscale=70 voffset=-0 hoffset=-0 psfile=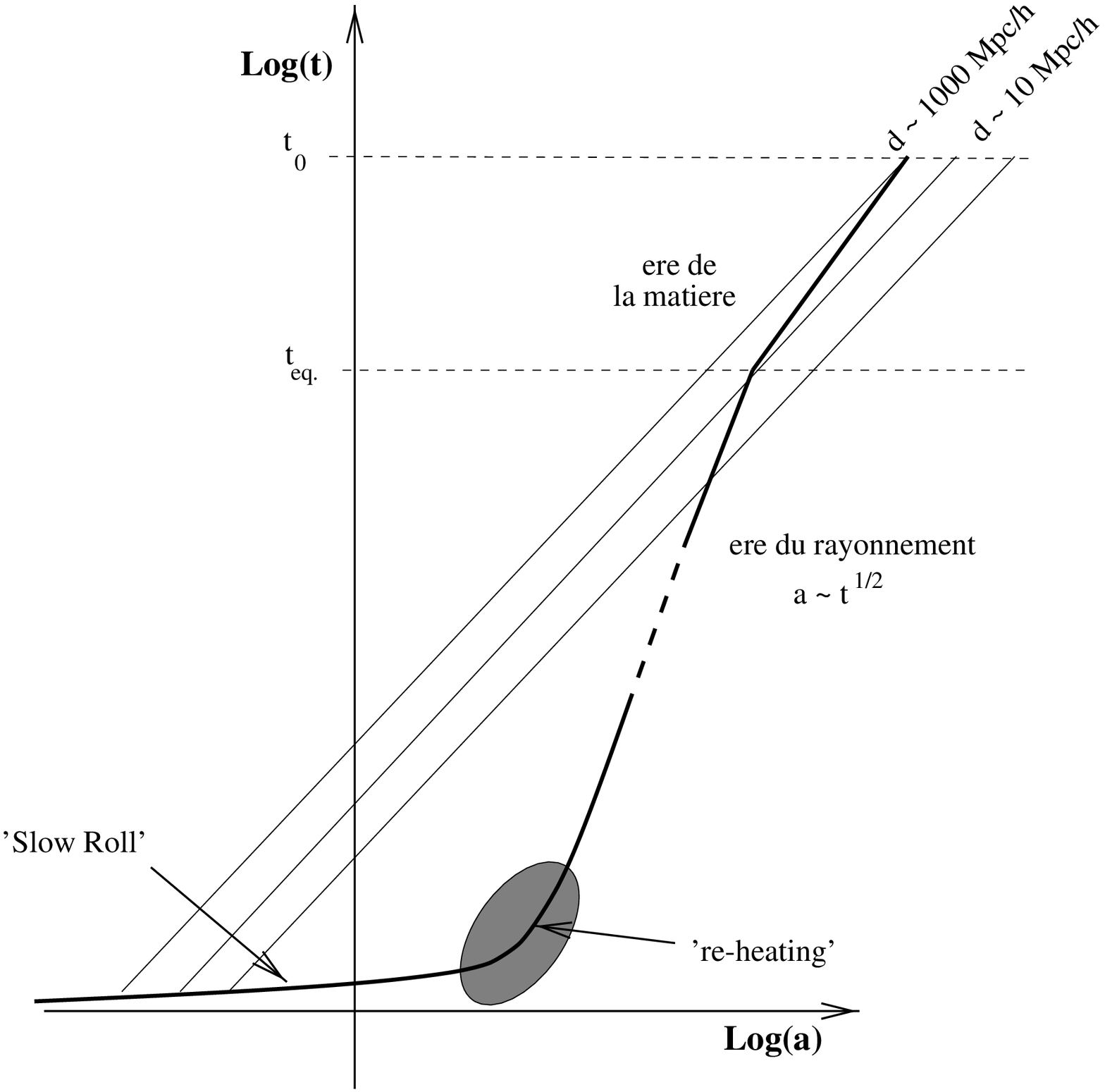}
\caption{Repr\'esentation sch\'ematique du sc\'enario de l'inflation}
\label{Inflation2}
\end{figure}

Comment marche l'inflation?
L'id\'ee c'est qu'\`a un moment donn\'e, apr\`es l'\'echelle de
Planck, la densit\'e d'\'energie est concentr\'ee dans un champ scalaire,
l'inflaton $\varphi$, et que le potentiel de ce champ est \`a l'origine non
nul: cela conduit \`a une croissance rapide du facteur d'expansion.

Dans la suite, je ne vais pas essayer de faire une pr\'esentation
exhaustive de la th\'eorie de l'inflation. Je vais m'int\'eresser
\`a  un cas simple correspondant \`a ,
\begin{itemize}
\item
un champ simple pour l'inflaton;
\item 
cas o\`u l'approximation dite de 'Slow Roll' est valable.
\end{itemize}

Rappelons que l'action pour un champ quantique s'\'ecrit,
\be
S=\int\d^4\vx\,\sqrt{-g}\,\left({\cal L}-
{{\cal R}\,m_{\rm pl.}^2\over 16\,\pi}\right)
\ee
o\`u $\cal{L}$ est la densit\'e de Lagrangian,
\be
{\cal L}={1\over2}\left(\partial_{\mu}\varphi\,\partial^{\mu}\varphi\right)-
V(\varphi).
\ee
Le terme suppl\'ementaire qui appara\^\i t dans l'action redonne les
\'equation d'Einstein.
Dans cette partie, pour reprendre l'habitude de la physique des
hautes \'energie, la constante de gravitation $G$ sera not\'ee,
\be
G\equiv {1\over m_{pl.}^2}\ \ {\rm quand}\ \ \hbar=c=1.
\ee

Dans la suite, je vais illustrer mon propos avec le cas simple, 
\be
V(\varphi)\propto {\lambda\over 4} \varphi^4,
\ee
correspondant \`a l'inflation chaotique ch\`ere \`a Linde. Il y a deux
dynamiques qui sont li\'ees,
l'une du facteur d'expansion qui d\'epend de la densit\'e d'\'energie,
l'autre  du champ $\varphi$.
L'\'equation d'\'evolution pour $\varphi$ s'obtient en cherchant un
extremum de l'action et on trouve,
\be
\ddot{\varphi}+3\,{\dot{a}\over a}\,\dot{\varphi}-{1\over a^2}\,
\Delta\varphi=-{\d\,V\over \d\varphi}.
\ee
L'\'equation d'\'evolution du facteur d'expansion s'obtient en 
\'ecrivant les \'equations d'Einstein. La densit\'e locale
d'\'energie est donn\'ee par,
\be
\rho(\vx)={\dot{\varphi}^2\over2}+{(\nabla\varphi)^2\over
2}+V(\varphi).
\ee
On a donc,
\be
\dot{a}^2={8\pi\,a^2\over 3\,m_{\rm pl.}^2}\,
\left[{\dot{\varphi}^2\over2}+{(\nabla\varphi)^2\over
2}+V(\varphi)\right]-k.
\ee

En fait d'abord consid\'erer la valeur moyenne du champ
qu'on \'ecrit donc,
\be
\varphi(t,\vx)=\varphi_0(t)+\delta\varphi(t,\vx)
\ee
et on fait l'hypoth\`ese que la relativit\'e g\'en\'erale s'applique 
'classiquement' \`a la
valeur moyenne. Cela suppose qu'on s'int\'eresse \`a une \'epoque
post\'erieure \`a l'\'epoque de Planck (jusqu'\`a preuve du contraire
on ne sait pas faire de la gravit\'e quantique).
La quantification du champ se fera sur la partie fluctuante
uniquement, et on suppose que la densit\'e
d'\'energie port\'ee par $\varphi_0$ est bien plus importante. Avec
cette hypoth\`ese on peut traiter
la dynamique de $\varphi_0$ ind\'ependamment de ses fluctuations.

L'\'equation d'\'evolution donne pour $\varphi_0$,
\be
\ddot{\varphi_0}+3\,{\dot{a}\over a}\,\dot{\varphi_0}=
-{\d\,V\over \d\varphi}(\varphi_0),
\ee
et celle pour le facteur d'expansion donne,
\be
\dot{a}^2={8\pi\,a^2\over 3\,m_{\rm pl.}^2}\,
\left[{\dot{\varphi_0}^2\over2}+V(\varphi_0)\right]-k.
\ee
On reconna\^\i t pour l'\'equation d'\'evolution de $\varphi$,
les termes qu'on attend dans une m\'etrique de Minkowski
plus un autre terme qui vient de la d\'ependance en temps du facteur 
d'expansion, qui agit comme un terme de frottement. 
Pendant l'inflation, c'est 
ce terme qui domine: on a un r\'egime stationnaire o\`u l'\'energie 
du champ se dissipe dans une expansion rapide,
\ba
\dot{a}^2&=&{8\pi\,G\,a^2\over 3}\,V(\varphi),\\
3\,{\dot{a}\over a}\,\dot{\varphi_0}&=&
-{\d\,V\over \d\varphi}(\varphi_0).
\ea

La solution de ces \'equations donnent pour le potentiel de d\'epart,
\be
\varphi_0(t)=\varphi_0\,\exp\left(-\sqrt{\lambda\over 6\,\pi}\,m_{\rm
pl.}\,t\right),
\ee
et,
\be
a(t)\sim \exp(H\,t),
\ee
avec,
\be
H(t)=\sqrt{8\,\pi\,V\over 3\,m_{\rm pl.}^2}.
\ee
Cette solution en $a(t)$ implique que 
\be
{\ddot{a}\over a}\sim\left({\dot{a}\over a}\right)^2
\ee
ce qui implique une \'equation d'\'etat effective pour le fluide 
cosmique $p=-\rho$.
Une cons\'equence est la rapide d\'ecroissance du terme
de courbure dans l'\'equation qui donne la constante de Hubble.
En effet $\Omega_k\equiv-k/\dot{a}^2$ cro\^\i t pour le big bang 
standard (comme $t^{1/2}$ ou $t^{1/3}$)
mais d\'ecro\^\i t tr\`es rapidement (comme $\exp(-Ht)$)
pendant l'inflation.

La validit\'e de l'approximation de 'Slow Roll'
implique un certain nombre de propri\'et\'es de platitute
sur la forme du potentiel de l'inflaton. On peut exprimer ces
conditions comme des conditions
sur la forme du potentiel. En particulier on doit
avoir $\dot{\varphi_0}^2\ll V$ ce qui implique,
\be
\epsilon={m_{\rm pl.}^2\over 16\,\pi}\,\left(V'\over V\right)^2\ll 1,
\ee
et la condition $\ddot{\varphi_0}\ll V'$ implique,
\be
\vert\eta\vert\ll 1
\ \ {\rm avec}\ \ 
\eta={m_{\rm pl.}^2\,V''\over 8\,\pi\,V}.
\ee

Cela dit on peut
quand m\^eme faire marcher l'inflation dans de tels cas. Par exemple
Starobinsky a examin\'e un certain nombre de cas o\`u certaines des
conditions de 'Slow Roll' ne sont pas v\'erifi\'ees, par exemple quand
on a une non-analyticit\'e dans la forme du potentiel.

Mais \'evidemment l'inflation n'est pas seulement int\'eressante
pour r\'esoudre un certain nombre de probl\`emes li\'es \`a l'Univers
homog\`ene, cela donne aussi un moyen de produire des fluctuations
de la m\'etrique \`a l'origine des grandes structures.

Dans ce cas l\`a, ce sont les fluctuations de l'inflaton que 
l'on doit examiner.
L'hypoth\`ese physique sous-jacente est qu'il est l\'egitime de quantifier
les fluctuations de l'inflaton dans une m\'etrique d\'ependante du temps.
Cela implique que l'on puisse \'ecrire,
\be
\delta\varphi=
\int\d^3 \vk\left[a_{\vk}\,\psi_k(t)\,\exp(\ii\vk.\vx)+
a^{\dag}_{\vk}\,\psi_k^*(t)\,\exp(-\ii\vk.\vx)\right]
\ee
o\`u $a^{\dag}$ et $a$ vont \^etre les op\'erateurs de cr\'eation de
d'annihilation du champ. Ils ob\'eissent donc \`a la relation de
commutation suivante,
\be
[a_{\vk},a^{\dag}_{-\vk'}]=\delta(\vk+\vk').
\ee
L'\'equation du mouvement v\'erifi\'ee par ce champ est,
\be
\ddot{\delta\varphi}+3\,H\,\dot{\delta\varphi}+{
\Delta{\delta\varphi}\over a^2}=-V''\delta\varphi\approx 0.
\ee
Dans l'approximation du 'Slow-Roll' le terme de source
est n\'egligeable et $H$ est constant. 
On a l\`a une \'equation de champ libre.
Cette \'equation donne
la forme fonctionelle de $\psi_k$ en fonction du temps.

Pour avoir la normalisation on peut imposer qu'\'a petite \'echelle
on retrouve les m\^emes r\`egles de quantification que
pour un espace de Minkowski, c'est \`a dire que
\be
\left[\varphi,\pi\equiv{\partial\over \partial \varphi}{\cal L}=
{\partial\over \partial t}\varphi\right]
=\ii\,\delta^{(3)}(a\,\vx_1-a\,\vx_2)
={\ii\over a^3}\,\delta^{(3)}(\vx_1-\vx_2)
\ee

Pour une m\'etrique de Minkowski les fonctions $\psi_k$ sont simplement 
proportionelles \`a des exponentielles et leur amplitude est donn\'ee par
\be
\psi_k(t)={1\over a^{3/2}}\,\left({a\over 2\,k}\right)^{1/2}\,
\exp[\ii\,k\,t/a].
\ee
Pour un espace de Sitter il faut tenir compte de l'\'evolution
du terme suppl\'ementaire en $H$ et on a,
\be
\psi_k(t)={H\over (2\,k)^{1/2}}\,
\left(\ii+{k\over \,a\,H}\right)\,\exp\left[{\ii\,k\over a\,H}\right].
\ee
A petite \'echelle on retrouve bien s\^ur les solutions correspondant
au cas Minkowski. Cela fixe au passage la normalisation
de ces modes: c'est parce que les fluctuations sont \`a l'origine quantique
qu'on peut en calculer l'amplitude. A la travers\'ee de l'horizon on a un
mode croissant tr\`es simple. Au bout du compte on a un mode simple de la
forme,
\be
\hat\varphi_{\vk}\approx {H\over
k^{3/2}}\,\left(a_{\vk}+a_{-\vk}^{\dag}\right),\ \ 
\delta\varphi=\int\d^3\vk\,\hat\varphi_{\vk}\,e^{\ii\,\vk.\vx}.
\ee
Les cons\'equences sont nombreuses et tr\`es int\'eressantes:
\begin{itemize}
\item
Toutes les observables qu'on peut construire \`a partir de ces champs
vont commuter entre elles. On passe donc de fluctuations quantiques \`a des
variables qui se comportent comme des grandeurs stochastiques classiques.
On peut facilement v\'erifier que les variables al\'eatoires 
$\left(a_{\vk}+a_{-\vk}^{\dag}\right)$ induisent un champ stochastique 
Gaussien. Pour \^etre plus sp\'ecifique on a 
\ba
<0\vert\hat\varphi_{\vk}\hat\varphi_{\vk'}\vert 0>
&=&\delta^{(3)}(\vk+\vk')\,P_{\varphi}(k)\\
<0\vert\hat\varphi_{\vk_1}\dots\hat\varphi_{\vk_{2p}}\vert 0>
&=&\sum_{\rm combinaisons}\prod_{\rm paires\ (i,j)}
<0\vert\hat\varphi_{\vk_i}\hat\varphi_{\vk_j}\vert 0>\\
<0\vert\hat\varphi_{\vk_1}\dots\hat\varphi_{\vk_{2p+1}}\vert 0>
&=&0.
\ea
Par la suite on identifie les moyennes sur le vide \`a des moyennes
d'ensemble sur les variables 'classiques stochastiques' $\varphi_{\vk}$.

\item
C'est la valeur de $H$ qui va fixer les variations de l'amplitude
des fluctuations de densit\'e par la suite. Dans l'approximation
du 'Slow roll' la valeur de $H$ ne change que tr\`es peu pour les
\'echelles accessibles \`a l'observation, et du coup on va avoir
un spectre proche d'un spectre Harrison-Zeldovich.
Pour \^etre plus pr\'ecis, le spectre de puissance
du potentiel est en $k^{-3}$, celui de la densit\'e sera en
$k^{n_s}$ avec $n_s\approx 1$.

\item
Comme on va le voir par la suite, dans ces approximations on a des
fluctuations dites adiabatiques.
\end{itemize}

Conditions pout avoir un mod\`ele d'inflation valide:

Il faut bien sur que la pente du potentiel soit suffisamment faible
pour que \c ca marche. En effet pour
que l'inflation remplisse son r\^ole, i.e. remette causalement en relation
des  r\'egions de l'Univers
observable, il faut que cette expansion ait dur\'e suffisamment
longtemps. La sortie de l'inflation n'est pas forc\'ement tr\`es bien 
comprise: les fluctuations de l'inflaton doivent \^etre prises
en compte dans la densit\'e d'\'energie. On a cr\'eation de
particules. Mais en fait cela n'affecte pas forc\'ement le probl\`eme
de la formation des grandes structures (voir figure \ref{Inflation2}).

L'amplitude des fluctuations sera de l'ordre de $H$, donc
\`a peu de chose pr\`es constante sur les \'echelles qui nous 
int\'eressent: de 1 \`a $1000\,h^{-1}$Mpc on ne couvre qu'une petite
fraction du temps d'inflation.

Cela a un certain nombres de cons\'equences:
\begin{itemize}
\item Univers \`a courbure plane;
\item Fluctuations initiales gaussiennes;
\item Un spectre de fluctuation invariant d'\'echelle dans les
cas g\'en\'eriques, c'est \`a dire quand le potentiel de l'inflaton
ne pr\'esente pas d'accident dans le domaine d'\'energie qui nous 
int\'eresse.
\end{itemize}

\section{\'El\'ements de cosmologie, l'Univers inhomog\`ene}

\subsection{ La croissance des fluctuations}

On s'int\'eresse ici \`a la croissance des fluctuations dans diff\'erents
r\'egimes, aussi bien pour des \'echelles plus grandes que l'horizon,
plus petites, au moment de l'\`ere du rayonnement que de l'\`ere de
la mati\`ere. Les calculs pr\'esent\'es dans cette section 
ont \'et\'e fortement inspir\'es de la th\`ese de Wayne Hu (1995).
 
D'une mani\`ere g\'en\'erale il faut se donner une m\'etrique qui puisse
d\'ecrire les fluctuations scalaires. On est confront\'e au probl\`eme 
du choix des jauges, c'est \`a dire qu'il est possible d\'ecrire les m\^emes
fluctuations physiques de differentes fa\c cons. 
Pour les fluctuations scalaires
on peut choisir la jauge suivante (jauge Newtonienne ou longitudinale,
voir Bardeen 1980),
\ba
g_{00}&=&-a^2\ \left[1+2\,\psi(t)\,Q(\vx)\right]\\
g_{0i}&=&0\\
g_{ij}&=&a^2\ \left[1+2\,\phi(t)\,Q(\vx)\right]\,
\gamma_{ij}
\ea
Ici la composante 0 d\'ecrit le temps conforme $\eta$ avec,
\be
\d t=a\d\eta,
\ee
de telle mani\`ere qu'on puisse factoriser $a$ dans la m\'etrique.
On va s'int\'eresser \`a l'\'evolution d'un certain mode de fluctuation
de vecteur d'onde $\vk$, c'est \`a dire d'un mode propre du Laplacien
\`a 3D,
\be
\nabla^2\,Q \equiv \gamma_{ij}\,Q=-k^2\,Q.
\ee
Ce mode propre est un champ scalaire. Il d\'efinit un champ
vectoriel $Q_i$ et un champ tensoriel $Q_{ij}$ d\'efinis par
\ba
Q_i&\equiv&Q_{,i}\\
Q_{ij}&\equiv& {1\over k^2}\,Q_{\vert ij}+{1\over 3}\gamma_{ij}\,Q.
\ea

L'id\'ee est alors de lin\'eariser les \'equations d'Einstein 
en $\phi$ et $\psi$ en supposant qu'on a des fluctuations de densit\'e
dans chacun des fluides cosmiques.

Evidemment on peut compliquer \`a l'infini les mod\`eles cosmologiques 
en introduisant des tas de composantes, mais pour se fixer les id\'ees
je vais supposer qu'on a essentiellement 3 composantes, des photons
(et des neutrinos), des baryons et de la mati\`ere noire.
Cette mati\`ere noire sera de plus suppos\'ee massive, autrement dit 
ce sont des 
mod\`eles de type CDM que je vais consid\'erer en priorit\'e.

La partie la plus difficile des calculs \`a mettre en \oe uvre
concerne la description du couplage des photons.
On doit \'ecrire l'\'equation d'\'evolution de la
densit\'e de photons $f(\vx,\vp)$ dans l'espace des phases.
Formellement cela s'\'ecrit sous la forme d'une \'equation de
Boltzmann,
\be
{\d f\over \d t}=C[f],
\ee
o\`u $C[f]$ est le terme de collisions, diffusion Thomson des photons
sur les \'electrons libres.

Il faut noter que, la diffusion Thomson et
les effets Doppler n'affectent pas la nature thermique de la 
distribution d'\'energie des photons. Cela justifie que l'on identifie
les fluctuations d'\'energie des photons avec les fluctuations locales
de temp\'erature (\`a un facteur 4 pr\`es). On d\'efinit,
\be
\theta(\vx,\gamma)=\disp{{1\over 4}
\left({1\over \pi^2\rho_{\gamma}}\int p^3\d p\,f-1\right)}.
\ee
C'est la temp\'erature des photons au point $\vx$ dans la direction $\gamma$.
L'\'equation des g\'eod\'esiques s'\'ecrit,
\be
{\d\over\d\eta}(\theta+\psi)={\partial\over\partial\eta}(\psi-\phi)
\ee
auquel il faut rajouter un terme de collisions.
On peut exprimer le r\'esultat sous forme d'une hi\'erarchie 
sur les coefficients d'une d\'ecomposition multipolaire
de $\theta$,
\be
\theta(\vx,\gamma)=
\sum_l\,\theta_l(\vk)\,P_l(\hat\vk,\gamma)\,e^{\ii\vk.\vx}
\,(-i)^l.
\ee
L'\'equation des g\'eod\'esiques avec terme de collisions donne alors,
\ba
\dot\theta_0&=&{-k\over 3}\theta_1-\dot\phi\\
\dot\theta_1&=&k\,\left[\theta_0+\psi-{2\over 5}\theta_2\right]-
n_e\,x_e\,\sigma_T\,(\theta_1-V_b)\\
\dot\theta_2&=&k\,\left[{2\over 3}\theta_1-{3\over
7}\theta_3\right]-{9\over 10}\,n_e\,x_e\,\sigma_T\,\theta_2\\
&\dots&\nonumber\\
\dot\theta_l&=&k\,\left[{l\over 2l-1}\theta_{l-1}-{l+1\over
2l+3}\theta_{l+1}\right]-\,n_e\,x_e\,\sigma_T\,\theta_l\ {\rm pour}\ l>2.
\ea
On reconna\^\i t un terme de couplage entre les photons et la
mati\`ere baryonique o\`u $n_e$ est la densit\'e d\'electrons, $x_e$
la fraction d'ionisation et $\sigma_T$ est la section efficace 
de la diffusion Thomson. 

La partie sans couplage correspond \`a la diffusion libre.
C'est le syst\`eme d'\'equations qu'on a pour les
neutrinos.

Si je veux pouvoir \'ecrire les \'equations d'Einstein je dois,
pour chaque esp\`ece $x$, \'ecrire le tenseur \'energie impulsion,
\ba
T^0_0&=&-[1+\delta_x(t)\,Q(\vx)]\rho_x(t)\\
T^0_i&=&\left[\rho_x(t)+p_x(t)\right]\,V_x(t)\,Q_i(\vx)\\
T^i_j&=&p_x(t)\,\left[\delta^i_j+{\delta p_x(t)\over p_x(t)}\,\delta^i_j
\,Q(\vx)+\Pi_x(t)\,Q^i_j(\vx)\right].
\ea
On peut alors \'ecrire les \'equations de continuit\'e pour chacun des
fluides jusqu'\`a l'ordre 1 par rapport aux fluctuations de densit\'e.
A l'ordre 0 on obtient,
\be
\dot{\rho_x}=-3\,\left(1+{p_x\over \rho_x}\right)\,{\dot{a}\over a}\,\rho_x
\ee
Cette une \'equation d\'ej\`a connue.

A l'ordre 1 des perturbations on a, \`a partir de $T^{\mu\,0}_{;\mu}=0$,
\be
\dot{\delta_x}=-\left(1+{p_x\over \rho_x}\right)\left(k\,V_x+3\,\dot{\phi}\right)-
3\,{\dot{a}\over a}\delta\left({p_x\over \rho_x}\right).
\ee
Si on applique cette relation aux photons on a,
\be
\dot{\delta_{\gamma}}=-{4\over 3}\,\left(k\,V_{\gamma}+3\,\dot{\phi}\right),
\ee
et si on l'applique \`a la mati\`ere,
\be
\dot{\delta_{\rm mat.}}=-\left(k\,V_{\rm mat.}+3\,\dot{\phi}\right).
\ee
Pour les photons on retrouve la premi\`ere \'equation
de la hi\'erarchie pr\'ec\'edente.
Cela implique en particulier que,
\be
{\d \over \d \eta}
\left(\delta_{\rm  mat.}-{3\over
4}\delta_{\gamma}\right)=k(V_{\gamma}-V_{\rm mat.}).
\ee

On a des \'equation du mouvement (d'Euler) pour chacun des fluides
s'obtiennent \`a partir de $T^{\nu\,i}_{;\nu}=0$.
Pour les photons l'\'equation est d\'ej\`a \'ecrite.
En appliquant cette contrainte au m\'elange photons+baryons
on a
\ba
\dot{V_b}&=&-{\dot{a}\over a}\,V_b+k\psi+{4\rho_b\over 3\rho_{\gamma}}\,
n_e\,x_e\,\sigma_T\,(V_{\gamma}-V_b).
\ea
Il reste pour la composante mati\`ere froide,
\ba
\dot V_{CDM}&=&-{\dot{a}\over a}\,V_{CDM}+k\psi.
\ea

Pour clore mon syst\`eme d'\'equation je dois \'ecrire les \'equations
de Poisson (i.e. les \'equations d'Einstein) qui donnent,
\ba
k^2\,(\psi+\phi)&=&-8\,\pi\,G\,a^2\,p_T\,\Pi_T;\\
{\dot{a}\over a}\,\psi-\dot{\phi}&=&
4\,\pi\,G\,a^2\,(\rho_T+3\,p_T)\,{V_T\over k};\\
k^2\,\phi&=&4\,\pi\,G\,a^2\,\rho_T\,
\left(\delta_T+3\left(\dot{a}\over a\right)\,
(1+p_T/\rho_T)\,{V_T\over k}\right),
\ea
ce qui donne un syst\`eme ferm\'e.

\begin{figure}
\vspace{12 cm}
\special{hscale=70 vscale=70 voffset=10 hoffset=-10 psfile=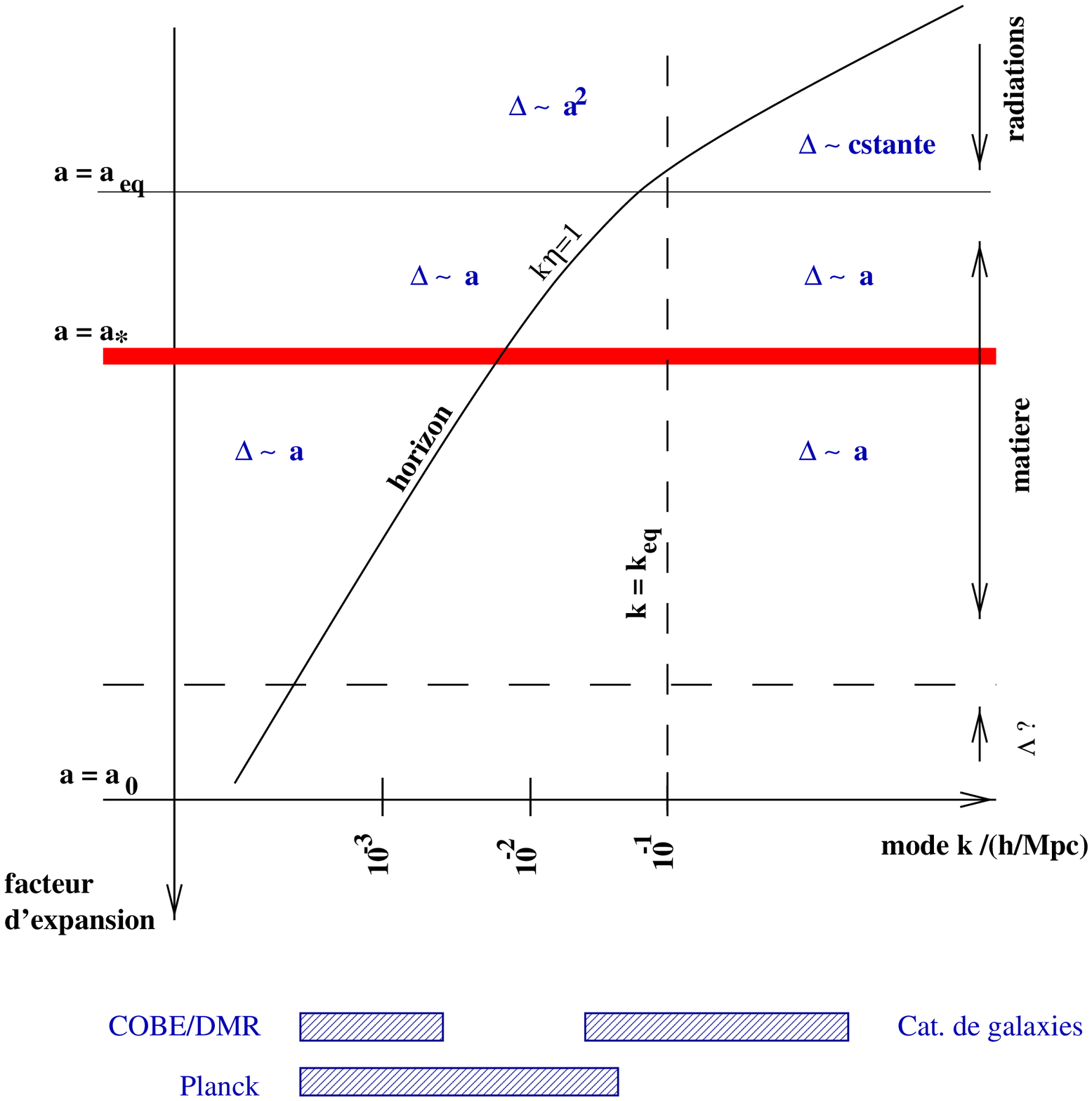}
\caption{Les diff\'erents r\'egimes de croissance gravitationnelle.
$\Delta$ donne la croissance du contraste de densit\'e
de la mati\`ere noire froide. Les zones hachur\'ees du bas montrent
les diff\'erents domaines accessibles aux observations.}
\end{figure}

Quelques remarques:
\begin{itemize}
\item
Les indices $T$ font r\'ef\'erence \`a des quantit\'es
totales (obtenues en sommant sur tous les fluides pond\'er\'es
par leur densit\'e moyenne respective).
\item
Ces \'equations d\'ependent de la jauge utilis\'ee. En particulier le
terme
de source de l'\'equation de Poisson proprement dite (la derni\`ere)
s'identifie dans une certaine jauge \`a la fluctuation de densit\'e
totale. 
\item
$\Pi_T$ est la partie anisotrope de la pression. Elle est nulle
pour la mati\`ere ($p=0$ de toute fa\c con). Elle est nulle
pour les photons quand ils sont coupl\'es aux baryons. La seule
contribution r\'esiduelle vient des neutrinos apr\`es qu'ils se soient
d\'ecoupl\'es de la mati\`ere.
\end{itemize}

A partir de l\`a je peux chercher \`a calculer les diff\'erents modes
de croissance. Le sch\'ema de la Fig. 9 montre les diff\'erents r\'egimes
et leur encha\^\i nement.

\begin{figure}
\vspace{6 cm}
\special{hscale=80 vscale=80 voffset=-10 hoffset=20 psfile=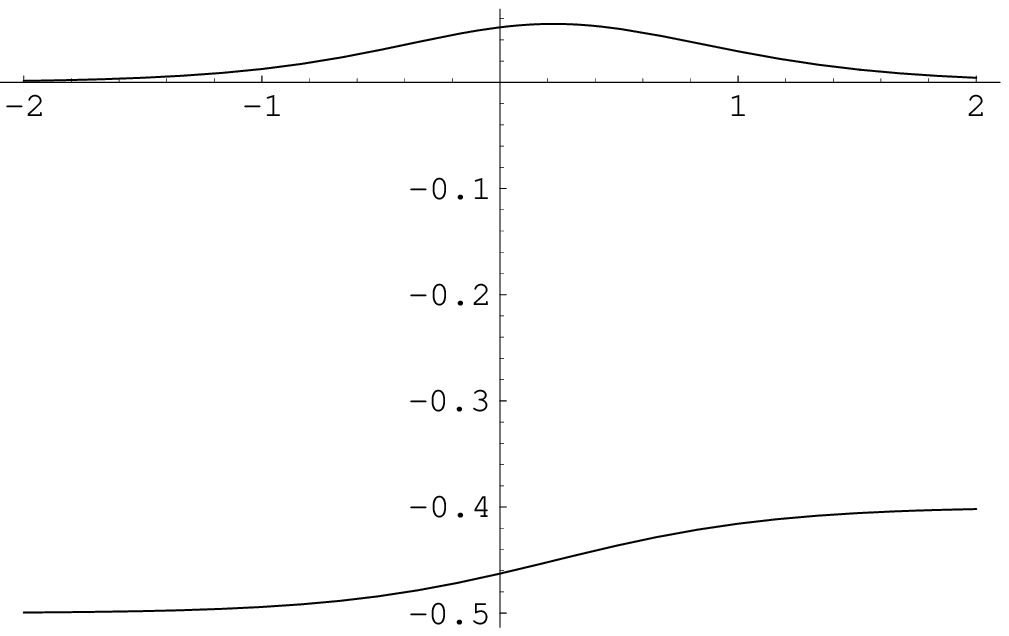}
\caption{Evolution de l'amplitude du potentiel gravitationnel
\`a grande \'echelle au moment du passage de l'\'equivalence. La
courbe du haut donne la variation de $\psi$ (source de l'effet ISW)
et la courbe du bas, $\theta_0+\psi$.}
\label{T0_superH}
\end{figure}

Une quantit\'e importante est la travers\'ee de l'horizon.
Dans l'espace de Fourrier cela est donn\'e par $k\,\eta=1$
($\eta$ est le temps conforme).

\begin{itemize}
\item
\`A une \'echelle plus grande que l'horizon, $k\eta\ll1$, Univers
domin\'e par le rayonnement.
On a
\ba
{\d\over \d\eta}(V_{\rm mat.}-V_{\gamma})&=&-{\dot{a}\over a}\,
(V_{\rm mat.}-V_{\gamma});\\
{\d\over \d\eta}\left(\delta_{\rm mat.}-{3\over
4}\delta_{\gamma}\right)
&=& k\,(V_{\rm mat.}-V_{\gamma}).
\ea

Dans une th\'eorie inflationnaire \`a champ simple
la mati\`ere tombe dans les potentiels ensemble. Il n'y a pas
de cr\'eation d'entropie, donc,
\ba
\delta_{\rm mat.}&=&{3\over4}\delta_{\gamma},\\
V_{\rm mat.}&=&V_{\gamma}.
\ea
Dans le cas contraire
(inflation hybride...) on peut avoir des fluctuations isocourbes:
fluctuation de la composition de l'Univers sans fluctuation
de courbure.

Finalement on doit r\'esoudre,
\be
{\dot{a}\over a}\psi+\dot\psi=-{1\over 2}
\left(\dot{a}\over a\right)\delta_{\gamma}; \ \ 
\dot\delta_{\gamma}=4\dot\psi,
\ee
ce qui donne apr\`es avoir \'eliminer $\delta_{\gamma}$
\be
{\d^2\psi\over \d\,a^2}+4{d\psi\over\d a}=0
\ee
dont la solution g\'en\'erale est,
\be
\psi\sim\psi_C+\psi_D\,{1\over a^3}.
\ee
On a un terme 'croissant' et un terme 'd\'ecroissant'. 
Pour ce terme dit croissant.
Les fluctuations de densit\'es (que j'identifie de mani\`ere g\'en\'erale
avec le terme de source de l'\'equation de Poisson)
croissent comme $k^2\,\eta^2$. 

\item
La travers\'ee de l'\'equivalence \`a une \'echelle plus grande
que l'horizon.
L'\'equation g\'en\'erale s'\'ecrit en fait,
\be
{\dot{a}\over a}\psi+\dot\psi=-{1\over 2}
\left(\dot{a}\over a\right){\delta_{\gamma}\over 4}\,
{4+3\,a/a_{\rm eq.}\over 1+a/a_{\rm eq.}}; \ \ 
\dot\delta_{\gamma}=4\dot\psi,
\ee
Il existe une solution explicite (voir Fig. ).
Finalement on a,
\ba
\psi(a\to \infty)&=&{9\over 10}\psi(a\to 0)\\
{\delta_{\gamma}(a\to \infty)\over 4}&=&-{2\over 3}\psi(a\to \infty),
\ea
et on remarquera que $\delta_{\gamma}/4$ s'identifie \`a la
fluctuation locale de temp\'erature.

\item
La travers\'ee de l'horizon. 
C'est l\`a que les choses deviennent plus compliqu\'ees puisqu'il
faut tenir compte des int\'eractions entre photons et baryons.
Tant qu'on est avant la recombinaison ces 2 fluides
restent coupl\'es, mais la mati\`ere non-baryonique n'est plus
coupl\'ee \`a ceux-ci.
Quand le rayonnement domine on a,
\be
\ddot{\delta_{\gamma}}+{k^2\over 3}
\delta_{\gamma}=4\,\ddot{\psi}-{4\over 3}\,k^2\,\psi.
\ee
Du coup les fluctuations sont pi\'eg\'ees dans des ondes acoustiques.
La mati\`ere non-baryonique est spectatrice. Quand le rayonnement
domine l'\'equation d'\'evolution des fluctuations de
mati\`ere sont alors,
\be
\ddot\delta_{CDM}+{\dot{a}\over
a}\dot\delta_{CDM}=-3\ddot\phi-k^2\psi-3
{\dot{a}\over a}\,\dot\phi.
\ee
Le membre de droite est un terme de source d\'etermin\'e par le
comportement
des photons. Comme les croissances des fluctuations pour les
photons sont interrompues, on a essentiellement une \'equation
homog\`ene dont la solution croissante est une croissance
logarithmique de la fluctuation de densit\'e.

\item Univers domin\'e par la mati\`ere.
Cette fois le terme de source de l'\'equation d'\'evolution est 
donn\'e par la mati\`ere elle-m\^eme. En fait je retombe sur une
\'equation d\'ej\`a \'ecrite pr\'ec\'edemment, et les potentiels
sont de nouveau constants: les fluctuations de densit\'e croissent
comme $\eta^2\sim a$.

Par contre le fluide photons+baryons conna\^\i t toujours des
oscillations acoustiques. La vitesse du son $c_s$ est maintenant
donn\'ee par,
\be
c_s^2={1\over 3\,(1+\rho_b/\rho_{\gamma})}.
\ee
On a essentiellement,
\ba
{\delta_{\gamma}\over 4}&\sim& \cos(k\,\int_0^{\eta}\,c_s\,\d \eta)\\
V_{\gamma}&\sim&\sin(k\,\int_0^{\eta}\,c_s\,\d \eta).
\ea
(si $c_s$ \'etait constant cela ferait $k\,\eta/\sqrt{3}$
dans l'argument des fonctions trigonom\'etriques).
\end{itemize}

\begin{figure}
\vspace{9 cm}
\special{hscale=90 vscale=90 voffset=-10 hoffset=10 psfile=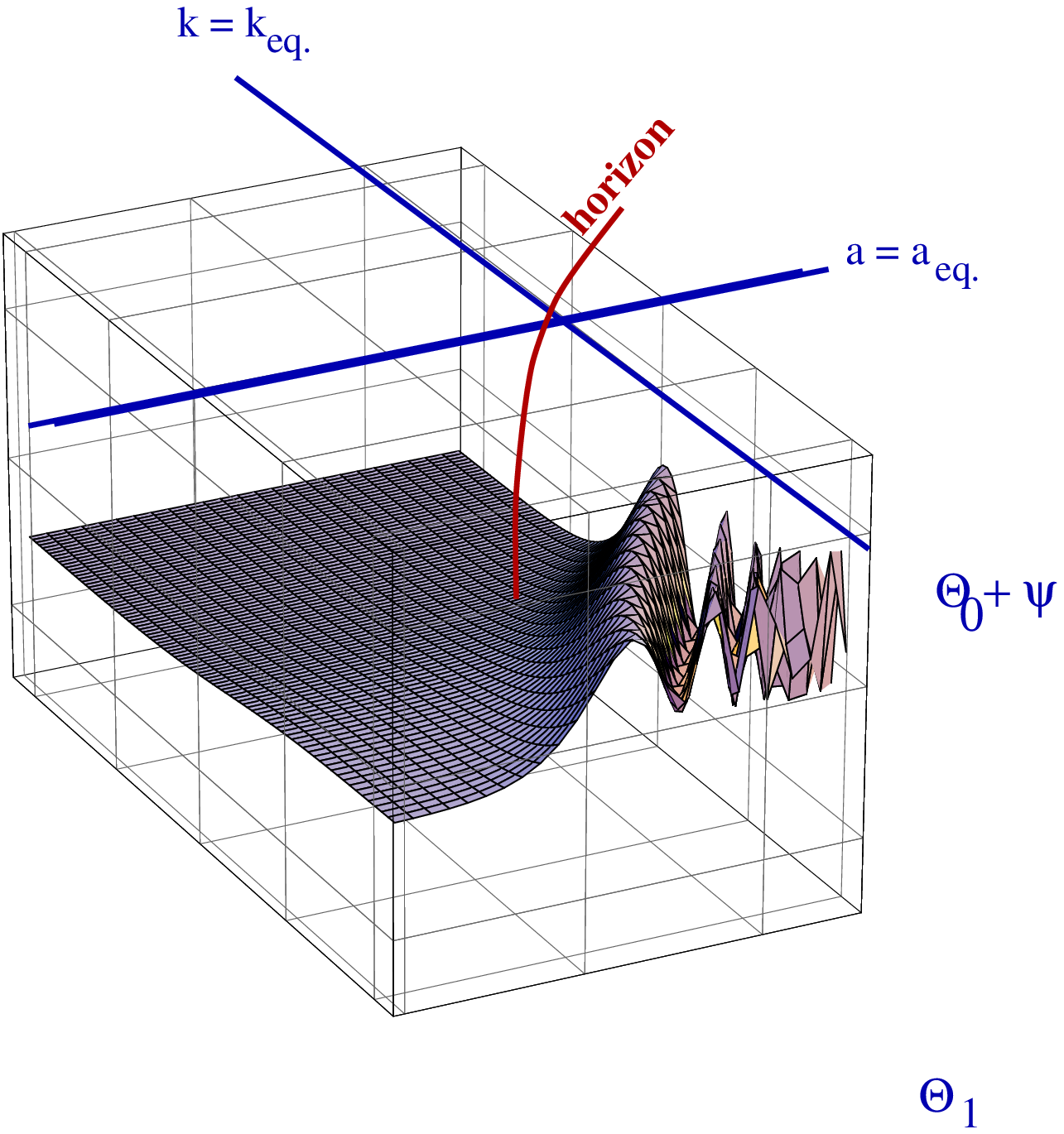}
\vspace{7 cm}
\special{hscale=90 vscale=90 voffset=30 hoffset=10 psfile=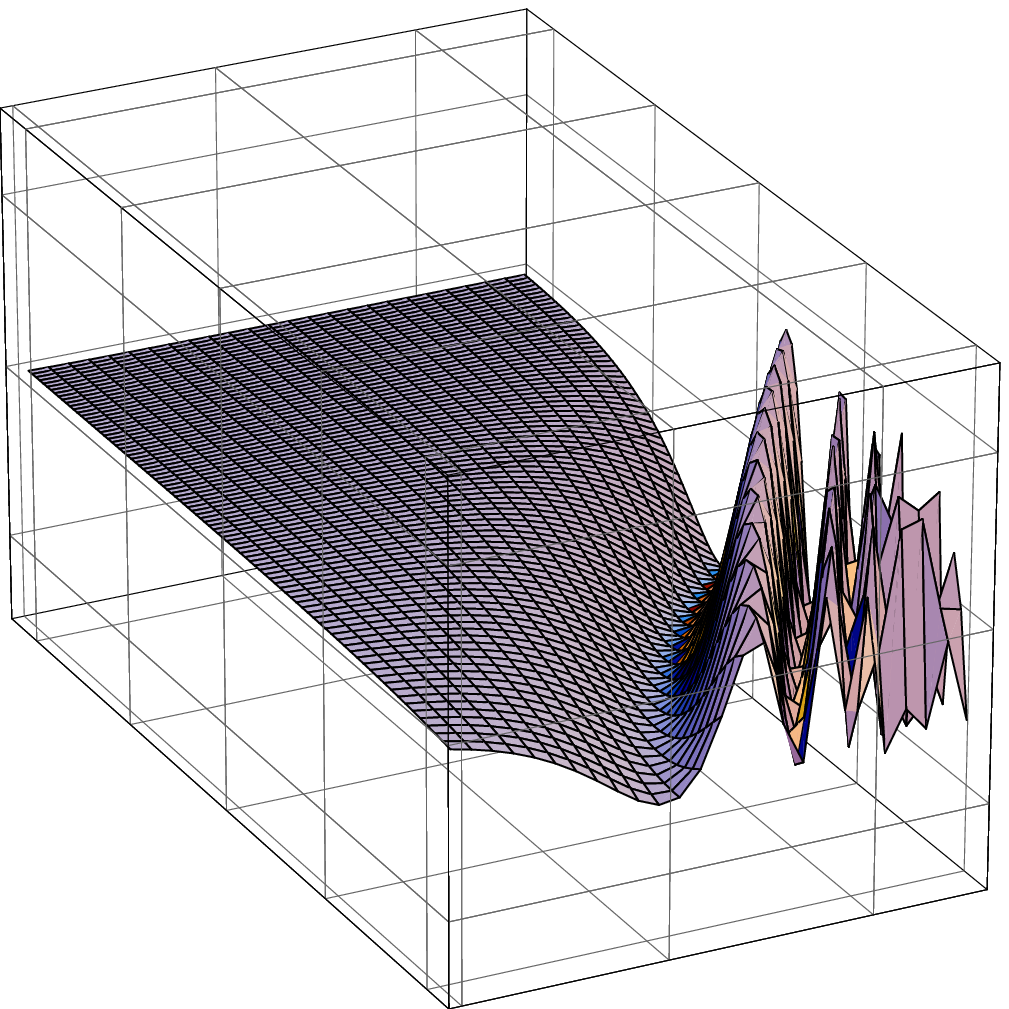}
\caption{Evolution de l'amplitude des fluctuations intrins\`eques
de temp\'erature (en haut) et de la vitesse particuli\`ere (en bas)
en fonction du facteur d'expansion et du mode $k$.}
\end{figure}

Quelles sont les cons\'equences observationnelles?
Il y a essentiellement 2 aspects qui sont importants d'un point de vue
observationnel. Les fluctuations de temp\'eratures sur le 3K
et les fluctuations de densit\'e qui vont donn\'ees naissance
aux grandes structures de l'Univers.

\begin{figure}
\vspace{7 cm}
\special{hscale=40 vscale=40 voffset=0 hoffset=10 psfile=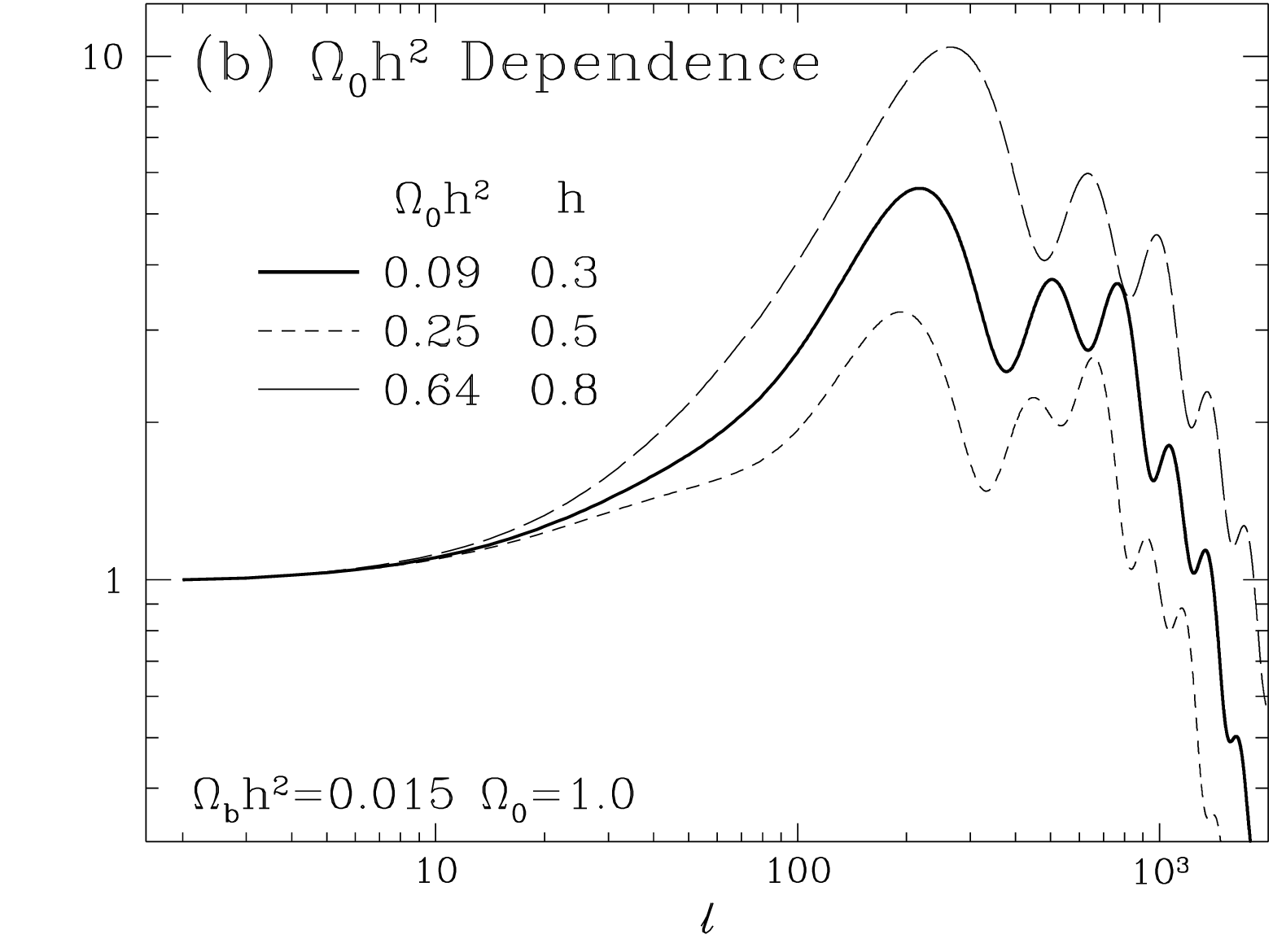}
\caption{Exemples de spectre de fluctuations de temp\'erature
pour des mod\`eles standards CDM. C'est la quantit\'e $C_l\,l(l+1)$
qui est donn\'ee en fonction de $l$. Diff\'erentes valeurs de 
$\Omega_0\,h^2$ sont propos\'ees (figure tir\'ee de Hu, 1995).}
\label{Cdel}
\end{figure}

\begin{itemize}
\item
Les fluctuations de temp\'erature.
Localement on peut d\'ecomposer les fluctuations de temp\'erature
du rayonnement dans son espace des phases par,
\be
\theta(\gamma)\equiv\left({\int\,p^3\d p\,f(\vx,\vp)\over
\pi^2\,\rho_{\gamma}}-1\right)/4
=\sum_{l,m}a_{l,m}\,Y_l^m(\gamma),
\ee
o\`u $\theta(\gamma)$ est la fluctuation de temp\'erature dans 
la direction $\gamma$, d\'ecompos\'ee en harmonique sph\'erique
(implicitement \`a la position de l'observateur).
Pour un champ de fluctuations isotropes on a
\be
<a_{l,m}\,a^*_{l',m'}>=\delta_{ll'}\,\delta_{mm'}\,C_l.
\ee
Cela d\'efinit les $C_l$.

Les anisotropies du 3K permettent de voir en quelque
sorte l'\'etat du plasma photons+baryons au moment de la recombinaison.
Apr\`es la recombinaison les photons se d\'ecouplent compl\`etement, leur 
libre parcours moyen devient infini, ils diffusent donc librement.
Les anisotropies de temp\'eratures sont alors la superposition de
3 effets diff\'erents: les fluctuations de temp\'erature intrins\`eques,
les fluctuations de potentiel qui induisent des effets de redshift
gravitationnel, enfin des effets Doppler d\^us aux mouvements du fluide
le long de la ligne de vis\'ee. Il y a un cut-off naturel 
aux petites \'echelles d\^u au fait que la surface de derni\`ere
diffusion a une \'epaisseur finie (ce qui induit un 'Silk damping').
L'amplitude du mode $l$ est donn\'ee essentiellement par,
\ba
{\theta_l\over 2l+1}\sim&\disp{\left[
(\theta_0(\eta_*)+\psi(\eta_*))j_l(k\eta_*)+\theta_1{1\over k}
{\d\over\d\eta}j_l(k\eta)+\right.}\\
&+\disp{\left.\int_{\eta_*}^{\eta_0}(\dot\psi(\eta')-
\dot\phi(\eta'))j_l(k\eta')\d\eta'\right]\,\exp(-k^2/k_D^2)},
\ea
pour un mode $\vk$. Cette expression est valable dans la limite
d'une surface de derni\`ere diffusion infiniment mince. Le premi\`ere
correction \`a cette limite donne un effet d'amortissement \`a petite
\'echelle, le 'Silk damping' \`a une \'echelle $k_D$ li\'ee
\`a la longueur de diffusion des photons. 
En g\'en\'eral il faut r\'esoudre un syst\`eme d'\'equations
coupl\'ees pour obtenir les diff\'erents moments.
La projection des fluctuations locales de temp\'erature
sur les harmoniques sph\'eriques fait intervenir des fonctions de
Bessel $j_l$ pour le cas d'un univers plat. Pour un univers avec
$k\neq0$, c'est plus compliqu\'e.
On a alors un spectre $C_l$ typiquement de la forme donn\'ee
par la figure \ref{Cdel}.

A ce jour, on ne conna\^\i t correctement que la partie
\`a grande \'echelle grace aux observations de COBE/DMR. La forme des
pics acoustiques est une chose que les gens cherchent activement
\`a mesurer parce qu'ils contiennent des informations pr\'ecieuses
sur les param\`etres cosmologiques. Il y a en pr\'eparation de nombreuses
exp\'eriences sol ou ballon, sans compter les projets satellitaires
am\'ericains avec MAP  ou europ\'eens avec Planck, qui doivent mesurer
ces anisotropies avec des sensibilit\'es et des pr\'ecisions diverses.

\item
La cons\'equence observationnelle qui va davantage m'int\'eresser
ici est le spectre de puissance $P(k)$  des fluctuations de densit\'e
de la mati\`ere. Dans les mod\`eles avec de la mati\`ere noire
en quantit\'e dominante les baryons ne vont contribuer \`a la forme
de ce spectre que de mani\`ere mineure. 

Du coup  $k_{\rm eq.}$, \'echelle qui traverse l'horizon
au moment de l'\'equivalence
qui est un param\`etre essentiel de la forme du $P(k)$.
En effet les modes $k$ inf\'erieurs \`a $k_{\rm eq.}$
croissent r\'eguli\`erement, comme $\eta^2$, alors que les
modes avec $k$ plus grand que $k_{\rm eq.}$ sont gel\'es
entre la travers\'ee de l'horizon et l'\'equivalence.

Pour les mod\`eles les plus r\'ealistes, le $P(k)$ a une forme
simple sans structures particuli\`eres. En particulier 
les oscillations plasma ne se
voient que lorsque la fraction de baryons est suffisante.

En r\'esum\'e on a 
\ba
P(k)&\propto& k\ \ {\rm pour}\ \ k\gg k_{\rm eq.},\\
P(k)&\propto& k^{-3}\ \ {\rm pour}\ \ k\ll k_{\rm eq.},
\ea
pour un spectre initial Harrison-Zeldovich.

\end{itemize}

\subsection{ Pourquoi un mod\`ele avec de la Mati\`ere Noire Froide ?}

La premi\`ere raison est de rendre compatible le niveau des anisotropies
du 3K avec les fluctuations de densit\'e dans l'Univers local.
Au moment de la recombinaison les
fluctuations de densit\'e de la mati\`ere noire peuvent
en effet \^etre beaucoup plus grandes que celles des photons+baryons.
Les baryons tombent dans les puits de potentiel de la mati\`ere
noire apr\`es la recombinaison.
Du coup, on peut tr\`es bien avoir des anisotropies
de temp\'erature \`a un niveau bas tout en ayant des fluctuations de
densit\'e de la mati\`ere noire relativement importantes.

\begin{figure}
\vspace{6.5 cm}
\special{hscale=40 vscale=40 voffset=0 hoffset=10 psfile=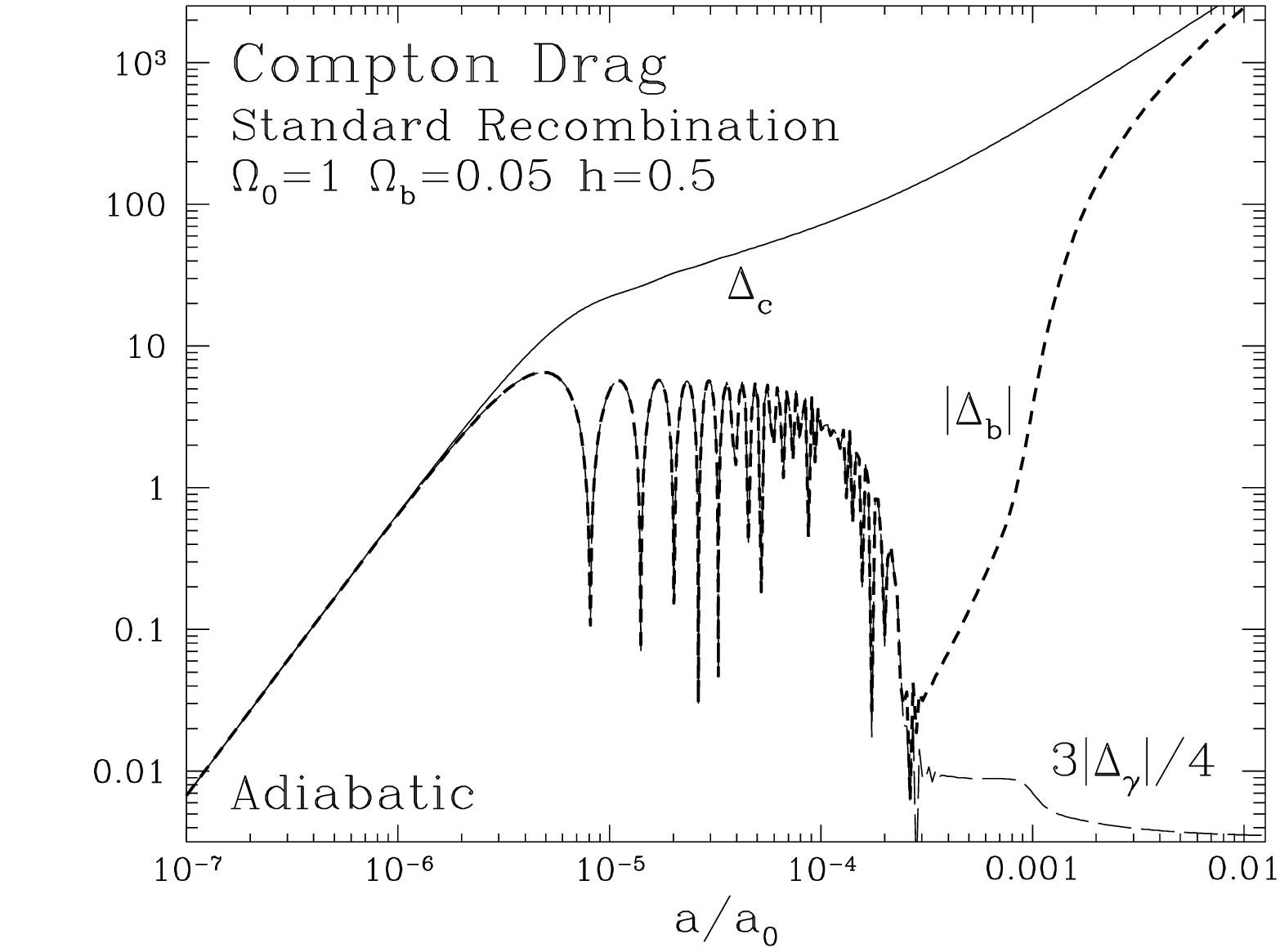}
\caption{Evolution \`a petite \'echelle du contraste en densit\'e
des baryons compar\'e \`a celui de la mati\`ere noire froide.
Les baryons restent coupl\'es tr\`es longtemps au photons. Au moment
de la surface de derni\`ere diffusion, le contraste en densit\'e
atteint par la mati\`ere noire froide peut \^etre beaucoup plus
important que celui des baryons (figure tir\'ee de Hu, 1995).
}
\label{drag}
\end{figure}

\subsection{ \'Evidences de l'existence de mati\`ere noire}

Evidemment cette belle id\'ee doit \^etre soutenue de mani\`ere 
exp\'erimentale
en essayant de d\'etecter la mati\`ere noire non-baryonique
de mani\`ere directe.

Les \'evidences sont les suivantes,
\begin{itemize}
\item
Les courbes de rotation des galaxies. Cela s'applique aux halos
des galaxies.
\item
Les \'etudes dynamique des amas de galaxies.
Un amas de galaxies est un immense puit de potentiel
dans lequel les galaxies sont li\'ees. 
La mesure de leur dispersion de vitesse oermet d'estimer
la masse de l'amas.
\item
Les mesures de masse par effet de lentille.
Exemple A2218 avec un
syst\`eme d'arcs.
C'est tr\`es propre puisque c'est une d\'etection directe de la masse 
projet\'ee, par d'hypoth\`ese sur l'\'equilibre hydrodynamique dans
l'amas.  C'est une m\'ethode tr\`es prometeuse
car elle peut s'appliquer \`a toutes les \'echelle, aussi bien pour
sonder les halos des galaxies, que pour faire des d\'etection de la 
mati\`ere gravitante \`a grande \'echelle.
\item Les champs de vitesse \`a grande \'echelle. L'id\'ee est de
mesurer les
vitesses particuli\`eres, \'ecarts des vitesses d'\'eloignement
avec le flot de Hubble, \`a l'aide d'indicateurs sp\'ecifiques
(de type dispersion de vitesse).
C'est un domaine potentiellement tr\`es int\'eressant, cependant il
souffre d'un d\'efaut important: les erreurs de mesure augmentent avec la
distance. Au del\`a de
50 $h^{-1}$ Mpc les erreurs d\'epassent les 100\% sur une galaxie
individuelle. Cela vient des
dispersions intrins\`eques des relations utilis\'ees et pas des erreurs
de mesures.

\end{itemize}

\subsection{ Mod\`eles alternatifs}

Les mod\`eles avec d\'efauts topologiques sont de moins en moins 
cr\'edibles en tant
que m\'ecanisme pour former les grandes structure. 
La principale raison est qu'il est tr\`es difficile de concilier le niveau
des anisotropies vues par COBE avec l'amplitude des fluctuations \`a grande
\'echelle (voir l'article recent de Albrecht 1998).

Le calcul de la croissance des fluctuations dans de tels
mod\`eles est rendu tr\`es difficile par l'existence de termes
de sources dans les \'equations d'Einstein. Il faut tenir compte de la
dynamique propre des sources, et des corr\'elations non-triviales
qu'elles induisent \`a des temps diff\'erents.

%% file: Cours456.tex
\section{ La dynamique gravitationnelle, les th\'eories
lin\'eaires}

A partir de maintenant je m'int\'eresse \`a la dynamique
de la mati\`ere noire en effondrement gravitationnel.

\subsection{\'Equation d'\'evolution dans l'espace des phases}

Sous l'horizon l'\'equation dynamique s'\'ecrit,
\be
{\partial \vu\over
\partial t}+{\dot{a}\over
a}\vu=\vg=G\rhob\,a\int\d^3\vx'{\delta(\vx')(\vx'-\vx)\over
\vert\vx'-\vx\vert^3},
\ee
o\`u $\vu$ est la vitesse particuli\`ere des particules (\'ecart
de la vitesse totale avec le flot de Hubble).
L'impulsion s'identifie \`a
(c'est aussi la variable conjugu\'ee de $\dot{\vx}$ dans le Lagrangien
d\'ecrivant la dynamique),
\be
\vp={\vu\, m\,a}.
\ee

J'\'ecris la densit\'e dans l'espace des phases,
\be
f(\vx,\vp)\,\d^3\vx\,\d^3\vp,
\ee
o\`u $\vx$ est la position comouvante et $\vp$
est l'impulsion. 
Le th\'eor\`eme de Liouville appliqu\'e \`a l'\'evolution
de la densit\'e $f$ donne,
\ba
{\d f\over \d t}=
{\dr\over\dr t}f(\vx,\vp,t)+{\vp\over ma^2}{\dr\over\dr\vx}f(\vx,\vp,t)
-m\grad_{\vx}.\Phi(\vx){\dr\over\dr\vp}f(\vx,\vp,t)=0\\
\Delta \Phi(\vx)={4 \pi G m\over a}\int f(\vx,\vp,t) \d^3\vp.
\ea
Dans toute sa g\'en\'eralit\'e cette \'equation est \'evidemment
tr\`es difficile \`a r\'esoudre.

On peut prendre diff\'erents moments de cette \'equation
par rapport \`a $\vp$.
La densit\'e de mati\`ere est par construction,
\be
\rho(\vx,t)={m\over a^3}\,
\int\d^3\vp\,f(\vx,\vp).
\ee
J'\'ecris cette densit\'e
\be
\rho(\vx,t)=\rhob(t)\left[1+\delta(\vx,t)\right],
\ee
et alors $\rhob(t)\propto a^3$.
La vitesse moyenne est $\vu$ donn\'ee par,
\be
\vu=\disp{{\int\d^3\vp\,\vp\,f(\vx,\vp)\over m\,a\,\int\,f\,\d^3\vp}},
\ee
et je vais \'ecrire de mani\`ere un peu cavali\`ere que,
\be
{\int\d^3\vp\,\vp_i\,\vp_j\,f(\vx,\vp)\over 
m^2\,a^2\,\int\,f\,\d^3\vp}=
\vu_i\,\vu_j+\big<\vu_i\,\vu_j\big>_c.
\ee

Quand on int\`egre l'\'equation de Liouville par rapport \`a $\vp$
on obtient,
\be
\rhob\,{\partial\delta\over\partial t}+{1\over
a}\nabla.\rho(\vx)\,\vu(\vx)
=0,\ {\rm soit}\ ,{\partial\delta\over\partial t}+{1\over
a}\nabla.(1+\delta(\vx))\vu(\vx)=0,
\ee
c'est \`a dire une \'equation de continuit\'e.

Quand on int\`egre l'\'equation par rapport \`a $\vp$
apr\`es l'avoir multiplier par $\vp$ on a,
\be
{\partial\over\partial t}
\int\,p_i\,f\,\d^3\vp+{1\over m\,a^2}
\partial_j\,\int p_i\,p_j\,f\,\d^3\vp+a^3\rho(\vx,t)\Phi_i=0.
\ee
Je peux r\'e\'ecrire cette \'equation en faisant appara\^\i tre
la vitesse particuli\`ere,
\be
{\dr\vu_i\over\dr t}+{\dot{a}\over a}\vu_i+{1\over
a}(\vu_j.\nabla_j)\vu_i=-{1\over a}\nabla_i\Phi-{1\over \rho\,a}
\left(\rho\,\big<u_i\,u_j\big>_c\right)_{,j}.
\ee

Faisons des hypoth\`eses suppl\'ementaires. Si localement
le fluide est thermalis\'e on a
\be
\big<u_i\,u_j\big>_c=\delta_{ij}{p\over \rho}.
\ee
La dispersion de vitesse locale s'identifie avec la
pression en supposant qu'elle est isotrope.
On a alors,
\be
{\partial\vu_i\over\partial t}+{\dot{a}\over a}\,\vu_i+{1\over a}
(\vu_i\grad_j)\vu_j=-{1\over a}\grad_i\phi-{p_{,i}\over \rho a}.
\ee
 Si on se donne
une \'equation d'\'etat on peut alors fermer le syst\`eme.

\subsection{La longueur de Jeans}

Une question qu'on peut se poser est de savoir \`a quelle
\'echelle la pression peut jouer un r\^ole et freiner
les effondrements gravitationnels. Pour faire ce calcul il faut
se donner une expression pour la pression.

Par exemple, pour un gaz mono-atomique on a,
\be
c_s^2\equiv{\d p\over \d \rho}={5\,kT\over 3\,m}
\ee

On peut ensuite lin\'eariser les \'equations d'\'evolution,
ce qui donne
\ba
{\dr\delta\over\dr t}+{1\over a}\nabla.\vu&=&0\\
{\dr^2\delta\over\dr\,t^2}+2\,
{\dot{a}\over a}{\dr\delta\over\dr\,t}&=&
{c_s^2\over a^2}\Delta\delta+4\,\pi\,G\,\rhob\,\delta.
\ea
puisque 
\be
p=\overline{p}+c_s^2\,\rhob\,\delta.
\ee

Cela d\'efinit une longueur caract\'eristique, la longueur de Jeans,
avec 
\be
\lambda_J=c_s(\pi/G\rhob)^{1/2}.
\ee
Les fluctuations de taille plus grande que $\lambda_J$
s'effondrent librement sans \^etre contrer par la pression.
Cela donne des \'echelles de masse de 
$10^5\ M_{\odot}$. Pour les \'echelles cosmologiques
(i.e. la masse d'une galaxie est typiquement de
$10^{12}\ M_{\odot}$, celle d'un amas de $10^{15}\ M_{\odot}$)
on peut donc en toute s\'ecurit\'e n\'egliger la
pression due aux fluctuations thermiques.

\begin{figure}
\vspace{7 cm}
\special{hscale=60 vscale=60 voffset=0 hoffset=10 psfile=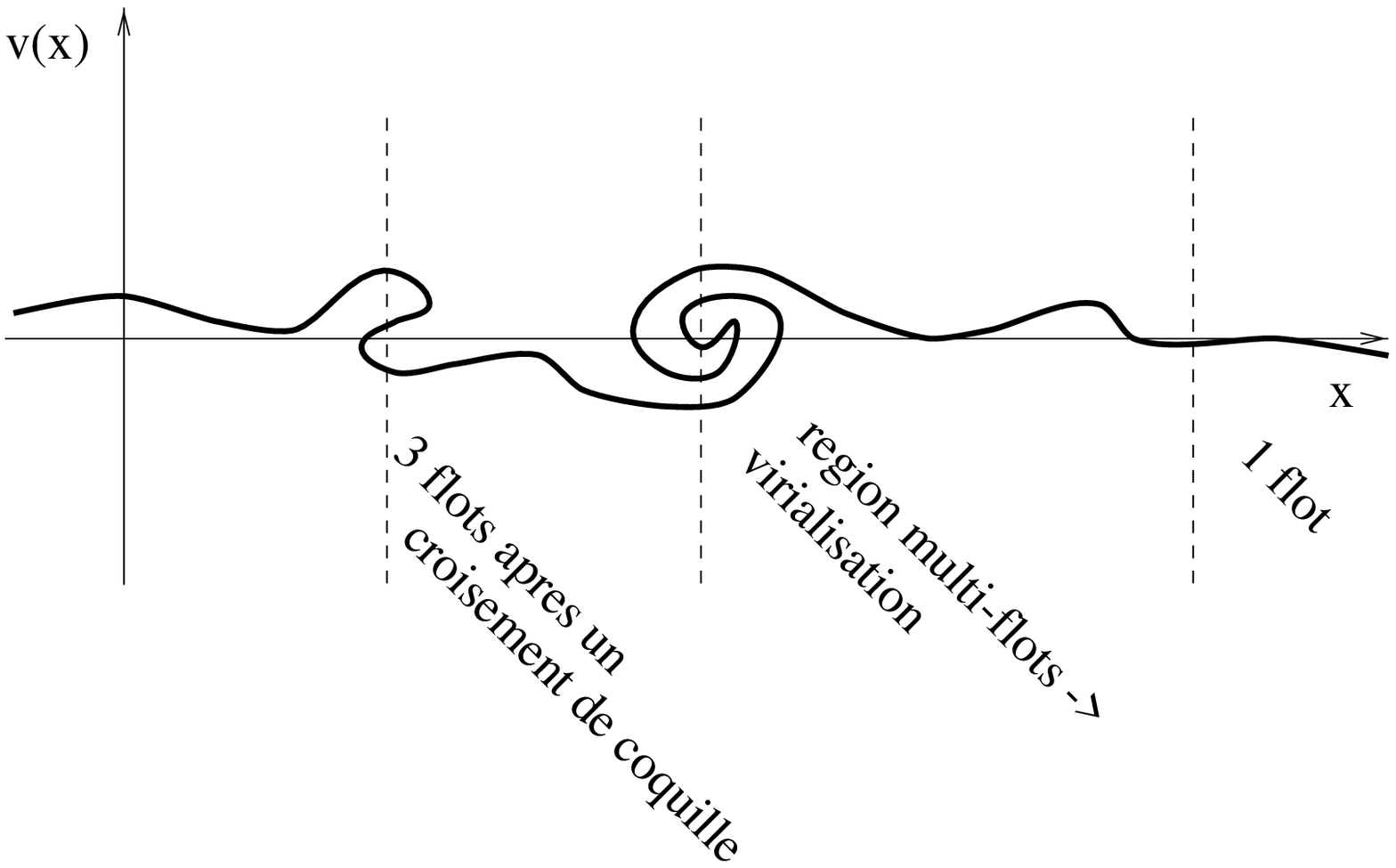}
\caption{Description sch\'ematique de l'espace des phases
apr\`es les premiers croisements de coquilles.}
\label{flots}
\end{figure}

Notons quand m\^eme qu'admettre que la pression est nulle c'est dire que
\be
\big<u_i\,u_j\big>_c=0
\ee
donc la densit\'e dans l'espace des phases peut s'\'ecrire
\be
f(\vx,\vp)=\rho(\vx)\,\delta^{(3)}[\vp-m\,a\,\vu(\vx)].
\ee
Cela revient \`a supposer qu'en tout point on a un seul flot.
Cela ne sera vrai qu'au d\'ebut de la dynamique. Ult\'erieurement
on s'attend \`a avoir des croisements de coquilles (voir fig \ref{flots}).

\subsection{ L'approximation Newtonnienne avec un seul flot}

Dans l'approximation \`a un seul flot, on a finalement
le syst\`eme d'\'equation suivant,

\ba
{\partial \over \partial t}\rho(\vx,t)+{1\over a}\gradx .\left[
\rho(\vx,t)\vu(\vx,t)\right]&=&0\\
{\partial \over \partial t}\vu(\vx,t)+{\dot a \over a}\vu(\vx,t)+{1\over a}
(\vu(\vx,t).\gradx)\vu(\vx,t)&=&-{1\over a}\gradx\psi(\vx,t)\\
\gradx^2\psi(\vx,t)-4\pi G\left[\rho(\vx,t)-\overline{\rho}(t)\right]a^2&=&0.
\ea
C'est ce qu'on appelle une description Eul\'erienne de
la dynamique.

\subsection{ Description Eul\'erienne ou description Lagrangienne}

On peut en faire une description \'equivalente en ayant
une approche Lagrangienne.
L'id\'ee est de d\'ecrire la trajectoire des particules
\`a partir de leur position dans l'espace de d\'epart $\vq$.
On \'ecrit donc qu'une particule initialement en $\vq$
se trouve \`a un instant $t$ \`a la position (comouvante)
$\vx$ apr\`es s'\^etre d\'eplac\'ee de $\Psi(\vq,t)$,
\be
\vx=\vq+\Psi(\vq,t).
\ee

Le champ d\'eplacement $\Psi(\vq,t)$ est donn\'e par
la m\^eme \'equation d'Euler simplement la densit\'e
est donn\'ee par l'inverse du Jacobien de la transformation
de $\vq$ vers  $\vx$. 
Plus explicitement on peut montrer que 
\ba
J(\vq,t)\equiv&\disp{\vert{\dr\vx\over\dr\vq}\vert}=
1+\nabla_{\vq}.\Psi(\vq,t)+{1\over 2}
\left[\left(\gradq.\Psi\right)^2-
\sum_{ij}\Psi_{i,j}\Psi_{j,i}\right]\\
&+{1\over 6}
\left[\left(\gradq.\Psi\right)^3-3\gradq.\Psi\sum_{ij}\Psi_{i,j}\Psi_{j,i}
+2 \sum_{ijk}\Psi_{i,j}\Psi_{j,k}\Psi_{k,i}\right].\nonumber
\ea

On va voir que selon le probl\`eme qui
nous int\'eresse on pourra prendre l'une ou l'autre
des formulations.

\subsection{ La croissance des fluctuations en th\'eorie lin\'eaire}

La th\'eorie lin\'eaire consiste \`a supposer que les fluctuations
sont de faible amplitude, $\delta\ll 1$.
Pour une description Lagrangienne, cela veut dire que 
les gradients du champ de d\'eplacement sont faibles, et donc
que le Jacobien de la transform\'ee est proche de 1.

Dans le r\'egime lin\'eaire il est plus simple 
de regarder la description Eul\'erienne.
La lin\'earisation des \'equations donne,
\ba
{\dr\delta\over\dr t}+{1\over a}\nabla.\vu&=&0,\\
{\dr^2\delta\over\dr\,t^2}+2\,
{\dot{a}\over a}{\dr\delta\over\dr\,t}&=&
4\,\pi\,G\,\rhob\,\delta.
\ea

La deuxi\`eme \'equation permet de rechercher les modes
de fluctuation. On trouve 2 modes,
\be
\delta(\vx,t)=D_+(t)\delta_+(\vx)+D_-(t)\delta_-(\vx),
\ee
dont la d\'ependance en $t$ d\'epend des param\`etres
cosmologiques (elle est cach\'ee dans la d\'ependance en temps
du facteur d'expansion).
Pour \^etre plus sp\'ecifique on a,
\be
\ddot{D}+2\,H\,\dot{D}-{3\over 2}\,H^2(t)\,\Omega(t)\,D=0
\ee
qui est valable pour tous les mod\`eles cosmologiques.

Pour un Univers Einstein-de Sitter on a,
\ba
D_+(t)&\propto& t^{2/3},\\
D_-(t)&\propto& 1/t.
\ea

Notons que $D_+$ peut s'exprimer analytiquement en fonction de $t$
si $\lambda=0$. Il doit \^etre calculer num\'erique pour des
mod\`eles avec constante cosmologique.

Une quantit\'e utile est la d\'eriv\'ee logarithmique de $D_+$
avec le facteur d'expansion, puisqu'elle appara\^\i t de mani\`ere
naturelle dans l'\'equation de continuit\'e.
On a un fit analytique de la forme (Lahav et al. 1991),
\be
f(\Omega,\Lambda)\equiv{\d \log D_+\over \d \log a}
=\Omega^{0.6}+{1\over 70}\Lambda\,(1+\Omega/2).
\ee
On voit que \c ca d\'epend tr\`es peu de la constante cosmologique.


Une relation int\'eressante en th\'eorie lin\'eaire est la
relation densit\'e-vitesse. Localement elle s'\'ecrit
\`a partir de la relation de continuit\'e,
\be
\theta(\vx)\equiv{1\over
a\,H}\nabla_{\vx}.\vu(\vx)=-f(\Omega,\Lambda)\delta(\vx).
\ee
Remarquons que $\theta$ est bien a priori une quantit\'e
observable qui ne d\'epend pas de la
constante de Hubble. Les flots \`a grande \'echelle sont connus
en $km/s$, \`a des distance qui sont d\'etermin\'ees en $km/s$.
Plus pr\'ecis\'ement $\theta$ exprime les fluctuations locales de la
constante de Hubble, i.e. du taux d'expansion.

Cette relation a \'evidemment une contrepartie non-locale
qui exprime la relation entre la vitesse locale et les fluctuations
de densit\'e environnantes,
\be
{\vu\over a\,H}={f(\Omega)\over 4\pi}\int\d^3\vx'\delta(\vx'){\vx'-\vx\over\vert\vx-\vx'\vert^3}.
\ee
Cette relation exprime que la vitesse acquise par une particule
en r\'egime lin\'eaire est proportionnelle \`a son acc\'el\'eration.

\`A ce niveau on peut facilement donner une interpr\'etation 
des modes croissants et d\'ecroissants. Le mode croissant correspond
au cas o\`u le champ de densit\'e et de vitesse sont en 'phase':
quand la mati\`ere tombe vers les puits de potentiel. On peut tr\`es
bien imaginer qu'on puisse construire un champ de vitesse 
qui soit tel que que les particules aient
tendance \`a s'\'echapper des puits de potentiel
et tendent ainsi \`a effacer les fluctuations initiales. Ce serait le
mode d\'ecroissant.

\subsection{ L'approximation de Zel'dovich}

Cela consiste \`a prendre la solution lin\'eaire pour le champ de
d\'eplacement mais \`a garder l'\'equation de continuit\'e dans toute
sa g\'en\'eralit\'e (Zel'dovich 1970). 
Dans une telle approximation, le champ de d\'eplacement
s'\'ecrit comme la somme de deux termes,
\be
\psi(\vq,t)=D_+(t)\psi_+(\vq)+D_-(t)\psi_-(\vq),
\ee
o\`u $D_+(t)$ et $D_-(t)$ sont les fonctions du temps solution de l'\'equation
pr\'ec\'edente.
On retrouve des termes croissants et des termes d\'ecroissants. Quand
on ne garde que le terme croissant, on a une factorisation dans la 
d\'ependance en temps et la d\'ependance en espace. Les
trajectoires des particules sont donc rectilignes et suivent la direction
de la force appliqu\'ee aux particules \`a l'instant initial.

La densit\'e locale peut alors s'exprimer en fonction des valeurs
propres de la matrice de d\'eformation.
Plus pr\'ecis\'ement on a,
\be
\rho={1\over (1-\lambda_1\,D_+)(1-\lambda_2\,D_+)(1-\lambda_3\,D_+)},
\ee
o\`u les $\lambda_i$ sont les valeurs propres de la
matrice $\psi_{i,j}$.

Pour des conditions initiales Gaussiennes on peut alors
calculer la distribution de densit\'e locale \`a partir
de la distribution des valeurs propres.
La distribution des valeurs propres est donn\'ee par 
(Doroshkevich 1970),
\ba
p(\lambda_{1},\lambda_{2},\lambda_{3})\,
\d\lambda_{1}\ \d\lambda_{2}\ \d\lambda_{3}=
\ \d\lambda_{1}\ \d\lambda_{2}\ \d\lambda_{3}\\
 {5^{5/2}\, 27 \over 8 \pi \sigma_0^6}\
(\lambda_{1}-\lambda_{2})(\lambda_{1}-\lambda_{3})(\lambda_{2}-\lambda_{3})\
{\rm exp}\left[-{1\over\sigma_0^2}
\biggl(3J_{01}^2-{15 \over 2} J_{02} \biggr)\right],
\ea
o\`u 
\ba
J_{01}&=&\lambda_{1} +\lambda_{2}+ \lambda_{3},\\
J_{02}&=&\lambda_{1}\lambda_{2}+\lambda_{1}\lambda_{3}+\lambda_{2}\lambda_{3}.
\ea

On peut alors calculer la fonction de distribution de probabilit\'e
de la densit\'e locale dans cette approximation
(Kofman et al. 1994),
\ba
P(\rho) \d \rho=&\disp{
 {{9\ 5^{3/2} \ \d\rho} \over {4\pi N_s \rho^3 \sigma^4}}
\int_{3 ({1 \over \rho })^{1/3}} ^{\infty}  \d s\
e^{-{(s-3)^2 / 2 \sigma^2}}}\\
&\times
\left( 1+ e^{-{6s/ \sigma^2}} \right)
\ \left( e^{-{\beta_1^2 / 2\sigma^2}}
   +e^{-{\beta_2^2 / 2\sigma^2}}
   -e^{-{\beta_3^2 / 2\sigma^2}}  \right)\ ,
\ea
\be
\beta_n (s) \equiv s\ 5^{1/2} \left( {1\over2}
+\cos\left[{2\over3}(n-1)\pi
+{1\over3} \arccos \left({54\over \rho s^3}
-1 \right)\right]\right) ,
\ee
o\`u param\`etre $\sigma(t)=D(t)\sigma_{0}$ est la d\'eviation
standard des fluctuations de densit\'e $\rho$ dans
la th\'eorie lin\'eaire et $N_s$ est le nombre
moyen de flots ($N_s=1$ dans le r\'egime \`a 1 flot).

\subsection{La vorticit\'e}

Il n'est pas inutile de remarquer que la vorticit\'e
va \^etre dilu\'ee par l'expansion. Si on d\'efinit
la vorticit\'e par,
\be
\vw_k=(\vu_{i,j}-\vu_{j,i})\,\epsilon^{ijk},
\ee
($\epsilon^{ijk}$ est le tenseur totalement antisym\'etrique)
on a (en r\'egime lin\'eaire),
\be
\dot{\vw}=-{\dot{a}\over a}\,\vw.
\ee
Il n'y a \'evidemment pas de terme de source. La vorticit\'e
est un mode d\'ecroissant: elle est dilu\'ee par l'expansion
(i.e. conservation du moment angulaire).

Cette propri\'et\'e est en fait vraie \`a n'importe quel
ordre de la th\'eorie des perturbations. 
(Notons que sa mise \oe uvre pour l'approche
Lagrangienne est loin d'\^etre triviale).

\subsection{Domaine de validit\'e de l'approximation lin\'eaire}

C'est valable au del\`a de 10 $h^{-1}$Mpc. En effet la r.m.s. des
fluctuations atteint 1 vers 8$h^{-1}$Mpc. Les exp\'eriences
num\'eriques
montre qu'une description lin\'eaire est correcte au del\`a de
cette \'echelle. On verra par la suite qu'\`a 
ces \'echelles on peut faire des choses plus sophistiqu\'ees
que la th\'eorie lin\'eaire, mais d'une mani\`ere g\'en\'erale
l'approche perturbative est valable \`a ces \'echelles.

Il est important d'avoir \`a l'esprit que
la th\'eorie lin\'eaire ne change pas la forme du spectre de
fluctuation mais uniquement son amplitude. Pour r\'esumer on a,
\be
P(k)^{\rm local}=\left({D_+(a_0)\over D_+(a_*)}\right)^2\,
P(k)^{\rm recomb.}=\left({D_+(a_0)\over D_+(a_*)}\right)^2\,
T^2(k)\,P(k)^{\rm prim.},
\ee
o\`u $T(k)$ d\'ecrit la fonction de transfert qui contient toute la
micro-physique de la recombinaison.

\subsection{Les galaxies comme traceurs du champ de densit\'e}

Un moyen \'economique (mais relativement dangereux)
de se faire une id\'ee de la forme du spectre est de consid\'erer
la distribution des galaxies dans notre Univers local.

On a maintenant des donn\'ees assez pr\'ecises sur la forme du spectre
dans notre Univers local.
Une forme ph\'enom\'enologique est donn\'ee par (spectre
obtenu \`a partir du catalogue angulaire APM par Baugh et Gazta\~naga 1996),
\be
P(k)\propto{k\over [1+(k/k_c)^2]^{3/2}}
\ee
avec 
\be
k_c={1\over 20}\ h\ Mpc^{-1}.
\ee
La normalisation du spectre est telle que la variance des contrastes
en densit\'e dans une sph\`ere de rayon $8\,h^{-1}$Mpc est de l'ordre
de l'unit\'e,
\be
\sigma_8=\big<\left(
{3\over 4\,\pi\,R_8^3}\int_{\vert\vx\vert<R_8}\delta(\vx)\d^3\vx
\right)^2\big>^{1/2}\approx 1\ \ {\rm pour}\ \ R_8=8\,h^{-1}\,{\rm Mpc}.
\ee
La valeur de $\sigma_8$ pour les galaxies varie un peu
avec la population d'objets s\'electionn\'es. On observe en
particulier une l\'eg\`ere augmentation de $\sigma_8$ avec la
luminosit\'e des objets.

Pour ces raisons historiques, la normalisation du spectre de $P(k)$
est souvent discut\'ee en terme de $\sigma_8$. D'un point de vue
dynamique il faut retenir que c'est une \'echelle de transition.

\section{ Vers la dynamique non-lin\'eaire: l'effondrement
sph\'erique}

 L'\'emergence de non-lin\'earit\'es est en soit un probl\`eme
difficile.
Un cas simple pour lequel on peut calculer l'\'evolution de mani\`ere
compl\`ete est de regarder l'effondrement d'une r\'egion \`a
sym\'etrie sph\'erique.

Je suppose donc que j'ai une fluctuation initiale de forme,
\be
\delta_i(\vx)=f(\vert\vx-\vx_0\vert).
\ee
Pour simplifier on peut penser \`a un profil en marche d'escalier
mais ce n'est pas n\'ecessaire.
\`A cette fluctuation on peut appliquer le th\'eor\`eme
de Birkhoff, donc on peut \'ecrire l'\'equation d'\'evolution
du rayon de cette perturbation,
\ba
\ddot{R}&=&-{G\,M(<R)\over R^2}\\
M(<R)&=&\left(\rho-{\lambda\over4\,\pi\,G}\right)\,
{4\,\pi\over 3}\,R^3.
\ea

Tant qu'on a pas de  shell-crossing, la masse (hors $\lambda$)
se conserve.
On a finalement une \'equation d'\'evolution tout \`a fait similaire
\`a celle du facteur d'expansion, mais avec une constante 
d'int\'egration diff\'erente. La solution de cette \'equation est
connue (quand $\lambda=0$). Il reste une difficult\'e technique
qui est d'exprimer la solution $R$ en fonction de $a$.
Quand $\Omega=1$, c'est relativement simple ($a$ et $t$ sont
simplement reli\'es) mais sinon la relation est formellement assez 
compliqu\'ee.

Les conditions initiales sont
d\'etermin\'ees par la valeur de la sur-densit\'e initiale,
\be
\dta_i={3\,M\over 4\pi R_i^3\rho_0(t)}-1,
\ee
o\`u $\rho_0(t)$ est la densit\'e moyenne de l'univers, 
et par la taille initiale $R_i$ de la fluctuation.

Dans le cas d'un univers plat sans constante cosmologique, la
sur-densit\'e ${3 M\over 4\pi R_i^3\rho_0}$ ne d\'epend que du
produit de la densit\'e initiale et du facteur d'expansion
dont l'univers a cr\^u depuis le d\'ebut de l'effondrement. La sur-densit\'e
s'\'ecrit, quand la sur-densit\'e initiale est positive,
\ba
\dta_i {a(t) \over a(t_i)}&=&\disp{
{3\over5}\left({3\over4}(\theta-\sin \theta)\right)^{2/3}}\\
{3M\over 4\pi R(t)^3\rho_0(t)}&=&\disp{
{9\over2}{(\theta-\sin\theta)^2 \over (1-\cos\theta)^3}},
\ea
et, quand la sur-densit\'e initiale est n\'egative,
\ba
\dta_i {a(t)\over a(t_i)}&=&\disp{
-{3\over5}\left({3\over4}(\sinh\theta-\theta)\right)^{2/3}}\\
{3M\over 4\pi R(t)^3\rho_0(t)}&=&\disp{
{9\over2}{(\sinh\theta-\theta)^2 \over (\cosh\theta-1)^3}}.
\ea
Il est int\'eressant de remarquer que la densit\'e
(exprim\'ee comme le rapport $M<(R)/R^3$) est donn\'ee en fonction
de la fluctuation lin\'eaire $\delta_i\,D_+$ 
approximativement par,
\be
\rho\approx{1\over \left(1-{2 \delta_i\,D_+\over3}\right)^{3/2}}-1.
\ee
Cette relation (exacte quand $\Omega\to 0$) 
est valable pratiquement pour
n'importe quelles valeurs des param\`etres cosmologiques.
Ce qu'on voit c'est qu'au bout d'un temps fini, on rencontre
une singularit\'e, i.e. la densit\'e devient infinie.
Cette singularit\'e est atteinte quand 
$\delta_i\,D_+(t)=\delta_c$ avec $\delta_c=1.69$ pour la dynamique
exacte dans un espace Einstein-de Sitter (dans la dynamique
approch\'ee ci-dessus $\delta_c=1.5$).
Si $\delta_i$ est n\'egatif on ne rencontre pas de singularit\'e, et
la densit\'e se comporte comme,
\be
\rho\sim(-\delta_i\,D_+)^{-3/2},\ \ \delta_i<0.
\ee
Les r\'egions initialement sous-denses se vident donc peu \`a peu.

\subsection{La virialisation}

La description de l'effondrement par les \'equations pr\'ec\'edentes
n'est valable que lorsqu'on n'a pas de croisement de coquilles.
D\`es que l'on a des croisements, la masse $M(<R)$ n'est plus
conserv\'ee. 

Ce qui va se passer alors est un processus de virialisation, la
mati\`ere va tendre vers un \'equilibre dynamique. Pour \'evaluer
l'\'etat final du syst\`eme on peut faire un simple
bilan \'energ\'etique. 
Pour un syst\`eme isol\'e on a l'\'equation
suivante,
\be
{\d(K+V)\over \d t}=-{\dot{a}\over a}(2K+V),
\ee
avec,
\ba
K&=&\sum_{\rm particules}{\vp^2\over 2\,m\,a^2},\\
V&=&\sum_{\rm paires}-{Gm^2\over 2\,a\,\vert\vx_i-\vx_j\vert}.
\ea
Si on suppose que l'\'equilibre est atteint quand le second membre s'annule, 
alors le rayon de la structure qui se forme ainsi doit
\^etre la moiti\'e du rayon de 'turn-around', rayon physique
(et non-comouvant) maximal atteint par la structure.

Cette id\'ee a \'et\'e v\'erifi\'ee dans des simulations num\'eriques.
Au moins pour des structures suffisamment isol\'ees, c'est une
bonne description de ce qui se passe. 
Le contraste en densit\'e qui en r\'esulte $\Delta$, est de l'ordre de
200,
\be
\Delta\equiv{\rho_{\rm objet.}\over\rho_0}\approx 200.
\ee

Les m\'ecanismes de virialisation sont cependant loin d'avoir \'et\'e
\'elucid\'es.

\subsection{La th\'eorie de Press et Schechter}


Comme application du mod\`ele du collapse sph\'erique,
on peut chercher \`a calculer le nombre d'objets
virialis\'es qui sont form\'es \`a une \'epoque donn\'ee.
L'id\'ee est la suivante, on cherche la fraction de mati\`ere
qui se trouve dans un objet virialis\'e de masse donn\'ee.

La masse d'un objet par conservation de la masse est donn\'ee par la
taille $R$ avec $M=4\pi\,R^3/3$ de la perturbation initiale. On va
donc chercher la probabilit\'e qu'un point donn\'e de l'Univers se
trouve initialement dans une fluctuation de densit\'e \`a l'\'echelle
$R$ qui soit telle qu'elle est virialis\'ee \`a une \'epoque
ult\'erieure donn\'ee. Cette probabilit\'e donne alors la fraction de
masse de l'Univers $f(>M)$ qui est dans des objets de masse plus
grande que $M$, 
\be
f(>M)=\int_{\delta_c\,D_+(t_i)/D_+(t_0)}\d\delta_i\,p(\delta_i,R) 
\ee
o\`u $p(\delta)$ est la fonction de distribution de probabilit\'e de la
sur-densit\'e initiale $\delta_i$ \`a une \'echelle de filtrage
$R$. Pour des conditions initiales gaussiennes  $p$ est simplement une
gaussienne de variance $\sigma_i(M)$ qui s'exprime en fonction du
spectre et de la fen\^etre de filtrage.  Alors,
\be 
f(>M)={1\over
2}\left(1-Erf\left[\delta_c\,D_+(t_i)\over
\sqrt{2}D_+(t_0)\sigma_i(M)\right] \right), 
\ee 
et donc  
\be
f(M)\d\,M={-\delta_c\over\sqrt{2\pi}\, {D_+(t_0)\over
D_+(t_i)}\sigma_i(M)}\,{\d\log(\sigma_i(M))\over\d M}\,
\exp\left[-{\delta_c^2\over 2\,\left(D_+(t_0)\over
D_+(t_i)\right)^2\sigma_i^2(M)} \right]\,\d\,M.  
\ee 
C'est une
relation s\'eduisante mais par bien des \'egards, tr\`es
suspecte. On peut d\'ej\`a remarquer que la fraction totale de
mati\`ere dans des objets virialis\'es atteint $1/2$ alors qu'on
s'attendrait \`a ce que cela atteigne $100\%$.  Il se trouve qu'en
corrigeant cette formule d'un simple facteur 2 on a un bon accord avec
les r\'esulats de simulations num\'eriques.

Le probl\`eme est que ce calcul ne prend pas en compte toutes les
structures qui sont finalement susceptibles d'entra\^\i ner un point
de mati\`ere donn\'e.

Bond et al. 1991 
ont donn\'e une interpr\'etation alternative
\`a cette formule en prenant comme fen\^etre de filtrage un filtre
carr\'e dans l'espace des phases. Alors \`a chaque changement
d'\'echelle la densit\'e locale est donn\'ee par une quantit\'e
suppl\'ementaire qui est ind\'ependante de la densit\'e locale \`a
l'\'echelle pr\'ec\'edente. Le probl\`eme finalement peut se formuler
sous forme d'une marche al\'eatoire.  Cette approche a le m\'erite de
redonner le facteur 2 (toute marche al\'eatoire croise une fronti\`ere
fixe \`a distance finie), mais il reste que le probl\`eme n'a pas
\'et\'e r\'esolu dans toute sa complexit\'e (voir Blanchard et al. 1992).

En plus de la complexit\'e du probl\`eme purement
statistique, la dynamique d'un \'el\'ement de volume pris
au hasard n'est certainement pas 
toujours locale, i.e. d\'ependante uniquement de la surdensit\'e
locale, ind\'ependante des effets de mar\'ees.

La th\'eorie de Press et Schechter est en tout \'etat de cause
plus pertinente pour des objets isol\'es, rares, comme les amas de galaxie.

\subsection{La densit\'e d'amas pour contraindre l'amplitude
du spectre de puissance $P(k)$}


La densit\'e des amas de galaxies est tr\`es utile
pour contraintre l'amplitude du spectre de
fluctuation, puisqu'elle teste le comportement de la
distribution de densit\'e initiale dans sa coupure exponentielle.
Typiquement un amas correspond \`a un \'ev\'enement \`a 2, 3
$\sigma$ par rapport aux fluctuations initiales.
M\^eme si il y a quelques incertitudes sur les pr\'efacteurs,
cela n'affecte pas trop les conclusions qu'on peut tirer sur
la valeur de l'amplitude des fluctuations donc de $\sigma_8$.

Comme on voit que la quantit\'e qui est effectivement test\'ee
est une combinaison de $D_+$ et de $\sigma$
on voit bien que l'on va contraindre une combinaison de $\sigma_8$
et des param\`etres cosmologiques (avec une d\'ependance
sur la forme du spectre que je vais ignorer dans la suite).

En pratique on a (voir par exemple Oukbir \& Blanchard 1997, Eke et al. 1996),
\be
\sigma_8\,\Omega_0^{0.5}\approx0.6\pm0.15.
\ee
Cette analyse d\'epend un peu de la forme du spectre.
L'\'echelle de masse des amas est bien 8 $h^{-1}$Mpc
pour un Univers Einstein-de Sitter mais ce n'est pas forc\'ement
vrai autrement. Notons que la d\'eg\'en\'erescence entre $\Omega_0$ et
$\sigma_8$ peut \^etre lev\'ee en examinant la d\'ependance
en $z$ de la distribution des amas (voir Oukbir et al. 1997).

\section{ Le r\'egime quasilin\'eaire, effets des couplages de
modes}

Dans cette partie je vais explorer la th\'eorie
des perturbations appliqu\'ee \`a la formation des grandes structures,
d'un point de vue th\'eorique mais aussi 
en illustrant ces techniques par des applications observationnelles
possibles.

Un des objectifs de ces calculs est de d\'eterminer des
quantit\'es statistiques qui tiennent compte aussi
bien des propri\'et\'es statistiques initiales que de la
dynamique.
Une quantit\'e plus particuli\`erement consid\'er\'ee 
(parce qu'elle est facile d'acc\`es observationnellement aussi 
bien que num\'eriquement) est la fonction de 
probabilit\'e de la densit\'e locale, $P(\delta)$. 

En pratique le champ de densit\'e est obtenu apr\`es filtrage 
d'une repr\'esentation discr\`ete de ce champ, par des points d'une
simulation num\'erique ou des galaxies dans un catalogue.
Alors $P(\delta)\d\,\delta$ est la probabilit\'e que la densit\'e
locale soit entre $\delta$ et $\d\delta$.

Initialement on sait que cette distribution doit \^etre gaussienne.
La largeur de cette distribution d\'epend de l'amplitude et de
la forme du spectre de puissance $P(k)$. Plus pr\'ecis\'ement,
\be
\sigma^2(R)=\int{\d^3\vk\over (2\pi)^3}
\,P(k)\,W^2(k\,R),
\ee
o\`u $R$ est l'\'echelle de filtrage et $W$ est la fonction
de filtrage dans l'espace des $k$.

\begin{figure}
\vspace{8 cm}
\special{hscale=60 vscale=60 voffset=0 hoffset=0 psfile=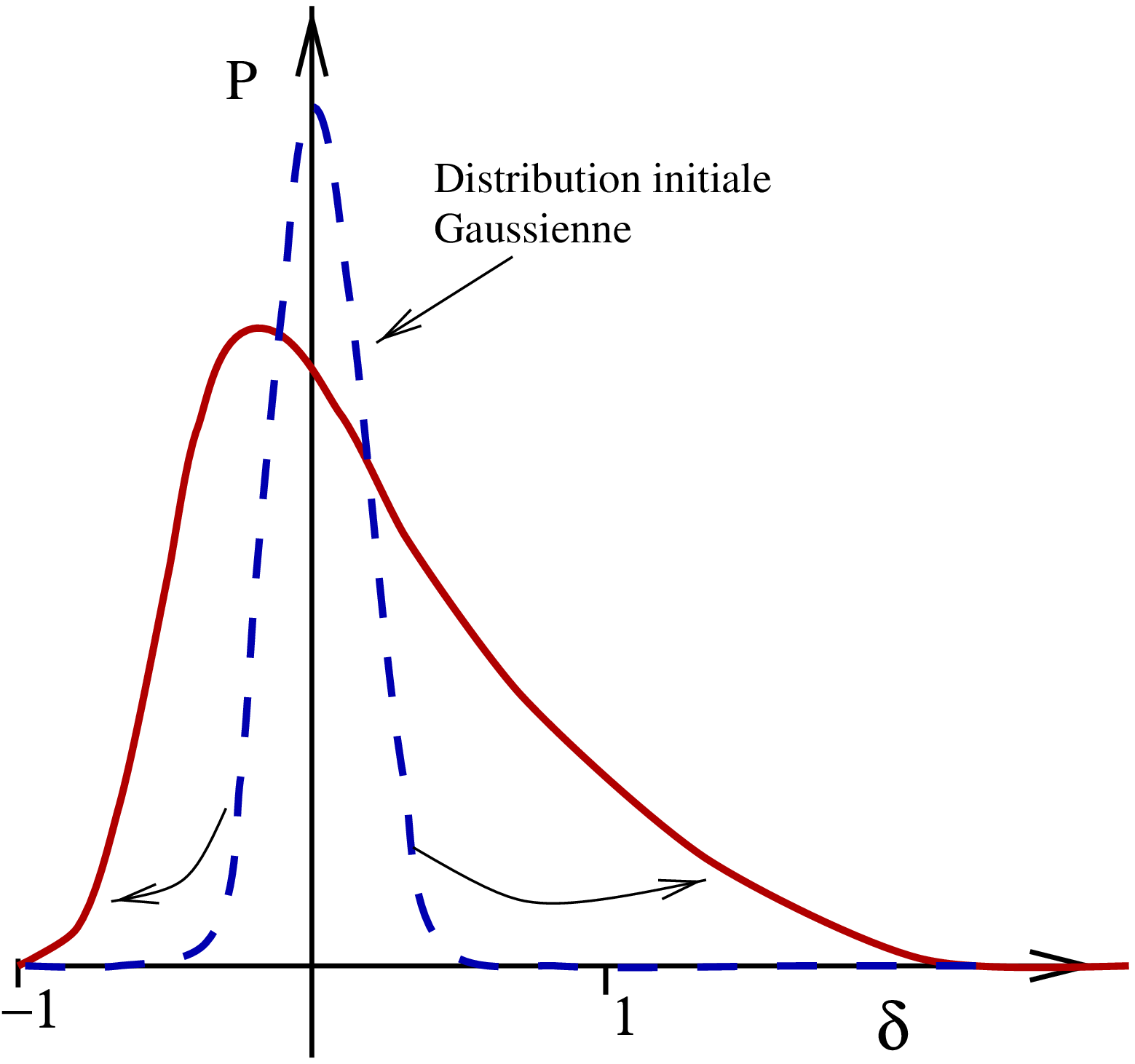}
\caption{Evolution quasi-lin\'eaire de la forme de la fonction de 
distribution de probabilit\'e $P$ quand on entre dans le r\'egime
non-lin\'eaire. L'\'evolution diff\'erentielle entre les parties
sur-denses et les parties sous-denses induit une asym\'etrie.}
\label{PDFsketch}
\end{figure}

Les r\'esultats obtenus 
sur l'effondrement sph\'erique donnent une id\'ee de ce qui
peut se passer quand les non-lin\'earit\'es 
commencent \`a jouer un r\^ole: l'effondrement des r\'egions 
surdenses est acc\'el\'er\'es alors que 
les r\'egions sous-denses se vident
de moins en moins vite. La figure (\ref{PDFsketch}) donne une
id\'ee de ce qui en r\'esulte: on s'attend \`a ce que 
les \'ev\'enements rares cr\'eent une asym\'etrie dans la distribution
de densit\'e locale. 
Le probl\`eme est que le collapse sph\'erique ne tient pas
compte des couplages non-locaux. Il ne saurait donc a priori donn\'e
une image fiable de ce qui se passe.
L'objet des calculs qui suivent est d'explorer aussi pr\'ecis\'ement
que possible ces couplages et leurs cons\'equences.

\subsection{ Propri\'et\'es g\'en\'erales du d\'eveloppement
perturbatif}

Je commence par donner des propri\'et\'es g\'en\'erales du
d\'eveloppement perturbatif en reprenant des r\'esultats
obtenus par Goroff et al. (1984).
Le point de d\'epart est le syst\`eme d'\'equation d\'ecrivant
la dynamique du fluide en espace Eul\'erien dans l'approximation
\`a un seul fluide. 
C'est d'ailleurs une hypoth\`ese fondamentale qu'on ne sait
pas lever.
Je me place dans un espace Einstein-de Sitter pour faciliter
la pr\'esentation (on peut le faire dans toute sa g\'en\'eralit\'e
mais c'est plus laborieux).

Je rappelle qu'on a,
\ba
{\partial \over \partial t}\rho(\vx,t)+{1\over a}\gradx .\left[
\rho(\vx,t)\vu(\vx,t)\right]&=&0\\
{\partial \over \partial t}\vu(\vx,t)+{\dot a \over a}\vu(\vx,t)+{1\over a}
(\vu(\vx,t).\gradx)\vu(\vx,t)&=&-{1\over a}\gradx\psi(\vx,t)\\
\gradx^2\psi(\vx,t)-4\pi G\left[\rho(\vx,t)-\overline{\rho}(t)\right]a^2&=&0.
\ea

A partir de l\`a je vais prendre la divergence de la deuxi\`eme
\'equation et utiliser l'\'equation de Poisson pour \'eliminer
le potentiel. On a,
\ba
{\partial \over \partial t}\rho(\vx,t)+{1\over a}\gradx .\left[
\rho(\vx,t)\vu(\vx,t)\right]&=&0\\
{\partial \over \partial t}\gradx .\vu(\vx,t)+
{\dot a \over a}\gradx .\vu(\vx,t)+&&\nonumber\\
{1\over a}
\sum_{ij}\vu_{i,j}(\vx,t)\,\vu_{i,j}(\vx,t)+{1\over a}\vu(\vx,t).
\gradx (\gradx .\vu(\vx,t))&=&-{3\over 2}{\dot{a}^2
\over a}\delta(\vx,t).
\ea
Enfin je vais introduire un nouveau gradient, 
$\nabla\equiv {1\over a\,H}\gradx$ et le champ $\Phi$
de telle mani\`ere\footnote{la partie rotationnelle du champ de 
vitesse correspondant \`a un mode d\'ecroissant, le champ
de vitesse qu'on regarde est a priori potentiel.} 
que $\vu=\grad\Phi$. Enfin
les d\'eriv\'ees par rapport au temps 
sont exprim\'ees comme des d\'eriv\'ees par rapport \`a $a$. 
Notons qu'alors $\Delta\Phi(\vx)$ s'identifie avec $\theta$
qui est, on s'en souvient, une observable ind\'ependante de la
constante de Hubble.
On obtient,
\ba
a{\partial\over \partial a}\delta(\vx)+(1+\delta(\vx))\Delta\Phi(\vx)+
\grad\delta(\vx).\grad\Phi(\vx)&=&0\label{cont}\\
a{\partial\over \partial a}\Delta\Phi(\vx)+{1\over2}
\Delta\Phi(\vx)+\grad\Phi(\vx)\cdot\grad\Delta\Phi(\vx)+&&\nonumber\\
\sum_{ij}\Phi_{,ij}(\vx)\Phi_{,ij}(\vx)+{3\over 2}\delta(\vx)&=&0.\label{eul}
\ea

Si on \'ecrit la densit\'e (et la divergence) comme une s\'erie
en fonction de la densit\'e initiale,
\ba
\delta(\vx)&=&\sum_n\delta^{(n)}(\vx),\\
\theta(\vx)&=&\sum_n\theta^{(n)}(\vx),
\ea
et en introduisant les modes de Fourier de ce champ initial,
\be
\delta(\vx)=\int{\d^3\vk\over(2\pi)^{3/2}}\delta(\vk)\,e^{\ii\vk.\vx},
\ee
alors
on peut facilement montrer que,
\ba
\delta^{(n)}(\vx)&=&\int{\d^3\vk_1\over(2\pi)^{3/2}}\delta(\vk_1)\dots
{\d^3\vk_n\over(2\pi)^{3/2}}\delta(\vk_n)\,a^n\,F_n(\vk_1,\dots,\vk_n)\\
\theta^{(n)}(\vx)&=&\int{\d^3\vk_1\over(2\pi)^{3/2}}\delta(\vk_1)\dots
{\d^3\vk_n\over(2\pi)^{3/2}}\delta(\vk_n)\,a^n\,G_n(\vk_1,\dots,\vk_n).
\ea
Les fonctions $F_n$ et $G_n$ sont des fonctions sans dimensions,
homog\`enes, des vecteurs d'onde $\vk_i$. Il existe des lois
de recursivit\'e liant ces fonctions entre elles.
Par exemple,
\ba
F_1&=&1\\
G_1&=&-1\\
F_2&=&{5\over 7}+{1\over 2}{\vk_1.\vk_2\over k_1^2}+
{1\over 2}{\vk_1.\vk_2\over k_2^2}+
{2\over 7}{(\vk_1.\vk_2)^2\over k_1^2\,k_2^2}\\
G_2&=&-\left[{3\over 7}+{1\over 2}{\vk_1.\vk_2\over k_1^2}+
{1\over 2}{\vk_1.\vk_2\over k_2^2}+
{4\over 7}{(\vk_1.\vk_2)^2\over k_1^2\,k_2^2}\right]\\
&&\dots\nonumber
\ea

Dans le cas d'un Univers qui n'est pas EdS, cette propri\'et\'e
n'est plus valable au sens stricte.
Mais si dans l'\'equation pr\'ec\'edente on remplace les facteurs
d'expansion $a$ par $D_+$, les fonctions $F_n$ qui en r\'esultent
ne d\'ependent que tr\`es faiblement du temps (et donc des
param\`etres cosmologiques). Les fonctions $G_n$ 
sont essentiellement toutes proportionelles \`a $f(\Omega)$.

La cons\'equence de ces propri\'et\'es est que les
d\'eveloppements perturbatifs qui se font formellement
par rapport \`a la densit\'e initiale, sont en fait
des d\'eveloppements par rapport \`a la solution {\it lin\'eaire}.

\subsection{Un exemple: le calcul de $F_2$ et $G_2$}

A partir des \'equations coupl\'ees, on peut facilement \'ecrire une 
relation r\'ecursive entre $F_1$, $G_1$, $F_2$ et $G_2$.
On a en effet,
\ba
2\,F_2(\vk,\vk')+G_2(\vk,\vk')+F_1(\vk)\,G_1(\vk')\left[1+
{\vk\cdot\vk'\over k'^2}\right]&=&0\\
{3\over 2}\,G_2(\vk,\vk')+G_1(\vk)\,G_1(\vk')\left[{\vk\cdot\vk'\over
k'^2}+{(\vk\cdot\vk')^2\over
k^2\,k'^2}\right]+{3\over 2}\,F_2(\vk,\vk')&=&0
\ea
A partir de $F_1=-G_1$ (qui s'obtient trivialement \`a partir
de l'\'equation de continuit\'e) on en d\'eduit $F_2$ et $G_2$.
Ce calcul n'est valable que dans le cas Einstein-de Sitter.
On voit facilement qu'on peut g\'en\'eraliser ce type d'approche
\`a un ordre arbitraire. La d\'ependance avec les vecteur d'onde
sera de plus en plus compliqu\'ee, mais fera toujours intervenir
les 2 m\^emes fonctions g\'eom\'etriques de base.

\subsection{ Effet du couplage de mode: la skewness}

Une premi\`ere cons\'equence de ce d\'eveloppement qu'il est facile
de voir est l'apparition de couplages de mode. 
La densit\'e \`a l'ordre 2 est une convolution des modes
lin\'eaires. D'un point de vue observationnelle
il est int\'eressant d'en d\'eduire des propri\'et\'es
statistiques a priori observables et un moyen de mettre
en \'evidence ces couplages de mode est d'examiner l'apparition de
propri\'et\'es non-gaussiennes.
Le premier moment non-trivial \`a appara\^\i tre est la skewness,
moment d'ordre 3 de la densit\'e.

On cherche donc \`a calculer $\mg\delta^3\md$ et son terme dominant.
On a
\be
\mg\delta^3\md=\mg\left(\delta^{(1)}+\delta^{(2)}+\dots\right)^3\md.
\ee
Si on r\'eordonne les termes qu'on obtient perturbativement on a,
\be
\mg\delta^3\md=\mg\left(\delta^{(1)}\right)^3\md+3\,
\mg\left(\delta^{(1)}\right)^2\,\delta^{(2)}\md+\dots
\ee
Les termes suivants sont d'ordre plus grand en th\'eorie
des perturbations.

Le premier terme de ce d\'eveloppement est identiquement nul
pour des conditions initiales gaussiennes.
Le terme suivant est donc a priori le terme dominant
pour cette quantit\'e.
On a donc,
\ba
\mg\delta^3\md&\approx&
3\,\mg\left(\delta^{(1)}\right)^2\,\delta^{(2)}\md\\
&=&
3\,\int{\d^3\vk_1\over (2\pi)^{3/2}}\,a\,
\int{\d^3\vk_2\over (2\pi)^{3/2}}\,a\,
\int{\d^3\vk_3\over (2\pi)^{3/2}}\,a\,
\int{\d^3\vk_4\over (2\pi)^{3/2}}\,a\,\times\nonumber\\
&&\mg\delta(\vk_1)\,\delta(\vk_2)\,\delta(\vk_3)\,\delta(\vk_4)\md
\,F_2(\vk_2,\vk_3)\,
\exp[\ii(\vk_1+\vk_2\vk_3+\vk_4)\cdot\vx]
\ea
Pour des conditions initiales gaussiennes il faut associer les
modes de Fourier par paires. Si on associe $\vk_2$ et $\vk_3$
ensembles on a 0 (\`a cause de $F_2$). Il reste deux termes
qui donnent,
\ba
\mg\delta^3\md&=&
6\,\int{\d^3\vk_1\over (2\pi)^{3}}\,
\int{\d^3\vk_4\over (2\pi)^{3}}\,a^4\,P(k_1)\,P(k_4)\,\times\nonumber\\
&&\left({5\over 7}+{1\over 2}{\vk_1\cdot\vk_4\over k_1^2}+
+{1\over 2}{\vk_1\cdot\vk_4\over k_4^2}
+{2\over 7}{(\vk_1\cdot\vk_4)^2\over k_1^2\,k_4^2}\right).
\ea
En int\'egrant sur les angles entre $\vk_1$ et $\vk_4$
on obtient finalement (Peebles 1980),
\be
\mg\delta^3\md={34\over 7}\mg\delta^2\md^2.
\ee
On voit qu'il appara\^\i t un nombre pur, $34/7$,
et la tradition est de d\'efinir,
\be
S_3\equiv{\mg\delta^3\md\over 
\mg\delta^2\md^2}={34\over 7}+{\cal O}(\sigma^2).
\ee
Ce calcul cependant est acad\'emique parce que la densit\'e
a \'et\'e prise ponctuellement. En pratique les champs
sont filtr\'es (que ce soit observationellement ou
dans des exp\'eriences num\'eriques). Il faut alors
tenir compte de cet effet dans le calcul de $S_3$.
La difficult\'e du calcul tient dans la complexit\'e qu'il y a \`a
int\'egrer la partie angulaire des vecteurs d'onde.
Pour avoir la skewness de la densit\'e locale filtr\'ee, $\delta_R$,
il faut en effet calculer,
\ba
\mg\delta_R^3\md&=&
6\,\int{\d^3\vk_1\over (2\pi)^{3}}\,
\int{\d^3\vk_4\over (2\pi)^{3}}\,a^4\,P(k_1)\,P(k_4)\,
W(k_1\,R)\,W(k_4\,R)
\times\nonumber\\
&&\left({5\over 7}+{1\over 2}{\vk_1\cdot\vk_4\over k_1^2}+
+{1\over 2}{\vk_1\cdot\vk_4\over k_4^2}
+{2\over 7}{(\vk_1\cdot\vk_4)^2\over k_1^2\,k_4^2}\right)\,W(\vert\vk_1+\vk_4\vert\,R).
\ea
La fonction $W$ est la fonction de filtrage dans l'espace des phases.
Elle d\'epend \'evidemment de la proc\'edure de filtrage utilis\'ee.
Il se trouve que le r\'esultat final peut prendre une forme simple
pour un filtrage top-hat dans l'espace r\'eel. Dans ce cas l\`a,
\be
W(k)=\sqrt{3\pi\over 2}{J_{3/2}(k)\over k^{3/2}}=
{3\over k^3}\left[\sin(k)-k\,\cos(k)\right]
\ee
Pour cette fen\^etre on a,
\ba
\int{\d\Omega_{12}\over 4\pi}\,W(\vert\vk_1+\vk_2\vert)\left[1-
{(\vk_1\cdot\vk_2)^2\over k_1^2\,k_2^2}\right]&=&
{2\over 3}\,W(k_1)\,W(k_2)\\
\int{\d\Omega_{12}\over 4\pi}\,W(\vert\vk_1+\vk_2\vert)\left[1+
{\vk_1\cdot\vk_2\over k_1^2}\right]&=&W(k_1)\left[W(k_2)+{1\over 3}
k_2\,W'(k_2)\right].
\ea
Ces propri\'et\'es s'obtiennent en utilisant le th\'eor\`eme de somme
des fonctions de Bessel.
On voit qu'on peut exprimer la fonction $F_2$ en utilisant
les 2 expressions polynomiales des relations pr\'ec\'edentes.
Finalement on obtient (Bernardeau 1994b),
\be
S_3={34\over 7}+{\d\log\sigma^2(R)\over \d\log R}.
\ee
Le r\'esultat final va d\'ependre de la forme du spectre du puissance
(essentiellement \`a l'\'echelle de filtrage). Pour un spectre
en loi de puissance,
\be
P(K)\propto k^n,
\ee
on a,
\be
S_3={34\over 7}-(n+3)
\ee
L'\'etude des amas de galaxies permet d'avoir des indications sur
la valeur de $n$. On trouve des valeurs de l'ordre de $n\approx -1.5$.
Les comparaisons avec les simulations num\'eriques ont montr\'e
que ce r\'esultat perturbatif sur $S_3$ \'etait tr\`es robuste.
Le domaine de validit\'e de ce r\'esultat est relativement grand
comme on peut le voir sur la figure.

\begin{figure}
\vspace{8 cm}
\special{hscale=60 vscale=60 voffset=0 hoffset=-20 psfile=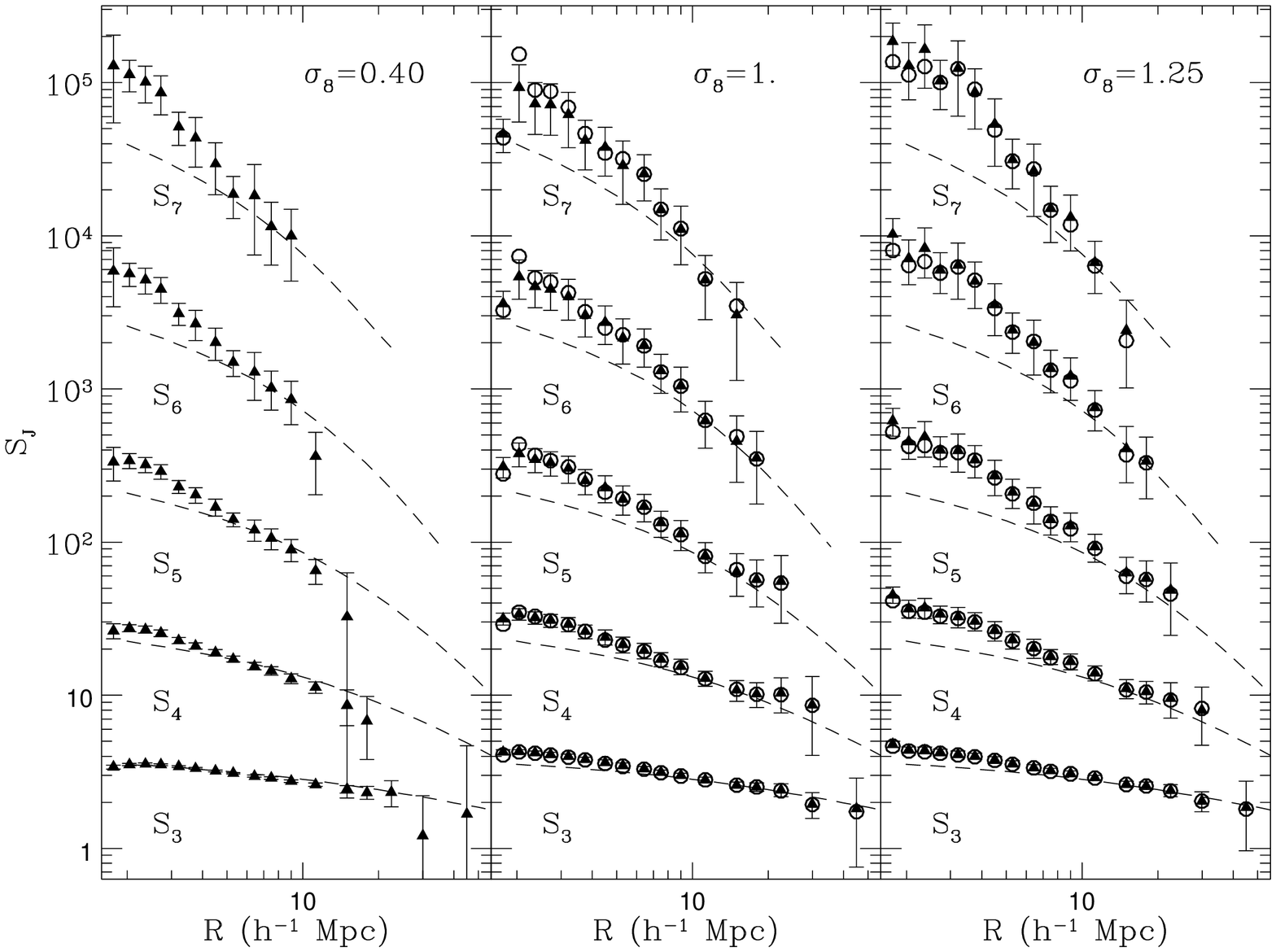}
\caption{Les param\`etres $S_p$ pour $3\le p\le7$. Comparaison des
pr\'edictions th\'eoriques (Bernardeau 1994c) avec les r\'esultats de
simulations num\'eriques (Baugh et al. 1994)}
\label{figSpGazta}
\end{figure}

\subsection{ D\'ependance avec les param\`etres cosmologiques}

Une question tr\`es int\'eressante qui se pose alors est de savoir si 
ce param\`etre $S_3$ qui quantifie l'\'emergence de propri\'et\'es
non-gaussiennes d\'epend des param\`etres cosmologiques.
Les calculs analytiques ont montr\'e que la d\'ependance attendue
\'etait tr\`es faibles. En particulier Bouchet et al. 1992 
ont obtenu,
\be
S_3={34\over 7}+{6\over 7}\left(\Omega_0^{-0.03}-1\right)-(n+3).
\ee
Ce r\'esultat est bas\'e sur le fait que la densit\'e au 2\`eme
ordre s'\'ecrit maintenant,
\ba
&\delta^{(2)}(\vx)=
\int{\d^3\vk_1\over (2\pi)^{3/2}}\,
\int{\d^3\vk_1\over (2\pi)^{3/2}}\,
D_+^2\,\delta(\vk_1)\,\delta(\vk_2)\,
\exp[\ii(\vk_1+\vk_2).\vx]\times\nonumber\\
&\left[{1\over 2}+{3\over 14}\Omega^{-2/63}+
{\vk_1.\vk_2\over k_1^2}+
\left({1\over2}-{3\over 14}\Omega^{-2/63}\right)
{(\vk_1.\vk_2)^2\over k_1^2 k_2^2}\right].
\ea
Ce r\'esultat a \'et\'e obtenu quand $\lambda=0$. On trouve
un r\'esultat similaire quand $\lambda\neq 0$.

\subsection{Interpr\'etation, effet du filtrage}

\begin{figure}
\vspace{14 cm}
\special{hscale=50 vscale=50 voffset=0 hoffset=40 psfile=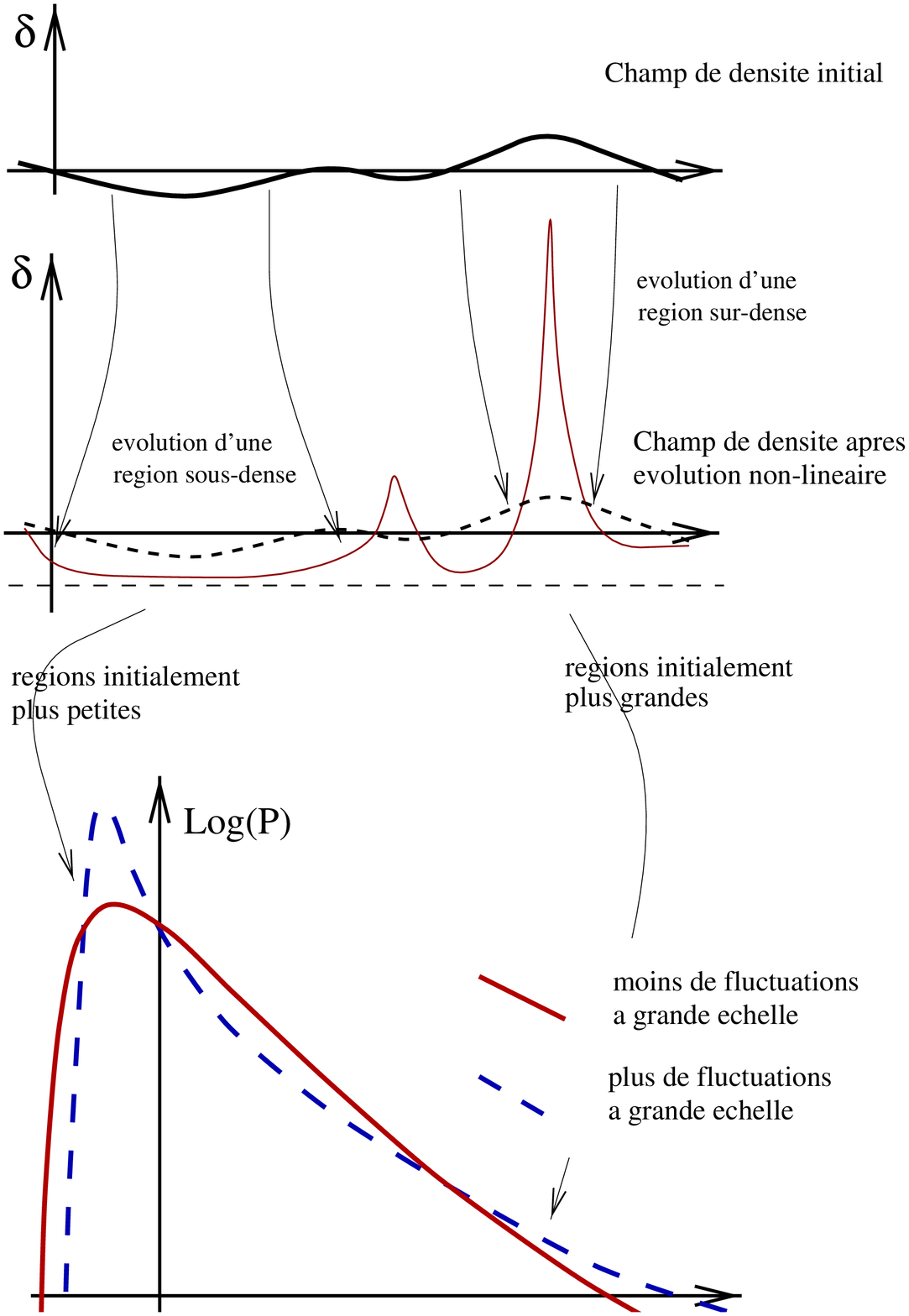}
\caption{La skewness est une mesure de l'asym\'etrie de la
distribution de densit\'e locale. Elle \'emerge parce que les
r\'egions sous-denses \'evoluent moins vite que les
r\'egions sur-denses d\`es que les non-lin\'earit\'es 
commencent \`a jouer un r\^ole. La
d\'ependance de la skewness avec la forme du spectre vient de
la correspondance entre
l'espace Lagrangien qui donne la taille initiale de la perturbation
et l'espace
Eul\'erien qui est sensible \`a la taille finale. \`A une \'echelle de
filtrage $R$ donn\'ee, les r\'egions sur-denses proviennent de
l'effondrement de r\'egions qui \'etaient initialement plus grandes,
alors que les r\'egions sous-denses d\'etect\'ees viennent de
r\'egions initialement plus petites. Du coup si on part d'un spectre
de puissance avec beaucoup plus de
fluctuations \`a petite \'echelle qu'\`a grande \'echelle,
on s'attend a ce que la skewness soit plus petite.}
\label{SmoothEff}
\end{figure}

La skewness  mesure la tendance du syst\`eme \`a cr\'eer une
distribution asym\'etrique avec des \'ev\'enements rares ayant un
contraste de densit\'e grand et positif (les proto-amas) et
beaucoup d'\'ev\'enements de contraste de densit\'e petit et
n\'egatif.
Un moyen de quantifier cet effet est de consid\'erer
le d\'eveloppement de Edgeworth de la distribution de densit\'e locale.
Ce d\'eveloppement est valable pour des distributions 
proches d'une gaussienne (dans un sens qui sera pr\'eciser par 
la suite). Il s'\'ecrit,
\be
p(\delta)\d\delta={1\over \sqrt{2\pi\sigma^2}}\,
\exp\left(-{\delta^2\over 2\sigma^2}\right)\,
\left[1+{S_3\over 6}\,\sigma\,H_3\left(\delta\over\sigma\right)+\dots\right]
\ee
Ce d\'eveloppement utilise les polyn\^omes de Hermite. Ici
\be
H_3(\nu)=\nu^3-3\,\nu.
\ee
Sa forme g\'en\'erale peut \^etre vue comme un d\'eveloppement
formel (et par indentification on peut relier les coefficients
des termes du d\'eveloppement 
avec les moments, Juszkiewicz et al. 1995) ou
\`a partir de la forme formelle de la distribution de densit\'e locale
obtenue comme transform\'ee de Laplace inverse de la fonction
g\'en\'eratrice des cumulants (Bernardeau \& Kofman 1995).

Pour comprendre la d\'ependance de ce param\`etre avec la forme 
du spectre il est tr\`es int\'eressant d'examiner en d\'etail 
la nature du terme qui est rajout\'e avec les effets de filtrage.

Un bon moyen d'appr\'ehender ce probl\`eme est de
regarder ce qui se passe dans l'espace Lagrangien. 
En effet si on calcule $J^{(2)}$ on obtient,
\ba
&J^{(2)}=\disp{\int{\d^3\vk_1\over (2\pi)^{3/2}}\,
\int{\d^3\vk_1\over (2\pi)^{3/2}}\,
a^2\,\delta(\vk_1)\,\delta(\vk_2)\,
\exp[\ii(\vk_1+\vk_2).\vq]}\times\nonumber\\
&\disp{{2\over7}\left[1-
{(\vk_1.\vk_2)^2\over k_1^2 k_2^2}\right]},
\ea
ce qui donne pour la densit\'e apr\`es que le Jacobien
(c'est \`a dire le volume) ait \'et\'e filtr\'e \`a une
\'echelle $R$ donn\'e,
\ba
&\delta_R^{(2)}=\disp{\int{\d^3\vk_1\over (2\pi)^{3/2}}\,
\int{\d^3\vk_1\over (2\pi)^{3/2}}\,
a^2\,\delta(\vk_1)\,\delta(\vk_2)\,
\exp[\ii(\vk_1+\vk_2).\vq]}\times\nonumber\\
&\disp{
\left[W(k_1\,R)\,W(k_2\,R)-{2\over 7}W(\vert\vk_1+\vk_2\vert\,R)\,\left(1-
{(\vk_1.\vk_2)^2\over k_1^2 k_2^2}\right)\right]}.
\ea
Il en r\'esulte que le param\`etre $S_3$ devient,
\be
S_3^{\rm Lag.}={34\over 7}.
\ee
Le fait que l'on ne trouve pas le m\^eme r\'esultat
ne doit pas surprendre. Dans ce cas on vient de faire un filtrage
\`a une \'echelle de {\it masse} donn\'ee. La diff\'erence entre les
2 c'est qu'un objet de masse donn\'ee a a priori une taille d'autant
plus petite que cette masse \'etait grande.
Un filtrage \`a une \'echelle Eul\'erienne donn\'ee m\'elange
donc diff\'erentes \'echelles de masse initiales. Or pour un spectre
hi\'erarchique les fluctuations de masse sont d'autant plus 
faibles que la masse est grande. On aura donc une asym\'etrie
moins importante que celle qu'on aurait pu escompter.

\section{ La hi\'erarchie des corr\'elations en r\'egime
quasi-lin\'eaire}

Peut-on faire mieux que le calcul de $S_3$?
Le calcul des fonctions homog\`enes $F_n$ ou $G_n$ est d'autant plus
difficile que l'ordre est \'elev\'e, et les int\'egrales
qu'il faut calculer deviennent elles aussi tr\`es fastidieuses.
En pratique on ne peut gu\`ere aller au del\`a du moment d'ordre
4 par un calcul direct. Fry (1984) avait explorer ce cas
mais sans prendre en compte les effets de filtrage.
Bernardeau (1994b), \L okas et al. (1995) ont fait ces calculs
directs pour $S_4$ dans le cas respectivement d'un filtre top-hat et
d'un filtre Gaussien.

Formellement le cumulant d'ordre 4 est donn\'e par,
\ba
\mg\delta^4\md_c\equiv&\mg\delta^4\md-3\,\mg\delta^2\md^2\\
&=12\,\mg\left(\delta^{(1)}\right)^2\,\left(\delta^{(2)}\right)^2\md_c+
4\,\mg\left(\delta^{(1)}\right)^3\,\delta^{(3)}\md_c.\nonumber
\ea
Dans ces \'equations le fait de prendre la partie connexe devient
essentiel. Les termes suppl\'ementaires qui appara\^\i ssent
sont des corrections \`a la variance. Ils doivent donc \^etre
enlev\'es (ils s'annulent naturellement quand on fait
la diff\'erence).

Une cons\'equence est que,
\be
\mg\delta^4\md_c\sim\mg\delta^2\md^3,
\ee
et on peut d\'efinir $S_4$ par 
\be
S_4\equiv {\mg\delta^4\md_c/\mg\delta^2\md^3}.
\ee
L'\'equation pr\'ec\'edente permet de calculer la partie
dominante de $S_4$ en r\'egime quasi-lin\'eaire.
D'une mani\`ere g\'en\'erale on d\'efinit
\be
S_n\equiv{\mg\delta^n\md_c/\mg\delta^2\md^{(n-1)}}.
\ee
Toutes ces quantit\'es sont finies \`a grande \'echelle pour des
conditions initiales Gaussiennes.
L'objet de cette section est d'en faire le calcul.

\subsection{ La fonction g\'en\'eratrice des cumulants de la
densit\'e locale}

\begin{figure}
\vspace{2 cm}
\special{hscale=50 vscale=50 voffset=0 hoffset=40 psfile=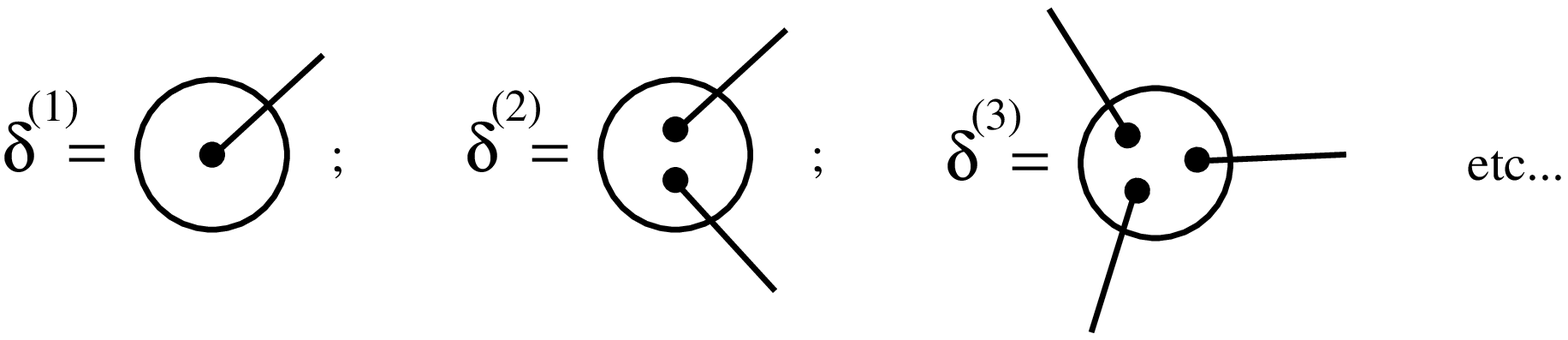}
\caption{Repr\'esentation diagramatique des $\delta_p$. Chaque point
repr\'esente un facteur $\delta(\vk)$.}
\label{deltap}
\end{figure}

On peut d\'ej\`a remarquer que 
\be
S_4=12\,\nu_2^2+4\,\nu_3,
\ee
avec
\ba
\nu_2&\equiv&\mg\delta^{(2)}\left[\delta^{(1)}\right]^2\md/
\mg\left[\delta^{(1)}\right]^2\md^2\\
\nu_3&\equiv&\mg\delta^{(3)}\left[\delta^{(1)}\right]^3\md_c/
\mg\left[\delta^{(1)}\right]^2\md^3.
\ea
d'une mani\`ere g\'en\'erale tous les $S_n$ peuvent s'exprimer
en fonction des seuls quantit\'es, $\nu_p$, avec
\be
\nu_p\equiv\mg\delta^{(p)}\left[\delta^{(1)}\right]^p\md_c/
\mg\left[\delta^{(1)}\right]^2\md^p.
\ee
Pour le voir on peut faire une repr\'esentation graphique des
termes qui contribuent \`a $S_n$. Dans chaque $\delta^{(p)}$
il y a un produit de
$p$ variables al\'eatoires gaussiennes $\delta(\vk)$. 
Je vais repr\'esenter chacune de ces variables par des points. 
Quand on doit calculer une valeur moyenne, l'application du
th\'eor\`eme de Wick fait que tous ces points doivent \^etre 
associ\'es par paires. Les quantit\'es $\delta^{(p)}$
ressemblent donc \`a ce qui est repr\'esent\'e sur la figure \ref{deltap}.

\begin{figure}
\vspace{2 cm}
\special{hscale=50 vscale=50 voffset=0 hoffset=40 psfile=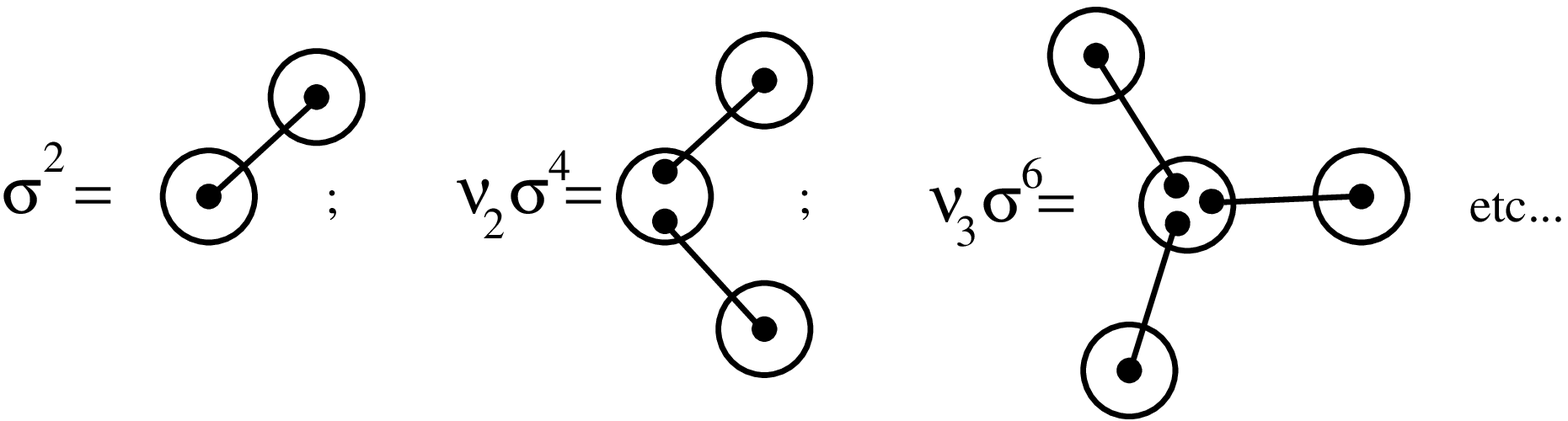}
\caption{Expression des graphes les plus simples.
Chaque ligne porte un facteur $\sigma^2$, Les vertex obtenus par
moyenne angulaire sur les vecteur d'onde portent $\nu_p$.}
\label{nup}
\end{figure}

\begin{figure}
\vspace{3 cm}
\special{hscale=50 vscale=50 voffset=0 hoffset=40 psfile=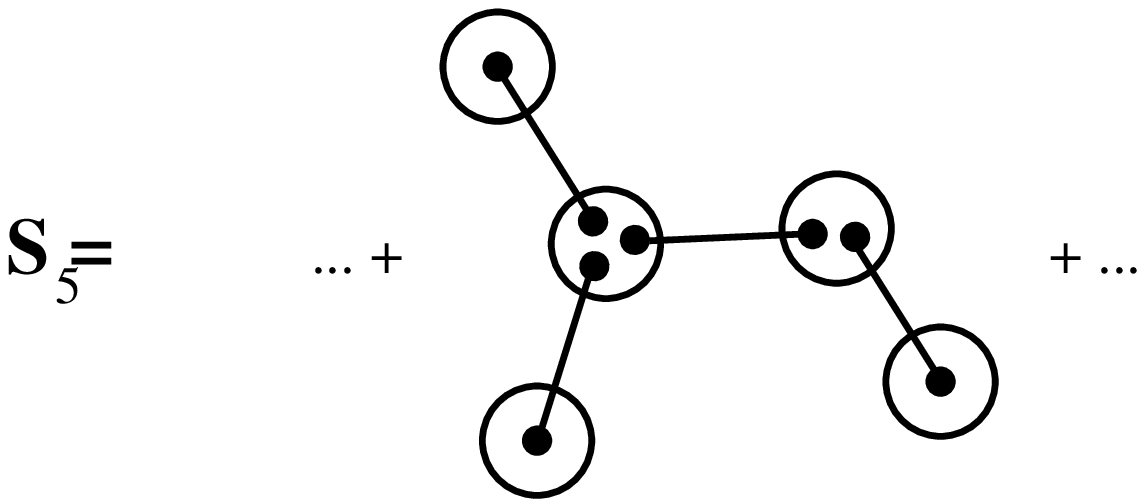}
\caption{Un exemple de graphe contribuant \`a $S_5$.}
\label{exS5}
\end{figure}

Pour calculer $S_n$ il faut calculer le terme dominant
de la partie connexe de produits de la forme 
$\delta^{(p_1)}\dots\delta^{(p_n)}$. Pour que le terme calcul\'e
contribue \`a la partie connexe il faut que tous les facteurs soient
li\'es par au moins une patte. Si ce n'est pas le cas on obtient un
terme qui entre dans les termes correctifs au cumulants d'ordre
inf\'erieur. Par ailleurs comme on cherche le terme dominant le nombre
de lignes utilis\'ees doit \^etre minimal et le nombre de lignes
n\'ecessaire pour relier $n$ points \'etant $n-1$ on aura,
\be
S_n=\disp{\sum_{\rm graphes,\ \sum_i p_i=2(n-1)}\mg
\delta^{(p_1)}\dots\delta^{(p_n)}\md_c/
\mg\left[\delta^{(1)}\right]^2\md^{n-1}}.
\ee
Un exemple d'un tel graphes pour $S_5$ est donn\'e sur la figure \ref{exS5}.

Il est utile de remarquer que tous ces diagrammes sont des arbres,
qu'ils ne contiennent pas de boucles. On peut alors int\'egrer 
sur les vecteurs d'onde de proche en proche. Chaque ligne
fait appara\^\i tre un facteur $\sigma^2$ et chaque vertex fait
appara\^\i tre un facteur $\nu_p$ tel que je les ai d\'efinnis
pr\'ec\'edemment. Graphiquement cela donne ce qui est repr\'esent\'e
sur la figure \ref{nup}.

\begin{figure}
\vspace{2 cm}
\special{hscale=50 vscale=50 voffset=0 hoffset=40 psfile=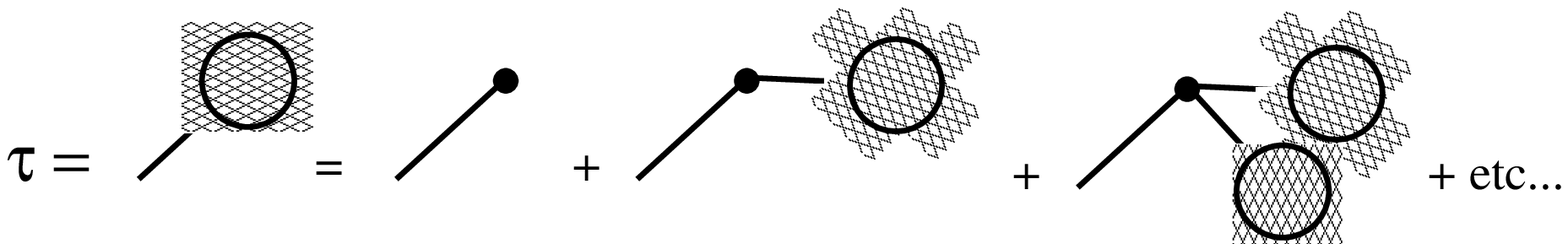}
\caption{Repr\'esentaion graphique de la relation (\ref{taueq}),
$\tau$ est la fonction g\'en\'eratrice des diagrammes \`a une patte externe.}
\label{taugraph}
\end{figure}

Ce qu'on aimerait pouvoir calculer c'est l'ensemble des 
coefficients $S_n$, et donc plus sp\'ecifiquement leur fonction
g\'en\'eratrice $\varphi(y)$, 
\be
\varphi(y)=\sum_{n=1}^{\infty}
-S_n\,{(-y)^p\over p!}.
\ee
Si on d\'efinit la fonction g\'en\'eratrice des vertex $\mGd(\tau)$
par,
\be
\mGd(\tau)=\sum_{p=1}^{\infty}{\nu_n}\,{(-\tau)^n\over n!},
\ee
il est possible de montrer $\varphi$ et $\mGd$ sont li\'es par le
syst\`eme,
\ba
\varphi(y)&=&y+y\,\mGd[\tau(y)]+{1\over2}\tau^2(y),\\
\tau(y)&=&-y\,\mGd'[\tau(y)].\label{taueq}
\ea
Pour d\'emontrer ce syst\`eme de relations il faut faire intervenir les
transform\'ees de Legendre des fonctions g\'en\'eratrices. 
Ces techniques ont \'et\'e
d\'evelopp\'ees initialement dans le domaine des polym\`eres
(voir Des Cloiseaux \& Jannink 1987). Dans un contexte de cosmologie
elles ont \'et\'e employ\'ees par Schaeffer (1984) et pr\'esent\'ees
en d\'etails par Bernardeau \& Schaeffer (1992).
S'il n'est pas possible de d\'emontrer cette relation en quelques
lignes on peut au moins se convaincre de la validit\'e de l'\'equation
(\ref{taueq}). En effet $\tau$ est la fonction g\'en\'eratrice des
graphes \`a une patte externe. Sur la figure (\ref{taugraph}) 
on voit que $\tau$
doit \^etre solution d'une \'equation implicite qui fait intervenir
la fonction g\'en\'eratrice des vertex.

Finalement la difficult\'e technique est dans le calcul de la fonction
g\'en\'eratrice des vertex $\mGd(\tau)$.
Il faut avoir bien \`a l'esprit que les vertex $\nu_p$ sont simplement
les moyennes g\'eom\`etriques des fonctions $F_p$. Pour faire les
calculs je vais aussi avoir besoin des moyennes g\'eom\'etriques
des fonction $G_p$ que je note $\mu_p$, soit,
\ba
\nu_p&=&\overline{F_p(\vk_1,\dots,\vk_p)}^{\rm angles}\\
\mu_p&=&\overline{G_p(\vk_1,\dots,\vk_p)}^{\rm angles}
\ea
et je d\'efini $\mGt(\tau)$ la fonction g\'en\'eratrice
des vertex $\mu_p$.
Pour obtenir $\mGd$ et $\mGt$ il faut partir des \'equations
(\ref{cont}-\ref{eul}) et les r\'e-\'ecrire dans l'espace de Fourier,
\ba
n\,F_n(\vk_1,\dots,\vk_n)+G_n(\vk_1,\dots,\vk_n)+
\sum_{p=1}^{n-1}F_p(\vk_1,\dots,\vk_p)\,
G_{n-p}(\vk_{p+1},\dots,\vk_n)\,&\times\nonumber\\
\left[1+{\left(\sum_{i=1}^p\vk_i\right)\cdot\left(\sum_{i=p+1}^n\vk_i\right)
\over \left(\sum_{i=p+1}^n\vk_i\right)^2}\right]=0&\\
\left(n+{1\over2}\right)\,G_n(\vk_1,\dots,\vk_n)+{3\over2}\,
F_n(\vk_1,\dots,\vk_n)+&\nonumber\\
\sum_{p=1}^{n-1}G_p(\vk_1,\dots,\vk_p)\,
G_{n-p}(\vk_{p+1},\dots,\vk_n)\,&\times\nonumber\\
\left[
{\left(\sum_{i=1}^p\vk_i\right)\cdot\left(\sum_{i=p+1}^n\vk_i\right)
\over \left(\sum_{i=p+1}^n\vk_i\right)^2}+
{\left[\left(\sum_{i=1}^p\vk_i\right)\cdot
\left(\sum_{i=p+1}^n\vk_i\right)\right]^2
\over \left(\sum_{i=1}^p\vk_i\right)^2\left(\sum_{i=p+1}^n\vk_i\right)^2}\right]=0.&
\ea
Quand on prend les moyennes angulaires les produits scalaires simples
tombent tandis que les produits carr\'es donnent $1/3$. On obtient
donc,
\ba
n\,\nu_n+\mu_n+\sum_{p=1}^n\nu_p\,\mu_{n-p}&=&0;\\
\left(n+{1\over2}\right)\mu_n+{3\over2}\nu_n+{1\over 3}
\sum_{p=1}^n\mu_p\,\mu_{n-p}&=&0.
\ea
En termes de fonctions g\'en\'eratrices cela implique que,
\ba
\tau{\d \over \d \tau}\mGd(\tau)+\mGt(\tau)\left[1+\mGd(\tau)\right]&=&0;\\
\tau{\d \over \d \tau}\mGt(\tau)+{1\over2}\mGt(\tau)+
{3\over2}\mGd(\tau)+{1\over 3}\mGt^2(\tau)&=&0.
\ea
On obtient un syst\`eme d'\'equations diff\'erentielles
coupl\'ees. Apr\`es quelques manipulations on peut montrer
que l'\'equation d'\'evolution du $\mGd(\tau)$
se r\'eduit \`a celle de l'effondrement sph\'erique
o\`u $\mGd(\tau)$ est le contraste de densit\'e de la structure
en effondrement et $\tau$ est la sur-densit\'e lin\'eaire.
D\`es lors il facile de calculer les $\nu_n$ \`a un 
ordre arbitraire,
\ba
\nu_2&=&{34\over 21};\\
\nu_3&=&{682\over189};\\
\nu_4&=&{446440\over 43659};\\
&&\dots
\ea
Les valeurs des $S_n$ en d\'ecoulent tout de suite,
\ba
S_3&=&{34\over7};\\
S_4&=&{60\,712\over1\,323}\approx 45.89;\\
S_5&=&{200\,575\,880\over 305\,613}\approx 656.3;\\
&&\dots
\ea
C'est int\'eressant mais il reste que ces r\'esultats ne tiennent pas
compte de l'effet de filtrage.

\subsection{ La fonction g\'en\'eratrice des cumulants de la
densit\'e filtr\'ee}

\subsubsection{Propri\'et\'es g\'eom\'etriques dans l'espace
Lagrangien}

La quantit\'e de base en espace Lagrangien est le Jacobien.
Je peux toujours imaginer un d\'eveloppement perturbatif
de $J$,
\be
J(\vq)=J^{(1)}(\vq)+J^{(2)}(\vq)+\dots
\ee
A un ordre donn\'e on aura,
\be
J^{(p)}(\vq)=\int{\d^3\vk_1\over (2\pi)^{3/2}}\dots
{\d^3\vk_p\over (2\pi)^{3/2}}\,a^p\,J_p(\vk_1,\dots,\vk_p)
\exp[\ii\vq.(\vk_1+\dots+\vk_p)]
\ee
La propri\'et\'e remarquable que l'on peut montrer est que
(Bernardeau 1994c)
\ba
j_p&\equiv&\overline{J_p(\vk_1,\dots,\vk_p)\,
W(\vert\vk_1+\dots+\vk_p\vert R)}^{\rm angles}\\
&=&
\overline{J_p(\vk_1,\dots,\vk_p)\,}^{\rm angles}\,
W(k_1\,R)\dots W(k_p\,R).\label{jpcom}
\ea
La d\'emonstration compl\`ete est assez ardue. Elle repose sur
les remarques suivantes. Le Jacobien s'exprime essentiellement
\`a partir de deux fonctions g\'eom\'etriques dans une construction
r\'ecursive. En effet 
\ba
J(\vq)&\equiv&\vert{\partial\vx\over\partial \vq}\vert\nonumber\\
&=&
1+\gradq.\Psi+{1\over 2}\left[\left(\gradq.\Psi\right)^2-
\sum_{ij}\Psi_{i,j}\Psi_{j,i}\right]\nonumber\\
&&+{1\over 6}
\left[\left(\gradq.\Psi\right)^3-3\gradq.\Psi\sum_{ij}\Psi_{i,j}\Psi_{j,i}
+2 \sum_{ijk}\Psi_{i,j}\Psi_{j,k}\Psi_{k,i}\right].
\ea
On a d\'ej\`a vu que le filtre top-hat en quelque
sorte commute avec $1-(\vk_1.\vk_2)^2/k_1^2/k_2^2$.
On peut aussi montrer la propri\'et\'e suivante,
\ba
&\int\d \Omega_1\d\Omega_2\d\Omega_3\
\WTHt\nonumber\\
&\times\left[1-\left({\vk_1.\vk_2\over k_1 k_2}\right)^2
-\left({\vk_2.\vk_3\over k_2 k_3}\right)^2
-\left({\vk_3.\vk_1\over k_3 k_1}\right)^2
+2\,{\vk_1.\vk_2\ \vk_2.\vk_3\ \vk_3.\vk_1\over k_1^2 k_2^2 k_3^2
}\right]
\nonumber\\
&=(4\pi)^3\,{2\over 9}\,W(k_1\,R)\ W(k_2\,R)\,W(k_3\,R).
\ea
L\`a encore, tout se passe comme si on prenait le r\'esultat
obtenu pour le jacobien ponctuel calcul\'e \`a partir du champ de
densit\'e initial tronqu\'e.

Il est relativement facile de voir que la fonction $J_p$
n'est construite r\'ecursivement qu'avec de telles fonctions
g\'eom\'etriques. La propri\'et\'e (\ref{jpcom}) s'obtient alors
par applications successives de ces propri\'et\'es g\'eom\'etriques.

Ce qu'on vient d'obtenir g\'en\'eralise ce que l'on avait
vu pour $S_3$ o\`u en effet $S_3$ dans l'espace Lagrangien 
n'\'etait pas affect\'e par le filtrage (pour un filtre top-hat
en tout cas).

\subsubsection{Le passage Lagrangien-Eul\'erien}

L'id\'ee est maintenant de trouver un truc pour r\'ecup\'erer
le terme dominant des $S_p$ \`a moindre frais \`a partir des
$S_p$ de l'espace Lagrangien.

La hi\'erarchie pr\'ec\'edente me donne des informations 
sur la fonction g\'en\'eratrice des cumulants de la distribution
de volume \`a \'echelle de masse fixe.
On peut alors faire la remarque suivante: la probabilit\'e
qu'une masse $M$ occupe un volume sup\'erieur \`a $V$
est aussi la probabilit\'e qu'un volume $V$ contienne
une masse inf\'erieure \`a $M$. Il suffit pour cela de consid\'erer
des sph\`eres concentriques autour d'un point $\vx_0$
donn\'e\footnote{La proposition n'est cependant rigoureuse que pour des
probabilit\'es dites centr\'ees}. Or une de ces probabilit\'es
s'exprime en espace Lagrangien, l'autre en espace Eulerien.
\`A travers cette proposition on peut alors trouver le comportement
dominant de la fonction g\'en\'eratrice des cumulants $\varphi(y)$
dans la limite des petites variances \`a partir de celle obtenue
pour l'espace lagrangien qui s'identifie \`a celle obtenu sans
tenir compte des effets de filtrage.

Rappelons d'abord que
\be
p(\delta)\d\delta=\int_{-\ii\infty}^{+\ii\infty}
{\d y\over 2\pi\ii\sigma^2}\exp\left[-{\varphi(y)\over \sigma^2}
+{y\dta\over \sigma^2}\right] \d\delta.
\ee
Il est alors int\'eressant de faire un calcul succinct 
par la m\'ethode du col qui est a priori  valable
dans la limite des petites variances.
On voit que la position du point col est donn\'ee par
l'\'equation
\be
{\d\over \d y}\varphi(y)=\delta,
\ee
et par ailleurs on a
\be
{\d\over \d y}\varphi(y)=\mGd(\tau),
\ee
quand $\tau$ est toujours donn\'ee par la m\^eme
relation implicite avec $y$. Autour de ce point col la 
partie sous l'exponentielle est simplement donn\'ee par
\be
\exp\left(-{\tau^2\over 2\,\sigma^2}\right).
\ee
La position du point col est donc essentiellement donn\'ee par un simple
changemement de variable passant de la densit\'e lin\'eaire $\tau$
au contraste densit\'e nonlin\'eaire $\delta$.

La diff\'erence entre une description Lagrangienne est que
$\sigma$ est pris dans un cas \`a une \'echelle de masse donn\'ee,
dans l'autre \`a une \'echelle physique $R$ donn\'ee.
Comme je cherche la partie dominante de $\varphi$ dans la limite
$\sigma\to0$ il est parfaitement l\'egitime de s'appuyer
sur cette approximation.

Pour rendre le calcul plus simple, je vais pr\'eciser un peu mes
notations. Je note $\varphi^L$ la fonction g\'en\'eratrice des
cumulants dans l'espace Lagrangien et $\mGd^L$ la fonction
g\'n\'eratrice des vertex. Dans l'espace Eul\'erien j'utilise
l'exposant $E$. Il est toujours possible de supposer qu'il existe
une fonction $\mGd^E$ associ\'ee \`a $\varphi^E$ (m\^eme
si il n'y a pas de repr\'esentation diagrammatique \`a partir de
vertex qui lui soit associ\'ee). La variable $\tau$ est de m\^eme prise
avec l'indice $L$ ou $E$ selon qu'elle concerne l'espace Lagragien
ou Eulerien.

Par identification des termes sous l'exponentielle
(les seuls qui comptent dans la limite consid\'er\'ee), on a
\be
-{\tau_E^2\over 2\sigma^2(R)}=
-{\tau_L^2\over 2\sigma^2\left[(1+\delta)^{1/3}\,R\right]}.
\ee
Le contraste de densit\'e est un param\`etre donn\'e a priori.
Les variables $\tau_E$ et $\tau_L$ d\'ependent de $\delta$
\`a travers les \'equations du col,
\ba
\mGd^L(\tau_L)&=&\delta,\\
\mGd^E(\tau_E)&=&\delta.
\ea
Dans le cas Lagrangien $\sigma$ (fonction de l'\'echelle
\`a priori connue) est pris \`a l'\'echelle de masse correspondant
\`a la surdensit\'e $\delta$ consid\'er\'ee. 
La fonction $\mGd^L(\tau_L)$ est connue. C'est elle qui d\'ecrit
la dynamique de l'effondrement sph\'erique.

\`A partir ce ces \'equations je peux \'eliminer $\tau_L$ pour
obtenir une \'equation implicite entre $\mGd^E$ et $\tau_E$,
\be
\mGd^E(\tau_E)=\mGd^L\left({\sigma\left[(1+\mGd^E(\tau_E))^{1/3}\,R\right]\over
\sigma(R)}\tau_E\right).
\ee
La fonction g\'en\'eratrice des cumulants $\varphi^E(y)$ 
est alors construite \`a partir de $\mGd^E(\tau_E)$ de la m\^eme 
mani\`ere que $\varphi^L(y)$ l'\'etait \`a partir de $\mGd^E(\tau_L)$.

En d\'eveloppant cette fonction autour de $y=0$ on a alors les
expressions explicites des prmi\`eres valeurs des $S_p$. Ils
s'expriment en fonction des d\'eriv\'ees logarithmiques successives
de la variance,
\be
\gamma_p\equiv{\d^p\log\sigma^2(R)\over\d\log^p R},
\ee
et alors,
\ba
S_3&=&{{34}\over 7} + {\gm_1},\\
S_4&=& {{60712}\over {1323}} + {{62\,{\gm_1}}\over 3} + 
   {{7\,{{{\gm_1}}^2}}\over 3} + {{2\,{\gm_2}}\over 3},\\
S_5&=& {{200575880}\over {305613}} + {{1847200\,{\gm_1}}\over {3969}} + 
   {{6940\,{{{\gm_1}}^2}}\over {63}} + {{235\,{{{\gm_1}}^3}}\over {27}}\nonumber\\
&&\ \ + {{1490\,{\gm_2}}\over {63}} + {{50\,{\gm_1}\,{\gm_2}}\over 9} + 
   {{10\,{\gm_3}}\over {27}},\\
S_6&=&12650 + 12330\,{\gm_1} + 4512\,{{{\gm_1}}^2} + 
   734.0\,{{{\gm_1}}^3} + 44.81\,{{{\gm_1}}^4}\nonumber\\
&&\ \  + 775.8\,{\gm_2} + 375.9\,{\gm_1}\,{\gm_2} + 
   45.56\,{{{\gm_1}}^2}\,{\gm_2} + 3.889\,{{{\gm_2}}^2}\nonumber\\
&&\ \  + 20.05\,{\gm_3} +  4.815\,{\gm_1}\,{\gm_3} + 0.1852\,{\gm_4},\\
S_7&=& 307810 + 383000\,{\gm_1} + 190700\,{{{\gm_1}}^2} + 
   47460\,{{{\gm_1}}^3}\nonumber\\
&&\ \  + 5914\,{{{\gm_1}}^4} +   294.8\,{{{\gm_1}}^5} + 27340\,{\gm_2}+   
   20300\,{\gm_1}\,{\gm_2}\nonumber\\
&&\ \  + 5026\,{{{\gm_1}}^2}\,{\gm_2} + 414.8\,{{{\gm_1}}^3}\,{\gm_2} + 
   358.1\,{{{\gm_2}}^2} +    88.15\,{\gm_1}\,{{{\gm_2}}^2}\nonumber\\
&&\ \  + 902.6\,{\gm_3} +  443.3\,{\gm_1}\,{\gm_3} + 
   54.44\,{{{\gm_1}}^2}\,{\gm_3} + 7.778\,{\gm_2}\,{\gm_3}\nonumber\\
&&\ \  + 14.20\,{\gm_4} + 3.457\,{\gm_1}\,{\gm_4}+ 0.08642\,{\gm_5},\\
&&\dots\nonumber
\ea

La figure (\ref{figSpGazta}) montre que ces coefficients sont en
parfait accord avec des r\'esultats de simulations num\'eriques
(ici pour un spectre CDM) d\`es lors que l'\'echelle de filtrage est
plus grande que $10h^{-1}$ Mpc. Il est assez remarquable de voir que
le domaine de validit\'e de ces r\'esultats ne se d\'et\'eriore pas
quand on augmente l'ordre des cumulants.

\subsection{ Application, construction de la distribution de
probabilit\'e de densit\'e locale}

Je vais reprendre la relation entre la fonction de distribution
de probabilit\'e et la fonction g\'en\'eratrice $\varphi(y)$.
Pour pouvoir utiliser cette relation 
il faut faire une hypoth\`ese suppl\'ementaire
non triviale. En effet $\varphi(y)$ est a priori une fonction de
$\sigma$  \`a travers la d\'ependance des $S_p$ avec $\sigma$.
Je vais alors supposer que la limite,
\be
\varphi(y,\sigma)\to\varphi(y)\ \ {\rm quand}\ \ \sigma\to 0,\label{PhiUni}
\ee
est {\rm uniforme}. Il n'y a pas
de preuve de cette propri\'et\'e, cependant elle est fortement
sugg\'er\'ee par les r\'esultats num\'eriques sur les $S_p$.

Quand la variance est suffisemment faible il est alors l\'egitime
de supposer que
\be
p(\delta)\d\delta=\int_{-\ii\infty}^{+\ii\infty}
{\d y\over 2\pi\ii\sigma^2}\exp\left[-{\varphi(y)\over \sigma^2}
+{y\dta\over \sigma^2}\right] \d\delta.
\ee
Je vais maintenant consid\'erer s\'erieusement cette \'equation
et rechercher les diff\'erentes formes de $p(\delta)$ qui en
r\'esultent (voir Balian \& Schaeffer 1989 pour une description plus
exhaustive des r\'esultats). Profitant de l'approximation $\sigma\ll0$ on peut
appliquer la m\'ethode du col, et on obtient 
\ba
p(\delta)\d\delta&=&{\d\delta\over -\mG_{\delta}'(\tau)}
\left[{1-\tau\mG_{\delta}''(\tau)/\mG_{\delta}'(\tau) 
\over 2\pi \sigma^2 }\right]^{1/2}\ 
\exp\left(-{\tau^2\over 2\sigma^2}\right),\\
\mG_{\delta}(\tau)&=&\dta.
\ea
\begin{table}
\caption{Valeur des param\`etres de la singularit\'e (\ref{phising}) pour
diff\'erentes valeurs de l'indice spectral $n$.}
\begin{tabular}{ccccc}
$n$&$\dta_c$&$y_s$&$a_s$&$\varphi_s$\\
-3  &0.656&-0.184&1.84 &-0.030\\
-2.5&0.804&-0.213&2.21 &-0.041\\
-2  &1.034&-0.253&2.81 &-0.058\\
-1.5&1.44 &-0.310&3.93 &-0.093\\
-1  &2.344&-0.401&6.68 &-0.172\\
-0.5&5.632&-0.574&18.94&-0.434
\end{tabular}
\label{tabSing}
\end{table}
Cette solution est valable quand $\dta\le\dta_c$
o\`u $\dta_c$ est la valeur du contraste de densit\'e
qui annule $1-\tau\mG_{\dta}''(\tau)/\mG_{\dta}'(\tau)$.
Quand $\dta$ est plus grand que $\dta_c$ on ne peut plus faire
l'approximation du point col. La forme de $p(\delta)$ est alors
plut\^ot d\'etermin\'ee par le comportement de $\varphi(y)$
pr\`es de sa singularit\'e,
\be
\varphi(y)-\varphi_s\simeq -a_s(y-y_s)^{3/2},\label{phising}
\ee
et on a
\be
p(\dta)\d\dta={3a_s\sigma \over 4\sqrt{\pi}}
(1+\dta)^{-5/2}\,
\exp\left[{-\vert y_s \vert \dta/\sigma^2+\vert \varphi_s 
\vert/\sigma^2}\right]\d\dta.
\ee
On voit que la forme du cut-off est tr\`es diff\'erente d'une forme
gaussienne. Il faut bien avoir \`a l'esprit que ce r\'esultat repose
en grande partie sur l'hypoth\`ese (\ref{PhiUni}), et 
en particulier que la position de la singularit\'e doit
rester \`a une distance finie de l'origine quand $\sigma$ est finie.
La table (\ref{tabSing}) donne les param\`etres de la singularit\'e
pour diff\'erentes valeurs de l'indice spectral. Je rappelle que
$n\approx -1.5$ est une valeur sugg\'er\'ee par les comptages d'amas.

\begin{figure}
\vspace{7 cm}
\special{hscale=70 vscale=70 voffset=0 hoffset=0 psfile=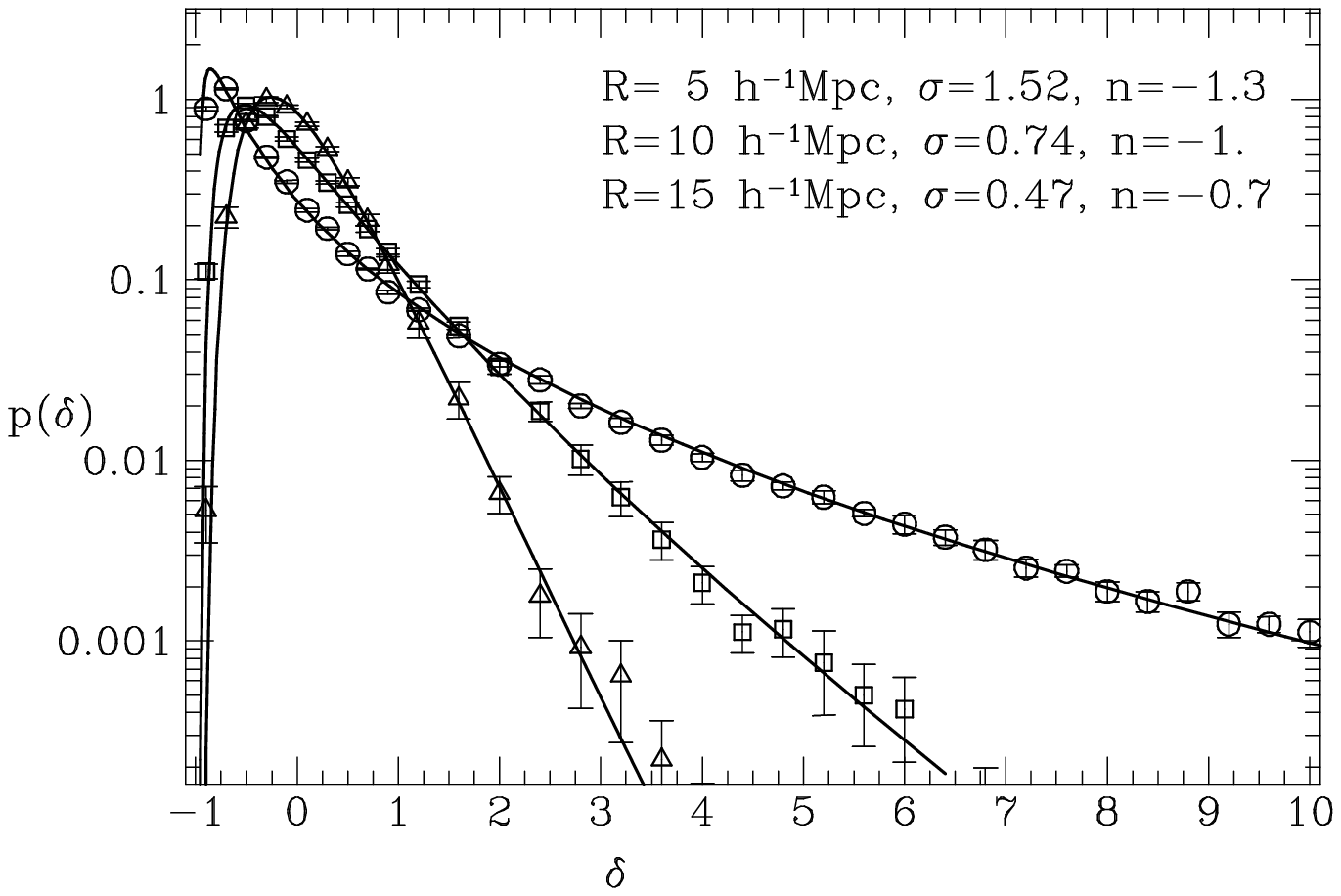}
\caption{Comparaisons des r\'esulats de simulations num\'eriques $N$
corps avec les pr\'edictions de la th\'eorie perturbative.}
\label{PDFnum}
\end{figure}

Num\'eriquement il est toujours possible d'int\'egrer sans faire
explicitement l'approximation du col. Les formes obtenues
sont en bon accord avec les r\'esultats de simulations
num\'eriques $N$-corps. Sur la figure
(\ref{PDFnum}) on a des fonctions de distribution pour diff\'erentes
\'echelles de filtrage, donnant diff\'erentes valeurs de la
variance. La forme pr\'edite pour la distribution
(calcul\'ee \`a partir de la donn\'ee de la variance et
de l'indice spectral local) est en remarquable accord
avec les r\'esultats num\'eriques.

\subsection{Les termes sous-dominants du d\'eveloppement perturbatif}

\begin{figure}
\vspace{2 cm}
\special{hscale=70 vscale=70 voffset=0 hoffset=0 psfile=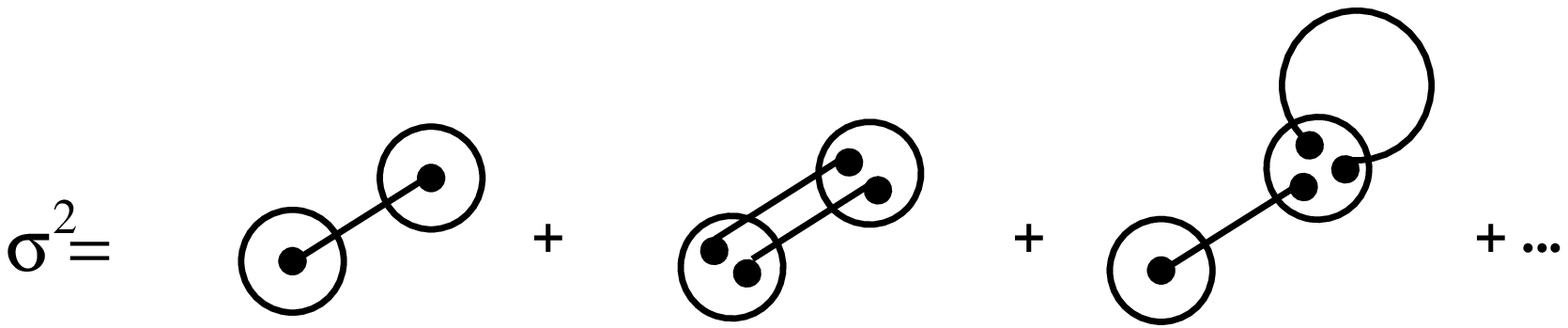}
\caption{Diagrammes \`a une boucle contribuant \`a la variance}
\label{sigLoops}
\end{figure}

\begin{figure}
\vspace{5.5 cm}
\special{hscale=70 vscale=70 voffset=0 hoffset=0 psfile=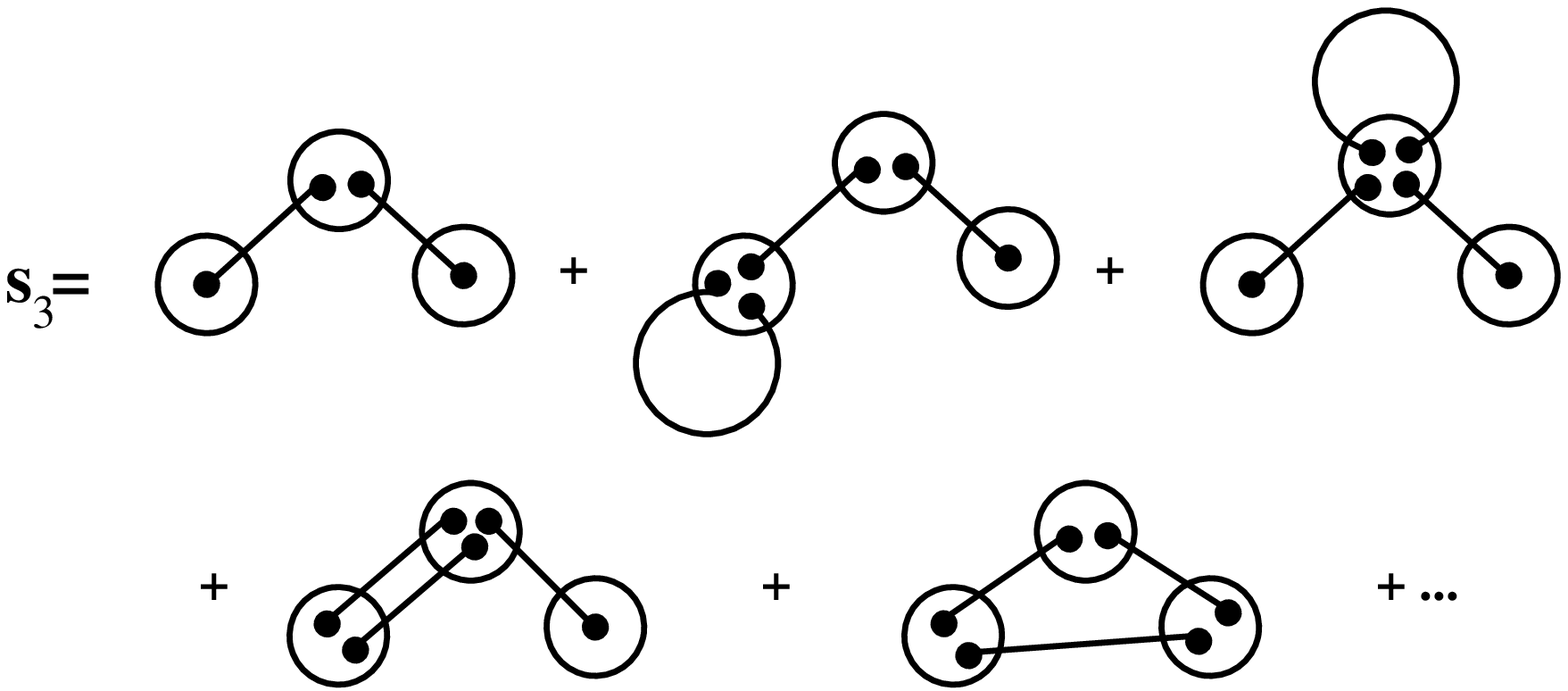}
\caption{Diagrammes \`a une boucle contribuant \`a $S_3$.}
\label{s3Loops}
\end{figure}

Les quantit\'es que l'on vient de calculer sont les termes dominants
(contributions en arbre) d'un d\'eveloppement plus g\'en\'eral.
On peut toujours se demander si les corrections en boucles permettent
d'\'etendre le domaine de validit\'e de la th\'eorie des
perturbations. Par exemple pour les termes correctifs \`a la variance
on a deux diagrammes suppl\'ementaires (voir figure \ref{sigLoops}), 
pour la skewness on en a quatre (figure \ref{s3Loops}).

C'est un domaine qui a \'et\'e relativement peu explor\'e.
On sait donc peu de chose sur ces termes correctifs.
La raison principale en est sans doute que les calculs sont
horriblement compliqu\'es. On ne peut plus simplement moyenner
sur les angles, et les effets de filtrage rendent les calculs
encore plus inextricables.

\begin{figure}
\vspace{12 cm}
\special{hscale=70 vscale=70 voffset=-100 hoffset=-50 psfile=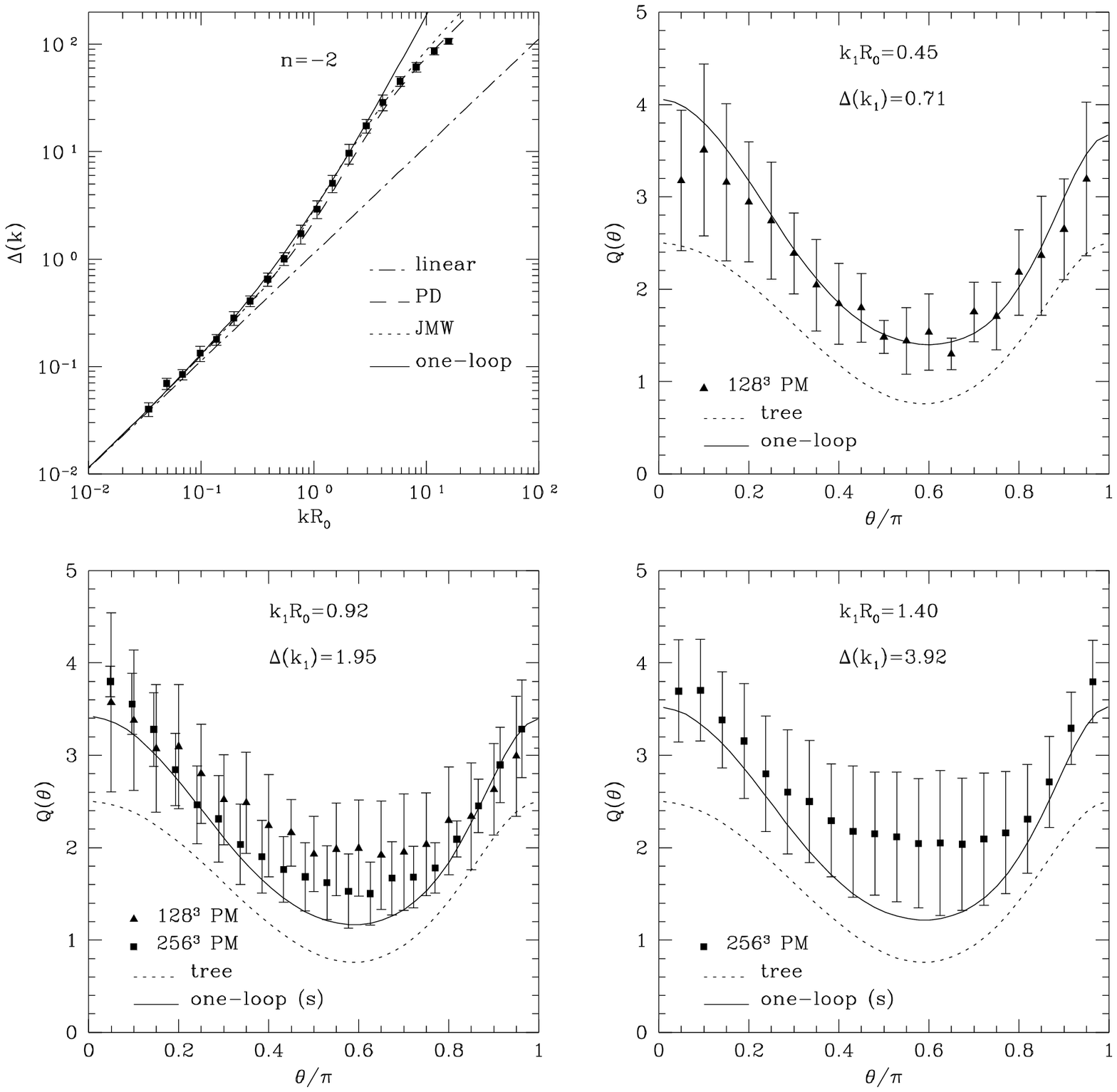}
\caption{R\'esultats de la th\'eorie des perturbations \`a une boucle
pour la variance et le bi-spectrum pour le cas $n=-2$. On voit que
les termes en boucles reproduisent correctement les r\'esultats 
num\'eriques. (figure extraite de Scoccimarro et al. 98)}
\label{Scoccim2}
\end{figure}

Les r\'esultats obtenus sur la variance et la skewness montrent que
(Scoccimarro \& Frieman 1996, Scoccimarro 1997, Scoccimarro et al. 1998)
\begin{itemize}
\item Les termes correctifs ne convergent que si l'indice spectral
$n$ est compris entre -3 et -1. A grande \'echelle \c ca ne converge
donc pas. Ce sont les 'petites boucles' qui induisent ces divergences.
\item  Quand on a convergence, on a une bonne description de
la transition vers le nonlin\'eaire (voir figure \ref{Scoccim2}).
\end{itemize}

En pratique on a un cut-off naturel parce que $n\to -3$ \`a petite
\'echelle. Il n'en reste pas moins qu'on ne sait pas si ce
cut-off doit \^etre explitement introduit pour r\'egulariser
les r\'esultats ou si une resommation (au moins partielle)
des boucles donnerait une contribution finie (mais \'eventuellement
avec une autre d\'ependance en $\sigma$).